\documentclass[a4paper,11pt]{article}
\pdfoutput=1 

\usepackage{jheppub} 

\usepackage[T1]{fontenc} 
\usepackage{amsmath}
\usepackage{comment}
\usepackage{float}
\usepackage[dvipsnames,table,xcdraw]{xcolor}
\usepackage{mathrsfs}
\usepackage{subcaption}
\usepackage{caption}
\usepackage{listings}
\newcommand{\be}{\begin{equation}}
\newcommand{\ee}{\end{equation}}
\newcommand{\bea}{\begin{equation}\begin{aligned}}
\newcommand{\eea}{\end{aligned}\end{equation}}

\newcommand{\oT}{\overline{T}}

\usepackage[normalem]{ulem}

\newcommand{\Ai}{\text{Ai}}

\title{$T\overline{T}$-deformed free energy of the Airy model}

\author[a]{Stephen Ebert,}
\author[b]{Hao-Yu~Sun}
\author[c]{and Zhengdi Sun}

\affiliation[a]{Mani L. Bhaumik Institute for Theoretical Physics,\\ University of California, Los Angeles, CA 90095-1547, USA}

\affiliation[b]{Department of Physics, University of Texas, Austin, TX 78712-1192, USA}

\affiliation[c]{Department of Physics, University of California, San Diego, CA 92093-0931, USA}

\emailAdd{stephenebert@physics.ucla.edu}
\emailAdd{hkdavidsun@utexas.edu}
\emailAdd{z5sun@ucsd.edu}

\abstract{Sharpening the correspondence of Jackiw-Teitelboim (JT) gravity and its dual matrix model description at a finite radial cutoff $\lambda$ through the $T\overline{T}$ deformation is of interest. To proceed, we simplify the problem by considering the Airy model and deform Airy correlators in the same way as in $T\oT$-deformed JT gravity. We use those correlators to compute the annealed and quenched free energies for both $\lambda > 0$ and $\lambda < 0$ from an integral representation of the replica trick. At the leading order in $\lambda$ and low temperatures, we confirm that the genus-zero quenched free energy monotonically decreases as a function of temperature when perturbation theory is valid. We then study the all-genus quenched free energy at low temperatures, where we discover and discuss subtleties due to non-perturbative effects in the Airy model, as well as the contributions from the non-perturbative branch under the $T\oT$ deformation.}

\begin{document} 
\maketitle
\flushbottom
\newpage
\section{Introduction}
An interesting question to ask is: does the AdS/CFT correspondence hold in a finite patch in spacetime? This question has thoroughly been investigated through the lens of the irrelevant $T\overline{T}$ deformation. Recently, the $T\overline{T}$ deformation \cite{Zamolodchikov:2004ce,Smirnov:2016lqw,Cavaglia:2016oda} has attracted much attention due to being a well-defined operator with complete solvability along the RG flow. Historically, unlike relevant and marginal deformations, irrelevant deformations modify the UV behavior of a theory and are mathematically challenging to determine their UV behavior along the RG flow as generically infinitely many operators are included. 

Given a seed action, for example Euclidean action $S_{\operatorname{E}}(0)$, the $T\overline{T}$ deformation is defined by point-splitting up to total derivatives of local operators\footnote{For a modern review of the $T\overline{T}$ deformation regarding applications to holographic systems as well as several other scenarios, see \cite{Jiang:2019epa,MG_CERN} and references therein.}
\begin{equation}
\label{definition}
   \frac{\partial S_{\text{E}}}{\partial \lambda}=8 \int d^{2} x\, \sqrt{\gamma} T \overline{T}\,,
\end{equation}
where the right-hand-side depends on the determinant of the two-dimensional stress tensor $T_{ij}(\lambda)$ of the deformed seed theory, and $T_{ij}(\lambda)$ is related to its trace via
\begin{equation}
 \label{Trace}
\operatorname{Tr} T_{ij} =-16 \lambda T \bar{T}=-2 \lambda\left(T_{i j} T^{i j}-\left(T_{i}^{i}\right)^{2}\right)\,.
\end{equation}

An application of the $T\overline{T}$ deformation \eqref{definition} relevant to this paper is probing the low temperature limit of JT gravity and its Airy model description.\footnote{Hereafter, we will refer to the Airy limit of Gaussian matrix models as the ``Airy model''.} We will now go over the salient features of JT gravity in the low temperature limit, and then discuss more of the $T\oT$ deformation in holography. 

JT gravity \cite{Jackiw:1984je, Teitelboim:1983ux} is a special case of two-dimensional dilaton gravity, and on AdS$_2$ which is of concern in this paper\footnote{On (nearly) dS$_2$, ``$+2$'' in \eqref{eq:action} will become ``$-2$''. The $T\overline{T}$ deformation in dS$_3$ has been studied recently by \cite{Gorbenko:2018oov,Lewkowycz:2019xse,Shyam:2021ciy,Coleman:2021nor} where the trace flow equation \eqref{Trace} is modified by an additive term $\propto \frac{1}{\lambda}$.}, it is described by the action in Euclidean signature with boundary terms 
\begin{equation}
\label{eq:action}
 S_{\mathrm{E}} =-\frac{1}{16 \pi G} \int_M d^2x \sqrt{g} \Phi\,(R+2)+ S_{\partial M}\,,
\end{equation}
where the dual description is a Hermitian random matrix model shown by Saad, Shenker and Stanford \cite{Saad:2019lba}. Here $R$ fixed to be $-2$ is the Ricci scalar of the metric $g_{\mu \nu}$, $\Phi$ is the dilaton, and $S_{\partial M}$ is the one-dimensional Gibbons-Hawking-York boundary action.

To further motivate the Airy model, we list a few reasons why it is interesting and useful to study. Firstly, the genus expansion can be summed via \cite{Okounkov:2001usa} allowing one to make definite statements on nonperturbative corrections. Secondly, JT gravity at low temperatures, or more precisely the 't Hooft limit:
\begin{equation}
\label{eq:'t Hooft}
    \hbar \rightarrow 0, \quad \beta \rightarrow \infty, \quad \operatorname{for ~ fixed}~ \hbar \beta\,,
\end{equation}
is dual to the Airy model at all genus known from \cite{Okuyama:2020ncd}. Here the authors of \cite{Okuyama:2020ncd} showed all the non-trivial information of the spectral curve of JT gravity is still preserved under the 't Hooft limit \eqref{eq:'t Hooft}. Additionally, there is an important caveat as \cite{Saad:2019lba} finds: the exact eigenvalue density has an exponential leakage when $E < 0$ which is denoted as the ``classically forbidden'' region making the system unstable. This non-perturbative instability is not special for the Airy model as the same phenomena occurs for the matrix model dual of JT gravity. However, a recent proposal by \cite{Johnson:2019eik} improves the non-perturbative behavior of JT gravity by removing the non-perturbative instabilities. Therefore, one should only expect this relation between JT gravity and the Airy model confidently holds perturbatively. Finally, there are subtleties for the Airy model's quenched free energy not monotonically decreasing as function of temperature using directly the replica trick as done by \cite{Engelhardt:2020qpv}. Fortunately, Okuyama \cite{Okuyama:2021pkf} showed this failure of monotonicity arises due to analytical continuation issues in the correlators $\langle Z^n\rangle$ to $\langle Z^{n=0}\rangle$ and proposed an alternative formulation of the replica trick to correctly give a monotonically decreasing quenched free energy in the low temperature limit after summing over all genus. With these motivations of the Airy model, we wish to investigate how some of these features change under the $T\oT$ deformation with a holographic picture in mind to understand JT gravity and its matrix model dual.

The first study of probing holographic systems under the $T\overline{T}$ deformation was initiated by \cite{McGough:2016lol} in AdS$_3$ gravity and inspired several follow-up investigations which computed partition functions and correlators of the gravitational and boundary field theory \cite{Kraus:2018xrn,Dubovsky:2018bmo,Datta:2018thy,Aharony:2018bad,Hartman:2018tkw,Caputa:2019pam,Cardy:2019qao,He:2019vzf,He:2019ahx,Mazenc:2019cfg,Brennan:2020dkw,He:2020udl,Ebert:2020tuy,Caputa:2020lpa,He:2020qcs,Hirano:2020ppu,Ebert:2022cle}, improving the $T\overline{T}$-deformed holographic correspondence. More specifically, through the lens of the $T\overline{T}$ deformation, it is interesting to probe observables between JT gravity and its double-scaled matrix model dual description at a finite cutoff governed by $\lambda$, as done in \cite{Gross:2019ach,Gross:2019uxi,Iliesiu:2020zld,Stanford:2020qhm,Rosso:2020wir,Griguolo:2021wgy}. The purpose of this work is to further sharpen the correspondence between JT gravity at low temperatures and its Airy model description by computing the correlators and quenched free energy. We now comment on the deformed energy spectrum in JT gravity.

Gross et al. \cite{Gross:2019ach,Gross:2019uxi} confirmed that the $T\overline{T}$-deformed energy spectrum of JT gravity and its dual one-dimensional Schwarzian quantum mechanics match.\footnote{JT gravity can be written in terms of a BF gauge theory. The deformed BF formalism and related supersymmetric quantum mechanics was found in \cite{Ebert:2022xfh,Ebert:2022ehb}.} Here the 2D definition in \eqref{definition} and \eqref{Trace} is dimensionally reduced via $T_{\tau \phi} = 0$ to yield the following differential equation for the one-dimensional stress scalar $T^\tau_\tau = f^{\pm}_\lambda (E)$:\footnote{An alternative perspective on deriving this flow equation may be found through the Wheeler-de Witt equation derived in \cite{Hartman:2018tkw} for higher-dimensional large-$N$ CFTs. For two-dimensions, results in \cite{Gross:2019ach} match \eqref{energyspectra} from the Wheeler-de Witt equation perspective. For family of dilaton gravity theories which are more general than JT gravity, e.g., \cite{Grumiller:2007ju}, the authors of  \cite{Grumiller:2020fbb} derived the deformed energy spectrum at a finite cutoff.}
\begin{equation}
\label{f(E)}
   \frac{\partial f^\pm_\lambda (E)}{\partial \lambda} = \frac{2f^\pm_\lambda (E)^2}{1-4\lambda f^\pm_\lambda(E)}
\end{equation}
with two solutions
\begin{equation}
\label{energyspectra}
    f^{\pm}_\lambda(E) = \frac{1\pm \sqrt{1-8\lambda E}}{4\lambda}\,,
\end{equation}
where $\lambda \in \mathbb{R}$ and $E$ is the undeformed energy.

As can be seen in the deformed energy spectrum \eqref{energyspectra}, the sign of $\lambda > 0$ violates unitarity when the undeformed energy is $E > 1/8\lambda$. \footnote{Throughout this paper, we frequently use the terminology ``bad sign'' for $\lambda > 0 $ as the deformed energy spectrum is complex-valued when $E > 1/8\lambda$. The ``good sign'' corresponds to $\lambda < 0$ and the deformed energy spectrum is always real for all energy values.} The authors of \cite{Iliesiu:2020zld} carefully dealt with this violation of unitarity and restored it in their rigorous non-perturbative treatment for the deformed partition function from a Wheeler-de Witt wavefunctional perspective. In short, one writes a linear combination of the wavefunctional of the two branches \eqref{energyspectra} such that the density of states stays real for all energies. 

Other treatments of the deformation parameter $\lambda$'s sign are addressed by \cite{Rosso:2020wir} from a double-scaled matrix model perspective of JT gravity with an attempt to define the dual deformed matrix model description at a finite cutoff. Unfortunately, the analysis of \cite{Rosso:2020wir} was unable to match the $T\overline{T}$-deformed correlators between JT gravity and the dual matrix model, but did provide several alternative methods on how one could properly make this correspondence well-defined. We will discuss the importance of \cite{Iliesiu:2020zld} and \cite{Rosso:2020wir} more throughout various places of the paper.\footnote{A recent proposal in \cite{He:2022bbb} successfully matched the $T\overline{T}$-deformed partition functions of $\mathcal{N} = 1$ type 0A and type 0B JT supergravity with the associated matrix models. Additional evidence of the duality from \cite{He:2022bbb} was calculating the deformed $T\overline{T}$-deformed matrix model correlators via topological recursion relations.}

Having laid out the motivations for studying the $T\overline{T}$ deformation in JT gravity and its dual matrix model, we will now summarize results found in this paper.

\section*{Summary of results and outline}
It is helpful to point out in this paper, we will think about JT gravity and its $T\oT$ deformation from the point of view of the matrix model description. This is because, as shown in \cite{Saad:2019lba}, higher topology contributions in JT gravity are captured entirely by a double-scaled matrix model. The partition function of JT gravity at any genus and with any number of boundaries can be computed from the correlators in its dual matrix model. From the point of view of boundary theory, we only know that JT gravity on a disk is dual to Schwarzian theory on the boundary. Also, the result of \cite{Saad:2019lba} showed the connected $n$-point function in JT gravity -- namely the partition function on a connected surface with $n$ boundaries -- does not factorize. The boundary theory was shown to be an ensemble of theories, not a single theory. Since the replica trick is essential to compute quenched free energy and requires knowledge on $n$-point correlators, we will think mostly from the point of view of matrix model and its correlators rather than directly from the boundary theory.

In \S \ref{review}, we first explain how the $T\overline{T}$-deformed partition function of various topologies in JT gravity are found through an integral transformation of the undeformed partition function. Additionally in \S \ref{review}, we then extend this integral transformation to find a relation between the deformed and undeformed correlators via: 
\begin{equation}
    \langle Z(\beta_1) \cdots Z(\beta_n)\rangle_\lambda = \int_{-\infty}^{\infty} dE_1 \cdots dE_n \, \rho(E_1,E_2,\cdots,E_n) e^{-\beta_1 f^-_\lambda(E_1)} \cdots e^{-\beta_n f^-_\lambda(E_n)}\,,
\end{equation}
where $\rho(E_1,\cdots,E_n)$ is defined by ensuring
\begin{equation}
    \langle Z(\beta_1) \cdots Z(\beta_n)\rangle_0 = \int_{-\infty}^{\infty} dE_1 \cdots dE_n \, \rho(E_1,E_2,\cdots,E_n) e^{-\beta_1 E_1} \cdots e^{-\beta_n E_n}\,.
\end{equation}
Only when the integration range is $E_i \in \left[\frac{1}{8\lambda},\infty \right)$ \textit{and} $\lambda < 0$, we can use the integration kernel $K(\beta,\beta') = \frac{\beta}{\sqrt{-8 \pi \lambda} \beta'^{\frac{3}{2}}} \exp \left( \frac{(\beta - \beta')^2}{8\lambda \beta'}\right)$ to be reviewed in \S \ref{review} to conveniently compute the deformed correlators:
\begin{equation}
\label{eq:2kernels}
\langle Z (\beta_1) \cdots Z (\beta_n) \rangle_\lambda =  \int^\infty_0  d\beta_1' K(\beta_1, \beta_1')  \cdots \int^\infty_0  d\beta_n' K(\beta_n, \beta_n') \langle Z(\beta_1') \cdots Z(\beta_n') \rangle_0\,,
\end{equation}
where $E\geq \frac{1}{8\lambda}$ and
\be 
\label{eq:trans}
\int_0^\infty d\beta' K(\beta,\beta')e^{-\beta' E} = e^{-\beta f_\lambda^-(E)}\,.
\ee
This is the case when we look at any particular order in the genus expansion. However, when we look at the exact solution of the Airy model, non-perturbative effects there extend the integration range to $E_i \in (-\infty,\infty)$, reflecting the translational invariance of the underlying effective Hamiltonian \cite{Johnson:2019eik}. As we will later encounter the non-perturbative effect of the $T\oT$ deformation, hereafter we  refer to this as the \textit{non-perturbative instability} to distinguish the two. 

With the density of states for the Airy model, the undeformed one-point function is given by
\be \langle Z(\beta) \rangle_{\text{Airy},0} = \int_{-\infty}^{\infty} dE\, \rho_{\text{Airy}}(E)e^{-\beta E}\,,
\ee
where, from \cite{Forrester:1993vtx,Ginsparg:1993is}, the density of states is
\be 
\label{eq:}
\rho_{\text{Airy}}(E) = \Ai'(-E)^2 + E \,\Ai(-E)^2\,,
\ee
and $\Ai(\cdot)$ is the Airy function of the first kind \cite{airy1838intensity}, defined as
\begin{equation}
    \Ai(x)\equiv\frac{1}{\pi}\int_0^\infty \cos\left(\frac{t^3}{3}+xt\right)dt\,.
\end{equation}
In this Airy model, we should use the following 
\be\label{airy1pt} \langle Z(\beta) \rangle_{\text{Airy},\lambda} = \int_{-\infty}^{\infty} dE\, \rho_{\text{Airy}}(E) e^{-\beta f^-_\lambda(E)}\,,
\ee
instead of \eqref{eq:2kernels}, which is invalid because the integral transformation \eqref{eq:trans} of $e^{-\beta E}$ diverges when $E<\frac{1}{8\lambda}$.

From determining the deformed correlators in \S \ref{review}, an immediate application is to calculate the annealed and quenched free energies in JT gravity at low temperatures where certain approximations are feasible. In \S \ref{Airy}, we review a new way of the computing the quenched free energy from an integral representation of replica trick due to Okuyama \cite{Okuyama:2021pkf}:
\begin{equation}
\label{okuyama1}
 \lim _{n \rightarrow 0} \frac{\left\langle Z^{n}\right\rangle-1}{n} =  \ln \langle Z\rangle-\int_{0}^{\infty} \frac{d x}{x}\left[\left\langle e^{-Z x}\right\rangle-e^{-\langle Z\rangle x}\right]\,,
\end{equation}
where $\left\langle e^{-Z x}\right\rangle$ is fully determined by connected correlators $\langle Z^n\rangle_c$. We elaborate later in \S \ref{Airy} why we use this integral representation \eqref{okuyama1} instead of directly using the replica trick directly in the low temperature regime and we comment more on the non-perturbative contributions from the $T\overline{T}$ deformation in \S \ref{sec:comments on non-perturbative}. 

However, it is generally difficult to evaluate the correlators of JT gravity under the $T\oT$ deformation let alone sum over all the connected correlators in order to use Okuyama's formula \eqref{okuyama1}. Since JT gravity is known to have a matrix model dual, we will simplify the problem by studying a simpler matrix model, the Airy model, to make progress. Here, it is important to notice that although the $T\oT$ deformation of random matrix models have been studied in \cite{Rosso:2020wir}, the correlators there do not match those of $T\oT$-deformed JT gravity studied in \cite{Iliesiu:2020zld}. Since ultimately our goal is to understand JT gravity and its dual matrix model at a finite cutoff, we will not follow the $T\oT$ deformation of the matrix model defined in \cite{Rosso:2020wir}. Instead, we simply require the correlators of the Airy model to transform in the same way as the correlators in JT gravity do under the $T\oT$ deformation, and we will take this as a working definition for our version of $T\oT$ deformation applied to the Airy model. We will not attempt to completely explore this version of deformation for matrix models and hope to revisit this problem in future work. 

It is also important to point out one caveat of our simplification. Naively, the double-scaled matrix model dual to JT gravity shares the same non-perturbative instability as the Airy model we considered; for instance, see \cite{Saad:2019lba,Johnson:2019eik}. However, there have been studies on how to improve the non-perturbative behavior of JT gravity and remove this undesired feature \cite{Johnson:2019eik}. Therefore, it would be interesting to extend our work to JT gravity at general temperatures with the improved non-perturbative behavior.

In \S \ref{sec:Airy}, we use \eqref{okuyama1} to numerically evaluate the quenched free energy in the Airy model for both $\lambda > 0$ and $\lambda < 0$. As a warm-up, we first compute the quenched free energy $F_{q,\lambda}(T)$ at genus-zero at the leading order of $\lambda$ in perturbation theory. We confirm that $F_{q,\lambda}(T)$ is a monotonic function in $T$ at low temperature when a leading order approximation is valid. Additionally, we find that the good sign $\lambda < 0$ deformation decreases the quenched free energy while the bad sign $\lambda > 0$ increases it. Intuitively, this sign of the deformation parameter $\lambda > 0$ corresponds to JT gravity in a finite box with Dirichlet boundary conditions at $r_c =\frac{\pi \lambda}{4G}$. The increase of the quenched free energy for the $\lambda > 0$ theory is related to the fact that the $T\overline{T}$ deformation cuts off the spectrum. On the gravity side, one can think of this phenomena as a gravitational redshift. For an object in a gravitational potential, the energy measured at the conformal boundary (which is the undeformed case) is reduced compared to the energy measured at the particle location. For example, in the extreme case where a particle is sent towards a black hole's horizon, the energy at infinity vanishes and is negative when the particle is inside the event horizon. Here, in the cutoff gravitational theory, we are measuring this local energy at the finite cutoff boundary, which is closer to the particle compared to the conformal boundary. Thus the amount the redshift decreases and the energy we measure increases.\footnote{We thank Per Kraus for explaining this gravitational interpretation.}

Next, we use the low temperature approximation $\langle Z(\beta)^n \rangle_c \simeq \langle Z(n\beta) \rangle$ of the Airy model to compute the quenched free energy for finite $\lambda$. Notice that this computation include contributions from all genus as well as non-perturbative effects in the Airy model, which makes $\rho_{\text{Airy}}(E) \neq 0$ even for $E < 0$.

For the bad sign $\lambda > 0$, as studied in \cite{Iliesiu:2020zld,Griguolo:2021wgy}, there are contributions from the non-perturbative branch. We compute the quenched free energy with this branch included and excluded. In both cases, we find the quenched free energy to be divergent and a careful analytical examination can be found in the Appendix \ref{app:badsigndetail}.

For the good sign of $\lambda < 0$, the issue of the complex-valued energy always arises regardless of the value of $\lambda$ since $\rho_{Airy}(E)$ has support on the entire real axis. Although the issue of complex energy of the good sign $\lambda < 0$ has been noticed before in \cite{Zamolodchikov:2004ce} for 2d CFT since the ground state energy is $-\frac{c}{12}$, we emphasize this is different from our case, as the energy spectrum is still bounded below by $-\frac{c}{12}$. One can avoid the complex energy by simply choosing $|\lambda | \leq \frac{3}{2c}$ in this case. However, for the Airy model, the issue of complex energy will always arise regardless of the value of $\lambda$.  This has not been noticed before for the good sign of the $T\overline{T}$ deformation. A simple solution will be to impose a hard cut-off in $E$, namely we simply remove these states with complex-valued energy. However, as we will see in \S \ref{goodsign}, this option would lead to a violation of the $T\overline{T}$ flow equation by a boundary term in the integral. Another option would be to include those states with complex-valued energy, however, to demand the partition function to be real-valued, one must include the other branch as well. By a careful choice of coefficient, one can make sure that the boundary terms cancel each other properly so the flow equation is satisfied. We then compute the quenched free energy for both cases: if we exclude the non-perturbative contribution, we find the quenched free energy to be finite and monotonic at low temperatures, and it decreases under the deformation. If we include the non-perturbative contribution, we find the quenched free energy diverges, and a careful analytical dissection is included in Appendix \ref{app:goodsigndetail}.

\section{$T\overline{T}$ deformation in JT gravity and matrix models}
\subsection{$T\overline{T}$ deformed correlation functions in JT gravity}
\label{review}
Here we review how one computes deformed partition functions in JT gravity via an integral transformation on the partition function in the undeformed theory. As already alluded to in the introduction, the deformed energy spectrum is given by \eqref{energyspectra} and given the density of states, one can immediately compute the deformed partition function systematically as
\begin{equation}
\label{DeformedZ}
    Z_\lambda(\beta) = \int ^\infty_{-\infty} dE ~ \rho(E) e^{-\beta f^-_\lambda (E)}\,.
\end{equation}
For the moment, we will only consider the contribution from the perturbative branch $f_\lambda^-(E)$. We will return to the potential contribution from the non-perturbative branch $f_\lambda^+(E)$ later in \S \ref{sec:comments on non-perturbative}.

The partition function $Z_\lambda (\beta)$ satisfies a flow equation derived by \cite{Iliesiu:2020zld}:
\begin{equation}
\label{flow}
    \left[ 4 \lambda \partial_\lambda \partial_\beta +2 \beta \partial^2_\beta + \left( 1 - \frac{4\lambda}{\beta}\right) \partial_\lambda \right] Z_\lambda (\beta) = 0\,,
\end{equation}
which is closely related to the usual inviscid Burgers' equation from the 2D $T\overline{T}$-deformed energy spectra \eqref{f(E)} (e.g., see \cite{Jiang:2019epa}). The deformed partition function \eqref{DeformedZ} may be written in terms of an integral transformation involving a kernel and the undeformed partition function \cite{Gross:2019ach} when $\rho \left(E < \frac{1}{8\lambda} \right) = 0$:
\begin{equation}
\begin{aligned}
\label{Direct}
  Z_\lambda(\beta) &=  \frac{1}{2\pi i} \int^\infty_0 d \beta' Z_0(\beta') \int^{i\infty}_{-i\infty} dE e^{-\beta f^-_\lambda (E) + \beta' E} \\&= \int^\infty_0 d\beta' K(\beta, \beta') Z_0(\beta')\,,
  \end{aligned}
\end{equation}
where $K(\beta, \beta')$ is a kernel determined from the deformed energy spectrum \eqref{energyspectra}:
\begin{equation}\label{kernel}
\begin{aligned}
    K(\beta, \beta') &= \frac{1}{2\pi i} \int^{i\infty}_{-i\infty} dE e^{-\beta f^-_\lambda (E) + \beta' E} = \frac{\beta}{\sqrt{-8 \pi \lambda} \beta'^{\frac{3}{2}}} e^{\frac{(\beta - \beta')^2}{8\lambda \beta'}}\,,
\end{aligned}
\end{equation}
namely the inverse Laplace transform of the Boltzmann factor after deformation for $\lambda<0$.

With the integral transform \eqref{Direct} and its kernel \eqref{kernel} at hand, one can proceed to compute the partition function of the deformed JT gravity on disk, trumpet and other topologies for $\lambda < 0$. For example, the map between the undeformed and deformed disk and trumpet partition functions respectively are 
\begin{equation}
    \begin{aligned}
    &Z_0 (\beta')_\text{D}=\frac{e^{\frac{\pi^2}{\beta'}}}{4\sqrt{\pi} \beta'^{\frac{3}{2}}} \implies  Z_\lambda (\beta)_{\text{D}} = \frac{\beta }{\sqrt{-8\lambda}\pi} \frac{e^{- \frac{\beta}{4\lambda}}}{\beta^2 + 8\pi^2 \lambda} K_2 \left( - \frac{\sqrt{\beta^2 + 8 \pi^2 \lambda}}{4\lambda} \right)\,, \\  &Z_0(b ,\beta)_{\text{T}} = \frac{e^{- \frac{b^2}{4\beta}}}{2\sqrt{\pi \beta}} \implies Z_\lambda(b ,\beta)_{\text{T}} = \frac{\beta}{2\pi \sqrt{-2\lambda}} \frac{e^{- \frac{\beta}{4\lambda}}}{\sqrt{\beta^2 - 2b^2 \lambda}} K_1 \left( - \frac{\sqrt{\beta^2 - 2b^2 \lambda}}{4\lambda} \right)\,,
    \end{aligned}
\end{equation}
where we have used an identity for the modified Bessel functions of the second kind 
\begin{equation}
\label{eq:BesselK}
    \int^\infty_0 d \beta' (\beta')^{-m-\frac{3}{2}} e^{\frac{a}{\beta'}} e^{ \frac{(\beta - \beta')^2}{8 \beta' \lambda}} =\frac{ 2 e^{-\frac{\beta}{4\lambda}}}{(\beta^2 + 8 a \lambda)^{\frac{2m+1}{4}}} K_{\frac{2m+1}{2}} \left( - \frac{\sqrt{\beta^2 + 8 a \lambda}}{4\lambda} \right)\,,\quad m\in\mathbb{R}\,,
\end{equation}
when $\lambda<0$ and $8a \lambda + \beta^2 > 0$. 

We comment on one of the convergence conditions  $8a \lambda + \beta^2 > 0$ in \eqref{eq:BesselK} implies that the disk partition function (where $m = 3/2, a = \pi^2$) is well-defined up to the Hagedorn temperature  $T_{\text{H}} =(-8\pi^2 \lambda)^{-1/2}$ as shown by \cite{Gross:2019ach}. Fortunately, for the trumpet partition function, there is no Hagedorn temperature (which takes place when $m=1/2, a = -b^2/4$). As pointed out in \cite{Iliesiu:2020zld} for the bad sign $\lambda > 0$, the trumpet partition function diverges when the total proper length of the boundary equals the geodesic length. This divergence is removed by taking into account the non-perturbative branch thus making the trumpet's density of states real and converts the modified Bessel function of the second kind $K_1 (\cdot)$ into the first kind $I_1 (\cdot)$ as done by \cite{Iliesiu:2020zld}. Likewise from \cite{Iliesiu:2020zld}, the same logic schematically holds for the disk partition function thus converting the modified Bessel for the second kind $K_2(\cdot)$ into $I_2(\cdot)$. We refer the reader to equations (4.7)-(4.8) and Appendix B in \cite{Iliesiu:2020zld} for more elaborate discussions.

As in \cite{Iliesiu:2020zld}, knowing $Z_\lambda(\beta)_{\text{D}}$ and $Z_\lambda(b,\beta)_{\text{T}}$ allows us to build general correlation functions in the deformed JT gravity. The connected correlators on a hyperbolic Riemann surface with $n$ boundary components and genus $g$ are
\begin{equation}
\label{connected cor}
    \left\langle Z(\beta)^{n}\right\rangle_{\mathrm{conn}, \lambda}=\sum_{g=0}^{\infty} e^{-S_{0}(2 g+n-2)} Z_{g, n; \lambda}(\beta)\,,
\end{equation}
where $S_0$ is the two-dimensional Einstein-Hilbert action, and $Z_{g,n;\lambda} (\beta)$ are defined from \cite{Griguolo:2021wgy} as follows:
\begin{equation}
\begin{aligned}
\label{eq:ingredientsZ}
Z_{0,1; \lambda}(\beta)&= Z_\lambda(\beta)_{\mathrm{D}}\,, \\
Z_{0,2;\lambda}(\beta_1,\beta_2)&= \int_{0}^{\infty} d b~ b Z_\lambda(b, \beta_1)_{\text{T}}Z_\lambda(b, \beta_2)_{\text{T}}\,, \\
Z_{g, n; \lambda}(\beta_1,\cdots,\beta_n)&= \int_{0}^{\infty}\Bigg(\prod_{j=1}^{n} d b_{j}~ b_{j} Z_\lambda\left(b_{j}, \beta_i\right)_{\text {T}}\Bigg) V_{g, n}\left(b_{1}, \ldots, b_{n}\right)\,,
\end{aligned}
\end{equation}
with the Weil-Petersson volume $V_{g, n}\left(b_{1}, \ldots, b_{n}\right)$ of a Riemann surface $\Sigma_{g,n}$ (i.e., with genus $g$ and $n$ distinct marked points $p_i$) defined in \cite{Mirzakhani:2006eta} as \footnote{A few of $Z_{g, n; \lambda}(\beta)$ has already been computed in \cite{Griguolo:2021wgy}. Also, see \cite{Dijkgraaf:2018vnm,Saad:2019lba} for a review of Weil-Petersson volumes in 2D topological gravity and matrix models.} 
\begin{equation}
    V_{g, n}\left(b_{1}, \ldots, b_{n}\right)= \frac{1}{(2\pi^2)^{3g-3+n}} \int_{\overline{\mathcal{M}}_{g, n}} \exp \left(\omega+\frac{1}{2} \sum_{i=1}^{n} \psi_{i} b_{i}^{2}\right)\,.
\end{equation}
Here $\overline{\mathcal{M}}_{g,n}$ is the Deligne-Mumford compactification of the moduli space $\mathcal{M}_{g,n}$ of $\Sigma_{g,n}$ of complex dimension $(3g-3+n)$, $\psi_i\equiv c_1(\mathcal{L}_i)$ is the first Chern class\footnote{$\psi_i$ are also called ``$\psi$-classes'', ``Witten classes'' or ``gravitational descendants''.} of the tautological line bundle $\mathcal{L}_i$ over $\overline{\mathcal{M}}_{g,n}$ whose fiber at the point $(C,x_1,\dots,x_n)\in\overline{\mathcal{M}}_{g,n}$ is the cotangent line to the curve $C$ at $x_i$, $\omega$ is the Weil-Petersson symplectic form on $\overline{\mathcal{M}}_{g,n}$, and $b_i$ is the length of the $i$th geodesic boundary component of $\Sigma_{g,n}$.

From the definition in \eqref{eq:ingredientsZ}, one might wonder if the Weil-Petersson volumes should flow under the $T\oT$ deformation? One possible way to see why Weil-Petersson volumes do not flow under the deformation\footnote{However, in the context of topological recursion, both the resolvent $R_{g,n;\lambda}$ and function $W_{g,n;\lambda}$ is deformed \cite{Griguolo:2021wgy}, while their relation to each other $W_{g,n;\lambda}(z_1,\cdots,z_n)\equiv(-2)^n z_1\cdots z_n R_{g,n;\lambda}(-z_1^2,\cdots,-z_n^2)$ is intact. So in the deformed theory, $V_{g,n}$ is no longer the Laplace transform of $W_{g,n;\lambda}$. The topological recursion formula in terms of $W_{g,n;\lambda}$ \cite{Eynard:2007fi} is covariant under the deformation, and retains the same form.} is that the flow equation \eqref{flow} should be satisfied on each \textit{asymptotic} boundary component with proper legnth $\beta_i$. But by definition, the flow equation only contains derivatives with respect to $\lambda$ and $\beta$, not $b_i$, the length of \textit{geodesic} boundary component to be glued together. This fact is also adopted in \cite{Rosso:2020wir,Griguolo:2021wgy}.

Then for generic $Z_{g,n;\lambda}(\beta)$, we can write the deformed partition functions as 
\begin{equation}
    Z_{g,n;\lambda}(\beta_1,\cdots,\beta_n) = \int_0^\infty dE_1 \cdots dE_n \, \rho_{g,n}(E_1,\cdots,E_n) e^{-\beta_1 f_\lambda^-(E_1)} \cdots e^{-\beta_n f_\lambda^-(E_n)}\,,
\end{equation}
where
\begin{equation}
    \rho_{g,n}(E_1,\cdots,E_n) = \int_0^\infty \bigg(\prod_{j=1}^n db_j \, b_j \, \rho_{\text{T}}(b_j,E_j)\bigg) V_{g,n}(b_1,\cdots,b_n)\,,
\end{equation}
and 
\begin{equation}
    \rho_{\text{T}}(b,E) = \frac{\cos(b\sqrt{E})}{2\pi\sqrt{E}}\quad \text{such that} \quad \int_0^\infty dE \, \rho_{\text{T}}(b,E) e^{-\beta E} = Z_0(b,\beta)_{\text{T}}\,.
\end{equation}
To conclude this section, we derive a differential operator presentation of the $T\oT$ deformation similar to \cite{Gross:2019uxi}, which is sufficient for computing perturbative expansions in $\lambda$ and will be used in \S \ref{genus0sec}, by rewriting the exponential
\bea \label{diffop} & e^{-\beta f^-_\lambda(E)} \\
=& e^{-\beta \sum^\infty_{m=1} c_m \lambda^m E^{m+1}} e^{-\beta E} \\
=& e^{-\beta \sum_{m=1}^\infty c_m \lambda^m (-\partial_y)^{m+1}}|_{y=\beta} \, e^{-y E} \\
=& \mathcal{D}_{y;\lambda}|_{y=\beta} \, e^{-y E}\,,
\eea
where we have used 
\begin{equation}
\label{tke}
    f^-_\lambda (E) =  \frac{1 - \sqrt{1 - 8 \lambda E }}{4\lambda} = \sum_{m=0}^\infty c_m \lambda^m E^{m+1}, \quad c_m = \frac{8^m \Gamma\left(\frac{1}{2}+m\right)}{\Gamma\left(\frac{1}{2}\right) \Gamma \left(m+2\right)}\,.
\end{equation}
It is then straightforward to compute partition functions of generic topologies using this differential operator. However, for multiple boundary components, one starts from the undeformed partition function with different inverse temperatures $\beta_i$ on each boundary component, and then applies the differential operator $\mathcal{D}_{y_i;\lambda}|_{y_i = \beta_i}$ to each boundary component separately
\begin{equation}
    \langle Z(\beta_1)\cdots Z(\beta_n) \rangle_\lambda = \left( \prod_{i=1}^n \mathcal{D}_{y_i;\lambda}|_{y_i = \beta_i} \right) \langle Z(y_1) \cdots Z(y_n) \rangle_0\,,
\end{equation}
which in particular implies
\begin{equation}
    \langle Z(\beta)^n \rangle_\lambda = \left( \prod_{i=1}^n \mathcal{D}_{y_i;\lambda}|_{y_i = \beta} \right) \langle Z(y_1) \cdots Z(y_n) \rangle_0\,.
\end{equation}
We will use this differential operator presentation to perform perturbation calculation in the leading order of $\lambda$ in \S \ref{genus0sec}.

\subsection{Quenched free energy}
\label{Airy}
In this subsection, we quickly review a recent novel way of performing the replica trick\footnote{From now on we will suppress the argument $\beta$ in $Z(\beta)$.}
\begin{equation}
\label{ReplicaTrick}
\langle \ln Z \rangle  =  \lim_{n \rightarrow 0}  \frac{\left \langle Z^n \right \rangle - 1}{n}\,,
\end{equation}
following \cite{Okuyama:2021pkf}. The replica trick \eqref{ReplicaTrick} has been shown in  \cite{Okuyama:2021pkf} to be written as a rather convenient integral representation
\begin{equation}
\label{OkuyamaF}
   \langle\ln  Z\rangle=  \ln \langle Z\rangle - \int_{0}^{\infty} \frac{d x}{x}\left[\left\langle e^{-Z x}\right\rangle-e^{-\langle Z\rangle x}\right]
\end{equation}
such that the analytical continuation from $\langle Z^n\rangle$ to $\langle Z^{n=0}\rangle$ remains unambiguous. From \eqref{OkuyamaF}, the first term is the annealed free energy while the second term encodes the contribution from Euclidean replica wormholes with the interpretation that the operator $e^{-Z x}$ creates spacetime boundary components, so-called ``spacetime D-brane'' or ``SD-brane'' introduced in the context of baby universes \cite{Marolf:2020xie}. Additionally, the term $\left \langle e^{-Zx}  \right \rangle$ can be rewritten as $e^{-\mathcal{Z}(x)}$ in terms of the following generating function of connected correlators
\begin{equation}
\label{calZdef}
    \mathcal{Z}(x) = \sum^{\infty}_{n=1} \frac{(-x)^{n+1}}{n!} \left \langle Z^n \right \rangle_c\,.
\end{equation}
This will turn out to be an important quantity when we compute the $T\overline{T}$-deformed annealed and quenched free energies of the Airy model in the upcoming section and the appendices. 

A motivation to study a simple observable, such as the free energy, is to see how do Euclidean replica wormholes contribute to the Euclidean gravitational path integral
\begin{equation}
\label{factorization}
    \langle Z(B) \rangle = \int_{B} ~[dg]~ e^{-S[g]}
\end{equation}
with spacetime boundary $B$, metric measure $[dg]$ and JT gravity action $S[g]$. An easy way to determine the presence of Euclidean wormholes is to see whether correlation functions of the partition function cease to factorize
\begin{equation}
\label{failure}
    \langle Z(B)^n \rangle \overset{?}{=}\langle Z(B) \rangle^n
\end{equation}
among $n$ boundary components. It turns out not to be the case due to the factorization failure as shown in \eqref{OkuyamaF}, and this fact can be confirmed by directly computing the annealed and quenched free energies\footnote{In \cite{Alishahiha:2020jko}, the quenched and annealed free energies in JT gravity with conical deficit angles were computed following the formalism developed by \cite{Mertens:2019tcm,Maxfield:2020ale,Witten:2020ert,Witten:2020wvy} and observed the same pathology of monotonicity failing at low temperatures as authors of \cite{Engelhardt:2020qpv} find.} as done in \cite{Engelhardt:2020qpv}:
\begin{equation}
    \begin{aligned}
        F_a(\beta) =- \beta^{-1}  \ln \langle Z \rangle\,, \quad F_q(\beta)= - \beta^{-1} \langle \ln Z \rangle\,,
    \end{aligned}
\end{equation}
at inverse temperature $\beta$. They are shown not to be the same indeed, clearly indicating the factorization failure already hinted at by \eqref{OkuyamaF}. 

Unfortunately, the authors of \cite{Engelhardt:2020qpv} computed $F_q(\beta)$ with direct usage of the replica trick \eqref{ReplicaTrick} and their analysis at low temperature found that $F_q(\beta)$ is \emph{not} monotonically decreasing as a function of temperature. This is fundamentally due to the non-uniqueness of analytically continuing $\langle Z^{n} \rangle$ to $\langle Z^{n = 0}\rangle$. Given this conundrum at low temperature, the correct analytical continuation was performed by the author of \cite{Okuyama:2021pkf} without directly using the replica trick \eqref{ReplicaTrick}, and there $F_q(\beta)$ was shown to be a monotonically decreasing function of temperature. We now review how this is done following the proof in \cite{Okuyama:2021pkf}. The correlator $\langle Z^n \rangle$ can be expanded in terms of connected correlators $\langle Z^k \rangle_c$ as \cite{Okuyama:2019xvg}: 
\be
\langle Z^n\rangle=\langle Z\rangle^n\left[1+\frac{1}{2}n(n-1)\frac{\langle Z^2\rangle_c}{\langle Z\rangle^2}+\dots\right],
\ee
so now the analytic continuation of  $\langle Z^n \rangle$ is unambiguous due to its being rewritten as a polynomial in $n$ up to an overall $\langle Z \rangle^n$. To further find the integral representation \eqref{OkuyamaF}, we generalize the above expansion to
\begin{equation}
\label{polynomial}
    \langle Z^n \rangle = \langle Z \rangle^n \sum_{j_i \geq 0} \frac{n!}{\left( n - \sum_{\ell \geq 2} \ell j_\ell \right)!} \prod_{k \geq 2} \frac{1}{j_k!} \left( \frac{1}{k!} \frac{\langle Z^k \rangle_c}{\langle Z \rangle^k} \right)^{j_k}\,, 
\end{equation}
where  $i \geq 2$ and $j_k$'s constitute an integer partition
\begin{equation}
    \sum_{k=1}^{n} k j_{k}=n\,.
\end{equation}
Now, using the standard prescription for analytical continuation
\begin{equation}
    \lim_{n \rightarrow 0} \frac{1}{n} \frac{n!}{(n-m)!}= (-1)^{m-1} (m-1)!
\end{equation}
and then the identity
\begin{equation}
\label{eq:trick}
    \int^\infty_0 dy ~ y^{k-1} e^{-y} = (k-1)!,
\end{equation}
the quenched free energy $F_q(\beta)$ can now be written as the integral representation \eqref{OkuyamaF}, whose integration range inherits that of \eqref{eq:trick}.

\subsection{Comments on non-perturbative contributions}
\label{sec:comments on non-perturbative}
\begin{figure}[h]
     \begin{subfigure}[b]{.5\textwidth}
         \centering
         \includegraphics[width=\textwidth]{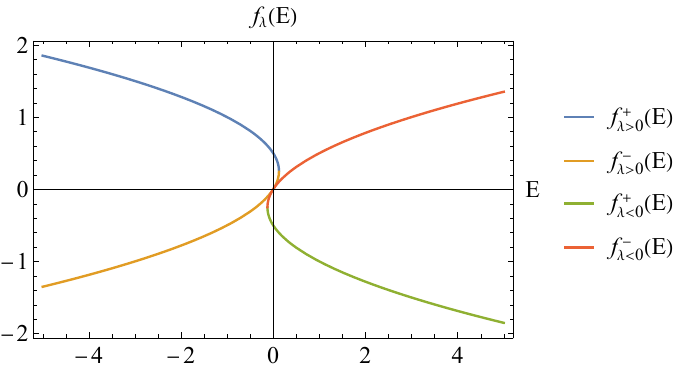}
         \caption{$|\lambda| = 1$}
         \label{fig:ncT1a}
     \end{subfigure}
     \begin{subfigure}[b]{0.5\textwidth}
         \centering
         \includegraphics[width=\textwidth]{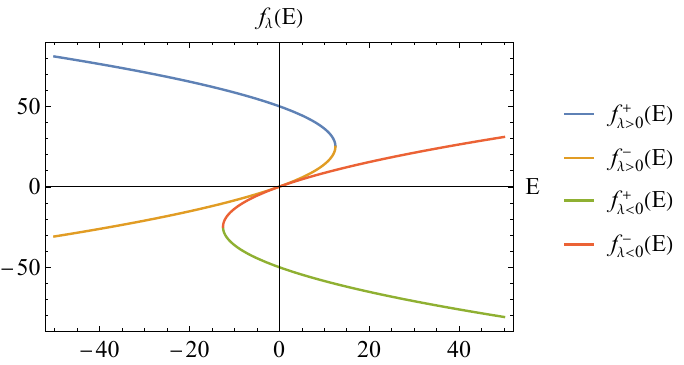}
         \caption{$|\lambda| = 0.01$}
     \end{subfigure}
     \caption{We plot the entire deformed energy spectrum as a function of the undeformed energy $E$ when $|\lambda| = 1$ and $|\lambda| = 0.01$. Here $f^{+}_{\lambda > 0}(E)$ (blue curve) and $f^{+}_{\lambda < 0}(E)$ (green curve) together describe the nonperturbative branch. While $f^-_{\lambda > 0}(E)$ (orange curve) and $f^-_{\lambda < 0}(E)$ (red curve) describe the perturbative branch. For either $\lambda>0$ or $\lambda<0$, we find a smoothly connected curve. We also see that when $|\lambda|$ decreases, the non-perturbative branch becomes further away from the origin.}
     \label{NPP}
\end{figure}

In this subsection, we clarify more of the \emph{non-perturbative} features arising from the $T\oT$ deformation appearing in this paper.\footnote{Another non-perturbative effects we will encounter is in the Airy model, whose density of states $\rho_{\text{Airy}}(E < 0) \neq 0$. This is already present in the undeformed theory and is unrelated to \cite{Griguolo:2021wgy}. Yet another non-perturbative effect takes place upon summing over all genus. We believe the non-perturbative effects in summing over genus results in the non-perturbative instability. We will often refer this effect as the non-perturbative instability in order to distinguish the two. But based on context, the readers should have not trouble in distinguishing the two.} The perturbative branch is denoted by the negative branch in the energy spectrum \eqref{energyspectra} since $\lim_{\lambda \rightarrow 0} f^-_\lambda(E) = E$. In contrast, the $\lambda \rightarrow 0$ limit of the positive branch for \eqref{energyspectra} diverges so, as expected, most papers omit this branch in their perturbative analysis and only consider the negative branch. We plot the entire deformed energy spectrum in Figure \ref{NPP} to show both perturbative and non-perturbative branches. Unfortunately, when $\lambda > 0$, the spectrum along the flow becomes complex-valued for large enough energies. To resolve this issue, as explicitly shown in \cite{Iliesiu:2020zld}, one is forced to include the non-perturbative contribution such that the partition function 
\begin{equation}
\label{eq:nonpert}
    Z_{\lambda > 0} (\beta) = \int^\infty_{-\infty} dE\, \rho_+ (E) e^{-\beta f^+_\lambda (E)} + \int^\infty_{0} dE \,\rho_- (E) e^{-\beta f^-_\lambda (E)} 
\end{equation}
is real with appropriate constraints on the density of states $\rho_\pm (E)$ for JT gravity and is a solution to the flow equation \eqref{flow}. The appropriate constraints on $\rho_\pm (E)$ are explained more in-depth by \cite{Iliesiu:2020zld}.

An alternative approach to naturally incorporate non-perturbative effects is through a resurgent analysis. In \cite{Griguolo:2021wgy}, the disk and trumpet deformed JT partition functions written as power series in $\lambda$ are Borel resummed to obtain non-perturbative results, which are used to further study how the partition functions summed over topologies (i.e., topological recursion) and spectral form factor are modified under the $T\overline{T}$ deformation. Alas, despite these technical advances of the $T\overline{T}$ deformed JT gravity, the resurgent analysis still contains a spectral density not positive-definite.

\section{Quenched free energy for deformed Airy model}
\label{sec:Airy}
In this section, we compute the $T\overline{T}$-deformed annealed and quenched free energies in the Airy model. It is important to point out that our deformation of the double-scaled matrix model is different from the ones considered in \cite{Rosso:2020wir}, whose $T\oT$ deformation of the double-scaled matrix model dual to JT gravity does not completely match the correlators with the $T\oT$-deformed JT gravity. 

Our ultimate goal is to understand the quenched free energy in $T\oT$-deformed JT gravity. However, this is a rather difficult problem since even for the one-point function $\langle Z(\beta)\rangle_{\operatorname{JT}}$ of the undeformed theory, there is no analytical expression which includes all-genus contributions, let alone non-perturbative effects. In contrast, there has been much progress with numerical calculations \cite{Johnson:2019eik,Johnson:2020exp,Johnson:2020heh,Johnson:2020lns,Johnson:2021rsh,Johnson:2021tnl}. Therefore, we want to make a simplification and study the Airy model instead, which is known to be the low energy approximation of JT gravity. Since it is a double-scaled matrix model, one could apply the $T\oT$ deformation defined in \cite{Rosso:2020wir}. However, we already know that the $T\oT$ deformation defined and studied there does not provide an exact match between JT gravity and its matrix model dual in terms of general correlators. Hence, this is not a good choice if our ultimate goal is to understand the quenched free energy in $T\oT$-deformed JT gravity. 

In order to investigate the Airy model with the hope of retaining some essential features of the $T\oT$-deformed JT gravity, we will have to work with a different deformation from the one considered in \cite{Rosso:2020wir}. Since one can construct $T\oT$-deformed JT correlators with any number of boundary components and genera from basic observables like the deformed disk and trumpet partition functions by gluing together pants decomposition \cite{Iliesiu:2020zld}, we can recast the $T\oT$ deformation of the correlators of JT gravity in various ways (e.g., using the differential operator $\mathcal{D}_{y;\lambda}|_{y=\beta}$) as reviewed in \S \ref{review}. Now in order to match the $T\oT$-deformed matrix model dual to JT gravity, correlators on both sides must be deformed in the same fashion. We can then apply the same recipe of deforming the matrix model correlators to the Airy model, instead of using the one defined in \cite{Rosso:2020wir}. This is the deformation we will adopt and study in this paper. 

To be more specific, we will take the deformed $n$-point functions in the Airy model as
\be 
\label{eq:deformednpt}
\langle Z(\beta_1) \cdots Z(\beta_n) \rangle_{\text{Airy},\lambda} = \int dE_1 \cdots  dE_n \, \rho_{\text{Airy},n}(E_1,\cdots,E_n) e^{-\beta_1 f_\lambda^-(E_1)}\cdots e^{-\beta_n f_\lambda^-(E_n)}
\ee
where $\rho_{\text{Airy},n}(E_1,\cdots, E_n)$ is such that
\be 
\label{eq:undeformednpt}
\langle Z(\beta_1) \cdots Z(\beta_n) \rangle_{\text{Airy},0} = \int dE_1 \cdots dE_n \, \rho_{\text{Airy},n}(E_1,\cdots,E_n) e^{-\beta_1 E_1} \cdots e^{-\beta_n E_n}\,.
\ee
It is important to notice that if we are interested in computing the deformation at any given genus, there will not be non-perturbative effects (from the Airy model itself before $T\overline{T}$ deforming) so the integration range of $E_i$ is $[0,\infty)$, and we can safely apply the integral transformation \eqref{Direct} on \eqref{eq:undeformednpt} to obtain \eqref{eq:deformednpt} for convenience. However, if we are interested in the exact result, for instance, $\langle Z(\beta)\rangle_{\text{Airy},\lambda}$, then we cannot use the integral transformation, no matter if $\lambda$ is positive or negative, because even for $\lambda<0$, $\rho_{\text{Airy}}$ has support over the entire real axis of $E$. 

There are a few other important caveats we need to mention. First, carrying over the $T\oT$ deformation of JT gravity correlators to matrix model correlators only determines the perturbative branch of the $T\oT$ deformation. There will be non-perturbative contributions of the $T\oT$ deformation that need to be addressed if we are interested in results with \emph{finite}, not infinitesimal, $\lambda$. We will analyze those non-perturbative effects case-by-case as we encounter them by matching the deformed Airy correlators with either the genus-zero result or the flow equation. A possible systematical treatment would be adapting the resurgent analysis performed in \cite{Griguolo:2021wgy}. However, in our study, eventually we will have to sum over not only all orders in $\lambda$, but all genus $g$ as well. As hinted in the introduction of this paper, it is well-known that there is non-perturbative instability in the exact spectral density in the Airy model related to the genus expansion. Hence, we expect this ``double'' resurgent analysis in $\lambda$ and $g$ (both variables on the same footing) to be rather complicated and will leave it for future work. 

Second, related to the non-perturbative instability of the Airy model, the same non-perturbative instability appears in the na\"ive matrix model dual of the JT gravity as well \cite{Saad:2019lba,Johnson:2019eik}. There have been works on how to improve the non-perturbative feature of the JT gravity and remove this undesired feature \cite{Saad:2019lba,Johnson:2019eik,Stanford:2019vob}. Hence, there are possibilities that the non-perturbative instability can qualitatively affect the behavior of the quenched free energy in the deformed theory and would be interesting as well as important to understand if this is the case. A possible way to figure this out is to extend our study to the JT gravity at general temperatures with the non-perturbative instability taken care of.

Third, it should be emphasized that we do not provide a complete description of our version of the $T\oT$ deformation for the matrix model. We simply require that in the deformed theory, every correlator has to transform in a way in order to match the gravity side. One can take this as the working definition of our version of the $T\oT$ deformation for matrix models.\footnote{One place where this is manifest is the matching between \eqref{eq:JTdisk} and \eqref{eq:Airyguess} in \S \ref{sec:bad} later.} To further illustrate the difference between our version and the one in \cite{Rosso:2020wir}, let us start with a Hermitian matrix integral:
\begin{equation}\label{MatrixIntegral}
    Z = \int [dM] e^{-\text{Tr}V(M)} = \int d^N x \prod_{1\leq i < j \leq N} (x_i - x_j)^2 e^{-\sum^N_{i=1} V(x_i)}\,,
\end{equation}
where $x_i$ are the eigenvalues of the $N \times N$ Hermitian matrix $M$ and the Vandermonde determinant appears from diagonalizing $M$. 

The undeformed correlation functions are computed by
\begin{equation}
    \langle O_1(M) \cdots O_n(M) \rangle_0 = \int [dM] e^{-\text{Tr}V(M)} O_1(M) \cdots O_n(M)\,.
\end{equation}
In \cite{Rosso:2020wir}, it is assumed that the deformed matrix model still takes the form of \eqref{MatrixIntegral}, and the only physical quantity that changes is the matrix model potential which shifts from $V(M)$ to $V_{\lambda}(M) = c_{\lambda} V\left(M-2 \lambda M^{2}\right)$. As a somewhat unfortunate consequence of this assumption, the $n$-point functions in the deformed matrix model still do not match the $n$-point functions in the deformed JT gravity. While in our case, we start by requiring that the $n$-point functions in the deformed matrix model match the results in deformed JT gravity leading to a different deformation on the matrix model side. In particular, naively the integration measure $[dM]$ of matrices will receive corrections from the $T\overline{T}$ deformation as well if we want to keep the potential $\operatorname{Tr}V(M)$ as a single trace operator as implicitly assumed in \cite{Rosso:2020wir}. To see this, consider the correlators 
\begin{equation}
\begin{aligned}
  &\left  \langle \text{Tr}\, e^{-\beta_1 M} \cdots \text{Tr} \,e^{-\beta_n M} \right \rangle_0 \\
  &\quad\quad\quad\quad\quad\quad=\int d^Nx \prod_{1\leq i \leq j \leq N} (x_i - x_j)^2 e^{-\sum^N_{i=1} V(x_i)} \bigg(\sum^N_{i_1 = 1} e^{-\beta_1 x_{i_1}}\bigg) \cdots \bigg(\sum^N_{i_n=1} e^{-\beta_n x_{i_n}}\bigg)
  \end{aligned}
\end{equation}
in the undeformed matrix model. To match the gravity result, its deformation should assume the following form:
\begin{equation}
\begin{aligned}
\label{half-diff1}
  &\left  \langle \text{Tr}\, e^{-\beta_1 f_\lambda^-(M)} \cdots \text{Tr}\, e^{-\beta_n f_\lambda^-(M)} \right \rangle_\lambda \\
  & \quad\quad\,=\int d^Nx \prod_{1\leq i < j\leq N}(x_i-x_j)^2 e^{-\sum^N_{i=1} V(x_i)} \bigg(\sum^N_{i_1 = 1}e^{-\beta_1 f^-_\lambda (x_{i_1})}\bigg) \cdots \bigg(\sum^N_{i_n =1} e^{-\beta_n f^-_\lambda (x_{i_n})}\bigg)\,.
  \end{aligned}
\end{equation}
To find the deformed matrix integral, we consider a change of variables for $x_i$, such that $f^-_\lambda (x_i) = x_i'$, so that we are computing the same correlators
\be
\label{half-diff2}
\left\langle \text{Tr}\,e^{-\beta_1 M'} \cdots \text{Tr} \,e^{-\beta_n M'} \right\rangle_\lambda
\ee
in terms of the new variable $M'$. This will not only change the potential $V(M)$, but the integration measure $[dM]$ as well if we want $e^{-\text{Tr}V(M)}$ to contain only single trace operators.

To further illustrate this fact, we consider the change of variables $y_i = f_\lambda^-(x_i)$ as well as neglect for the moment issues with the branch cut from the square root in $f^-_\lambda (x_i)$ and potential non-perturbative subtleties related to the change of variable for $M$.\footnote{This resembles the ``half-diffeomorphism'', which appears in the formulation of $T\oT$ deformation by coupling the 2D field theory to topological gravity in \cite{Coleman:2019dvf} where either the metric or coordinates change under the deformation, but not both.} Then $x_i = y_i(1-2\lambda y_i)$ and this implies
\begin{equation}
\begin{aligned}
   \left[ dM \right] e^{-\sum_{i=1}^N V(x_i)} = (d^N y)\prod_{1\leq i < j \leq N} (y_i - y_j)^2 e^{-\operatorname{Tr}V(y_i(1-2\lambda y_i))} \prod_{i,j=1,\cdots,N}(1-2\lambda(y_i+y_j))
\end{aligned}
\end{equation}
Since only the product of the matrix measure $[dM]$ and exponential $e^{-\operatorname{Tr}V(M)}$ are unambiguously defined, one could of course consider absorbing the extra piece
\begin{equation}
    \prod_{i,j = 1,\cdots,N}(1-2\lambda(y_i+y_j))
\end{equation}
into the definition of $e^{-\operatorname{Tr}V(M)}$ to retain the form of the measure $[dM]$. However, this will lead to an infinite sum over double-trace operators as following:
\bea
\label{eq:MMstartingpoint}
    & \prod_{i,j = 1,\cdots, N}(1-2\lambda(y_i + y_j)) \\ &= \exp \Bigg( \sum_{i,j = 1,\cdots,N}\log(1-2\lambda(y_i + y_j)) \Bigg)\\ 
    &= \exp \Bigg(\sum_{i,j = 1,\cdots,N}\sum_{m=1}^\infty \frac{(2\lambda)^m}{m} \sum_{p=0}^m C_m^p y_i^p y_j^{m-p} \Bigg) \\ 
    &= \exp \Bigg(\sum_{m=1}^\infty \frac{(2\lambda)^m}{m}\sum_{p=0}^m \operatorname{Tr}(M^p) \operatorname{Tr}(M^{m-p}) \Bigg)\,.
\eea
Either way, the deformation presented here violates the implicit assumptions in \cite{Rosso:2020wir} and would be the starting point on the matrix model side. However, it is unclear whether one should first $T\overline{T}$ deform and take the double-scaling limit or vice versa. Further analysis is beyond the scope of this paper, and we will leave the details of the $T\oT$-deformed matrix model to a future study.\footnote{We thank Per Kraus for this question which helped us clarify this point.}

\subsection{Genus-zero quenched free energy at leading order of $\lambda$}\label{genus0sec}
We begin our analysis on the deformed quenched free energy $F_{q,\lambda}(T)$ at genus-zero perturbatively. To determine $F_{q,\lambda}(T)$, we first compute the deformed genus-zero multi-boundary connected correlators $\langle Z(\beta_1)\cdots Z(\beta_n)\rangle_{c,\lambda}^{g=0}$. The undeformed connected correlators at genus-zero have been computed in \cite{Okuyama:2020ncd} using the genus-zero Korteweg–De Vries (KdV) flow:
\begin{equation}
    \langle Z(\beta_1) \cdots Z(\beta_n) \rangle_{c,0}^{g=0} = g_s^{n-2} \bigg(\sum^n_{i=1} \beta_i\bigg)^{n-3} \prod^n_{i=1} \left(\frac{\beta_i}{2\pi}\right)^{\frac{1}{2}}\,,
\end{equation}
where $g_s\equiv \sqrt{2}$ is the genus-counting parameter.\footnote{Here in fact $g_s = \sqrt{2} \hbar$ and we will set $\hbar$ to unity. Adopting the conventions by \cite{Okuyama:2021pkf}, $\hbar$ is the genus-counting parameter in the Airy limit of matrix models while $g_s$ is the natural genus-counting parameter in 2D topological gravity. Also refer to earlier discussions in \cite{Okuyama:2019xbv,Okuyama:2020ncd}.}

The deformed correlators can be computed directly using the integral transformation \eqref{Direct} and, by construction, solve the flow equation \eqref{flow}. For instance, for the good sign $\lambda < 0$, the deformed one-point, two-point, and $n$-point correlators are given by 
\begin{equation}
\begin{aligned}
\label{eq:Zcorrs}
	\langle Z(\beta)\rangle^{g=0}_{\lambda} &= g_s^{-1} \frac{e^{-\frac{\beta}{4\lambda}}}{2\pi\beta\sqrt{-\lambda}} K_2\bigg(-\frac{\beta}{4\lambda}\bigg)\,,\\
    \langle Z(\beta)^2 \rangle^{g=0}_{c,\lambda} &= - \frac{\beta_1 \beta_2}{16 \pi^2 \lambda} \int^\infty_0 d\beta_1' \int^\infty_0 d\beta_2' \, \frac{\sqrt{\beta_1' \beta_2'}}{\beta_1' + \beta_2'} \, \frac{1}{(\beta_1' \beta_2')^{\frac{3}{2}}}\exp \left(\frac{(\beta_1 - \beta_1')^2}{8 \lambda \beta_1'} + \frac{(\beta_2 - \beta_2')^2}{8 \lambda \beta_2'} \right)\,, \\
    \langle Z(\beta)^n \rangle^{g=0}_{c,\lambda} &=g_s^{n-2} \frac{e^{-\frac{n\beta}{4\lambda}}}{\beta^3} \bigg(\frac{\beta^2}{2\pi \sqrt{-\lambda}}\bigg)^n \, \sum_{i_j\geq 0}^{\sum_j i_j = n-3} \frac{(n-3)!}{i_1! \cdots i_n!} \prod_{j=1}^n K_{i_j}\left(-\frac{\beta}{4\lambda}\right)\,, \quad n\geq 3\,.
\end{aligned}	
\end{equation}
Despite the $n=2$ integral below can be easily integrated numerically, to our best knowledge, we do not know how to evaluate it in closed form; the $\beta_2'$-integral contains a piece
\begin{equation}
    \int^\infty_0\frac{d\beta_2'}{\beta_1'+\beta_2'}\exp\left(-\frac{(\beta_2-\beta_2')^2}{8\lambda \beta_2'}\right)
\end{equation}
not contained in standard tables of integrals, such as \cite{gradshteyn2014table}. A possible way to evaluate the $n = 2$ integral is by writing the Taylor series expansion
\begin{equation}
\label{eq:simpleseries}
        \frac{1}{\beta_1' + \beta_2'} = \frac{1}{\beta_2'} \sum_{m=0}^\infty  \left(-\frac{\beta_1'}{\beta_2'}\right)^m
\end{equation}
with the appropriate analytic coninuation such that the right-hand-side of \eqref{eq:simpleseries} is well-defined. Therefore, we see a similar pattern with modified Bessel functions of the second kind
\begin{equation}
        \langle Z(\beta_1) Z(\beta_2) \rangle^{g = 0}_{c, \lambda} = - \frac{\exp \left(- \frac{\beta_1 + \beta_2}{4\lambda}\right)}{4\pi^2 \lambda} \sum^\infty_{m = 0} \left( -1\right)^m \left( \frac{\beta_1^{m+1}}{\beta_2^m} \right) K_{m} \left( - \frac{\beta_1}{4\lambda} \right) K_{m+1} \left( - \frac{\beta_2}{4\lambda} \right)\,.
\end{equation}
\begin{figure}[h]
    \centering
    \includegraphics[scale = 0.1]{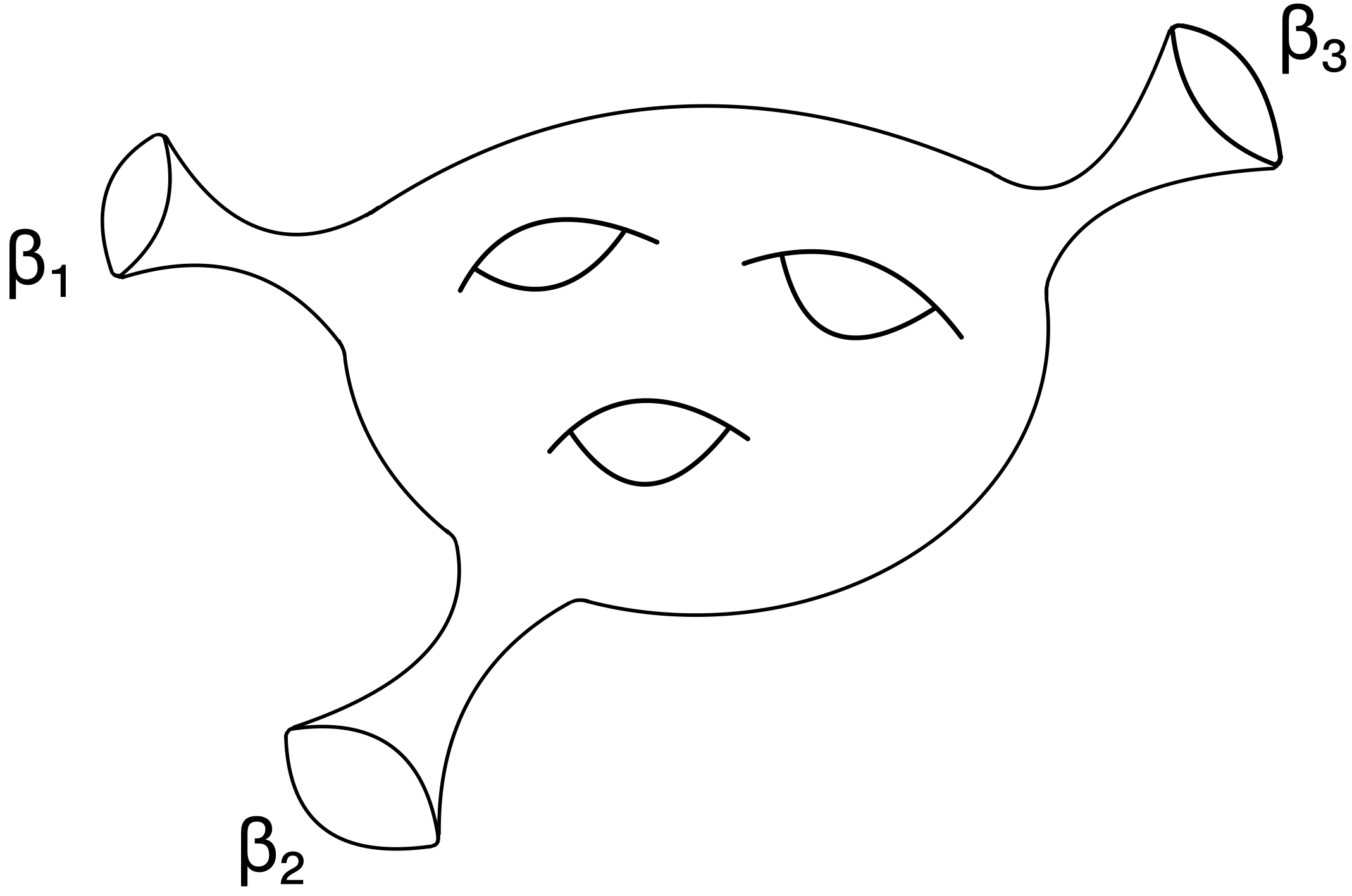}
    \caption{To merely illustrate the setup, we have drawn a genus-three Riemann surface with three boundary components, each of which is a circle of circumference $\beta_i$. It in fact has a constant negative scalar curvature $-2$. Here we restrict only to the genus-zero case.}
    \label{fig:surface}
\end{figure}
Generally, how one evaluates the above sums of products of modified Bessel functions remains unclear, let alone compute
\begin{equation}
    \mathcal{Z}^{g=0}_\lambda (x)= \sum_{n=1}^\infty \frac{(-x)^n}{n!}\langle Z(\beta)^n \rangle_{c,\lambda}^{g=0}
\end{equation}
in \eqref{calZdef} as in \cite{Okuyama:2021pkf}. 

Therefore, instead, we consider working with perturbation theory and just keep the leading order in $\lambda$. For this purpose, it is convenient (in comparison with our ``working definition'' introduced in the beginning of \S \ref{sec:Airy}) to express the $T\oT$ deformation in terms of differential operators and only keep the leading order term in $\lambda$ so that the deformation acts as 
\be 
\label{eq:diff}
\langle Z(\beta_1) \cdots Z(\beta_n)\rangle_\lambda \rightarrow \bigg(1 - \lambda \sum^n_{i=1} c_1 \beta_i \partial_{\beta_i}^2\bigg) \langle Z(\beta_1) \cdots Z(\beta_n)\rangle_0 + O(\lambda^2)\,,
\ee
where $c_1 = 2$ from \eqref{tke}.

Then, we find 
\be
    \langle Z(\beta)^n \rangle_{c,\lambda}^{g=0} = \langle Z(\beta)^n \rangle_{c}^{g=0} - \frac{c_1 \lambda}{2\beta^4} n^{n-4} \left(\frac{\beta^3}{\pi}\right)^{\frac{n}{2}}\left(\frac{7}{4}n^2 - 10n + 12\right) + O(\lambda^2)\,.
\ee
On the other hand, 
\begin{equation}
    \mathcal{Z}_\lambda^{g=0}(x) = - \sum_{n=1}^\infty \frac{(-x)^n}{n!} \langle Z(\beta)^n \rangle_{c,\lambda}^{g=0}
\end{equation}
can be evaluated using the similar trick as in \cite{Okuyama:2021pkf}. Specifically, focusing on the leading order piece in $\lambda$, we have the following sum:
\be \mathcal{Z}^{g=0}_\lambda(x) - \mathcal{Z}^{g=0}_0(x) = -\frac{c_1\lambda}{2\beta^4} \sum_{n=1}^\infty \frac{(-1)^{n-1}n^{n-4}}{n!} \left(x\sqrt{\frac{\beta^3}{\pi}}\right)^n \left(\frac{7}{4}n^2 - 10n +12\right)\,.
\ee
Let $z = x \sqrt{\frac{\beta^3}{\pi}}$, and the sum can be decomposed into three separate pieces:
\be 
\label{LambWDiff} 
\begin{aligned}
&A(W(z)) = \sum_{n=1}^\infty \frac{(-1)^{n-1} n^{n-2}}{n!} z^n\,,\quad B(W(z)) = \sum_{n=1}^\infty \frac{(-1)^{n-1} n^{n-3}}{n!} z^n\,,\\ &C(W(z)) = \sum_{n=1}^\infty \frac{(-1)^{n-1} n^{n-4}}{n!} z^n\,,
\end{aligned}
\ee
satisfying
\be (z\partial_z) A(W(z)) = W(z), \quad (z\partial_z)^2 B(W(z)) = W(z), \quad (z\partial_z)^3 C(W(z)) = W(z)\,,
\ee
where $W(z)$ is the Lambert function defined by the following Taylor series expansion:
\be W(z) \equiv \sum_{n=1}^\infty \frac{(-1)^{n-1}n^{n-1}}{n!}z^n\,.
\ee
The above differential equations \eqref{LambWDiff} can be solved by making the ansatz
\be A = A_2 W ^2 + A_1 W\,, \quad B = B_3 W^3 + B_2 W^2 + B_1 W\,, \quad C = C_4 W^4 + C_3 W^3 + C_2 W^2 + C_1 W\,,
\ee
and using the property of the Lambert function
\be z\partial_z W(z) = \frac{W(z)}{1+W(z)}\,.
\ee
We then find
\be A = \frac{1}{2}W^2 + W, \quad B = \frac{1}{6}W^3 + \frac{3}{4}W^2 + W, \quad C = \frac{1}{24}W^4 + \frac{11}{36}W^3 + \frac{7}{8}W^2 + W
\ee
and
\be \mathcal{Z}_\lambda^{g = 0}(z) = \frac{B(W(z))}{2} + \frac{c_1\lambda}{2\beta} \left(\frac{7}{4}A(W(z)) -10 B(W(z)) + 12 C(W(z)) \right)\,.
\ee
Then using \eqref{OkuyamaF}, we compute the $O(\lambda)$ corrections
\bea \label{intg0}
   & \langle \log Z\rangle^{g=0}_\lambda \\
  =& \log\langle Z\rangle^{g=0}_\lambda - \int_0^\infty dx \frac{e^{-\beta^{-3}\mathcal{Z}^{g=0}_\lambda(x)} - e^{-x\langle Z\rangle^{g=0}_\lambda}}{x} \\
=& \log \left[\frac{1}{\sqrt{4\pi\beta^3}}\bigg(1-\frac{15\lambda}{2\beta}\bigg)\right]  \\
&-\int_0^\infty dW\,\frac{1+W}{W}\left[e^{-\frac{1}{\beta^3}\left(\frac{B(W)}{2} - \frac{\lambda}{\beta}\left(\frac{7}{4}A(W) - 10B(W) + 12 C(W)\right)\right)} - e^{-\frac{1}{2\beta^3}We^W\big(1-\frac{15\lambda}{2\beta}\big)}\right] + O(\lambda^2) \\
=&\log\bigg(\frac{1}{\sqrt{4\pi\beta^3}}\bigg) - \frac{15\lambda}{2\beta} - \int_0^\infty \frac{dW}{W}(1+W)\bigg(e^{-\frac{B(W)}{2\beta^3}} - e^{-\frac{1}{2\beta^3} We^W}\bigg) \\
&- \frac{\lambda}{8\beta^4}\int_0^\infty dW (1+W)\bigg(e^{-\frac{1}{24\beta^3}W(12+9W+2W^2)}(30+31W+16W^2+4W^3)\\
&\quad\quad\quad\quad\quad\quad\quad\quad\quad\quad\quad\quad\quad\quad\quad\quad\quad\quad\quad\quad\quad\quad\quad\quad\quad - 30 e^{W\left(1-\frac{1}{2\beta^3}e^{W}\right)}\bigg) + O(\lambda^2)\,,
\eea
where we have changed the integration variable from $z$ to $W(z)$ and used
\be 
    \langle Z(\beta) \rangle_{\lambda}^{g=0} = \frac{1}{\sqrt{4\pi\beta^3}}\bigg(1-\frac{15\lambda}{2\beta} + O(\lambda^2)\bigg)\,.
\ee
Notice that in the last line of \eqref{intg0}, we expanded in $\lambda$ for the logarithm and exponential since our result is only valid in the leading order of $\lambda$.

This integral \eqref{intg0} can be evaluated numerically and below we plot the quenched free energy $F_q(\beta)$ at genus-zero with $T\overline{T}$ deformation coupling $\lambda = -1/20$ against temperature $T$.

\begin{figure}[h]
    \centering
    \includegraphics[scale = 0.68]{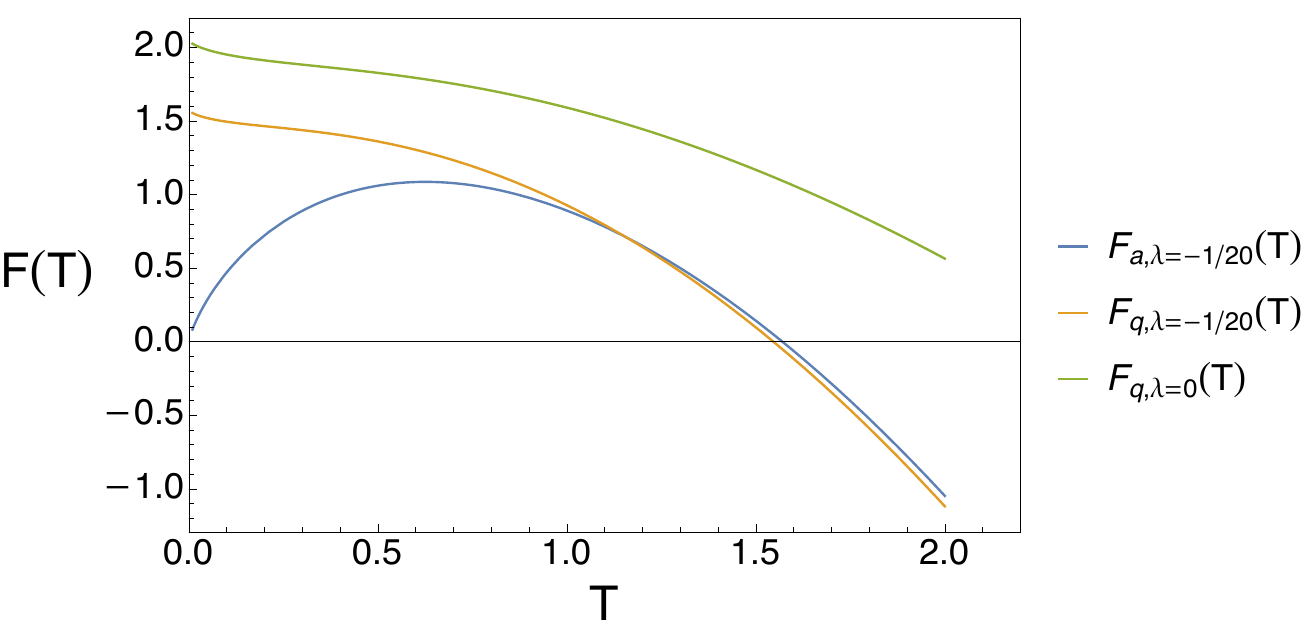}
    \caption{We plot the annealed and quenched free energies for the deformed Airy model as a function of temperature $T$. The blue and orange curves are quenched and annealed free energies respectively for the deformed theory at $\lambda = -1/20$ and the green curve is the quenched free energy for undeformed Airy model. The good sign of the $T\oT$ deformation lowers the quenched free energy.}
    \label{fig:genus0}
\end{figure}

\begin{figure}[h]
    \centering
    \includegraphics[scale = 0.68]{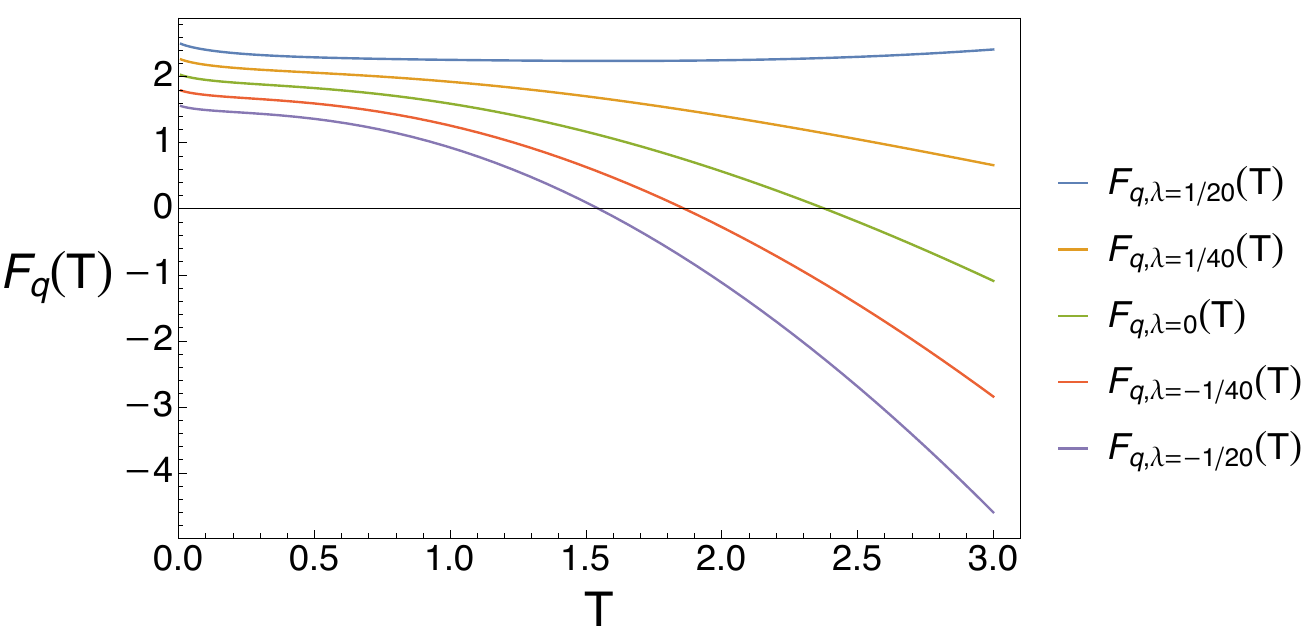}
    \caption{We plot the quenched free energy $F_{q,\lambda}(T)$ for the deformed Airy model as a function of temperature $T$. We notice that a deformation with the good sign $\lambda < 0$ lowers the quenched free energy while the one with the bad sign $\lambda > 0$ increases $F_q(T)$. For both signs of $\lambda$, $F_{q,\lambda}(T)$ monotonically decreases as $T$ increases for $T<1$; however, for the good sign of $\lambda$, this breaks down for $\lambda = 1/20$ and at around $T = 1$. Notice that the perturbative expansion of $\lambda$ always appears as $\lambda T$. In order for the first order approximation to hold, $\lambda T$ should be small. Therefore, this breakdown of monotonicity when $T$ becomes large is likely due to the fact that we go beyond the validity of perturbation theory.}
    \label{fig:badlambdas}
\end{figure}

We also plot the genus-zero quenched free energy for various different signs of $\lambda$, again using differential operators as in \eqref{eq:diff}, see Figure \ref{fig:badlambdas}. Notice that in the perturbative expansion of $\lambda$, $\lambda$ also always appears as $\lambda T$. Hence, for the leading order approximation to hold, $\lambda T$ must be small. In the numerical calculation, we find $F_{q,\lambda}(T)$ monotonically decreases as a function of $T$ for the good sign $\lambda > 0$. However, monotonicity can break down when $\lambda T$ is too large for the bad sign $\lambda > 0$ and this is likely due to $\lambda T$ exceeding the range of validity of leading order approximation in $\lambda$.

\subsection{All genus quenched free energy in low temperature limit}
Now using the low-temperature approximation as in \cite{Okuyama:2021pkf}, we compute the quenched free energy starting from the relation in the undeformed theory \cite{Okuyama:2019xvg,Okuyama:2021pkf}: 
\be 
\label{lowTapproximation} \langle Z(\beta)^n \rangle_c \simeq \langle Z(n\beta) \rangle, \quad T \lesssim 1\,.
\ee
Under this low temperature approximation, the deformation of $\langle Z(\beta)^n \rangle_c \simeq \langle Z(n\beta) \rangle $ is easily computable. This is seen from expressing the undeformed correlator for $n$ boundary components $\left \langle Z(\sum_{i=1}^n \beta_i) \right \rangle_0 $ as
\be \left \langle Z\left(\sum_{i=1}^n \beta_i\right) \right \rangle_0 = \int dE \, \rho(E) e^{-\sum_{i=1}^n \beta_i E}\,,
\ee
the term dependent on $\beta$ essentially factorizes. Thus, the deformed $n$-point function in low temperatures is
\be \left \langle Z(n\beta) \right \rangle_\lambda  = \int_{-\infty}^\infty dE \, \rho(E) e^{-n\beta f^-_\lambda (E)}\,.
\ee
However, we must show that the change from the $T\oT$ deformation for $\langle Z(n\beta) \rangle_{c,\lambda} - \langle Z(n\beta) \rangle_{c,0}$ is of lower order compared to the correction $\langle Z(\beta)^n \rangle_c - \langle Z(n\beta) \rangle$ to the approximation \eqref{lowTapproximation}. This is the case because the correction to the approximation $\langle Z(\beta)^{n}\rangle \simeq Z(n\beta)\rangle$ is exponentially suppressed as $e^{-c_0 \beta^3}$ in the low temperature limit $\beta\rightarrow \infty$, where $c_0$ is some positive constant. One can explicitly check this approximation for $n = 2, 3$, where the exact expression of the partition functions can be conveniently found in \cite{Okounkov:2001usa}.\footnote{However, they are originally due to Dijkgraaf and Zagier, respectively. See Appendix D and reference [Dij] in \cite{Faber:2000ma}.} For instance, with $n = 2$, we have
\be \label{n=2} \langle Z(\beta)^2 \rangle_{c,0} = \langle Z(2\beta) \rangle_0 \,\text{erf}\left(\sqrt{\frac{\beta^3}{2}}\right)\,. \ee 
Using the asymptotic expansion of the error function:
\be 
\label{eq:error}
\text{erf}(x) = 1 - \frac{e^{-x^2}}{x\sqrt{\pi}}\sum_{n=0}^\infty (-1)^n \frac{(2n-1)!!}{(2x^2)^n}, \quad \text{as\,\,}x\rightarrow \infty\,, \ee
it is clear that the correction is suppressed by $e^{-\beta^3/2}$ in the low temperature limit for \eqref{n=2}. 

Similarly, for $n=3$:
\be \langle Z(\beta)^3 \rangle_{c,0} = \langle Z(3\beta) \rangle_0 \left(1 - 12 T\left(\sqrt{3\beta^3},\frac{1}{\sqrt{3}}\right)\right)\,.\ee
From the definition of Owen's $T$ function:
\be
\label{eq:owen}
T(h,a) \equiv \frac{1}{2\pi} \int_0^a \frac{e^{-\frac{1}{2}h^2(1+x^2)}}{1+x^2} dx\,, 
\ee
one can see in the large $h$ limit (i.e., large $\beta$ limit), it is indeed suppressed as $e^{-\frac{1}{2}h^2} = e^{-\frac{3}{2}\beta^3}$. For instance, we can manipulate the integral \eqref{eq:owen} as
\bea T(h,a) &= \frac{e^{-\frac{h^2}{2}}}{2\pi} \int_0^a \frac{e^{-\frac{1}{2}h^2 x^2}}{1+x^2} dx \\
&= \frac{e^{-\frac{h^2}{2}}}{2\pi}  \int_0^{ha} e^{-\frac{1}{2}x^2} \frac{h}{h^2+x^2} dx \\
&\leq \frac{e^{-\frac{1}{2}h^2}}{2\pi} \int_0^\infty \frac{1}{h} e^{-\frac{1}{2}x^2} dx \\
&= \frac{1}{2\sqrt{2\pi} h} e^{-\frac{1}{2}h^2}\,.
\eea
One can also confirm this from numerical integration. In the following plot, we can see clearly, the correction is suppressed by $e^{-\frac{3}{2} \beta^3}$ in the low temperature limit. 

\begin{figure}[h]
    \centering
    \includegraphics[scale = 0.8]{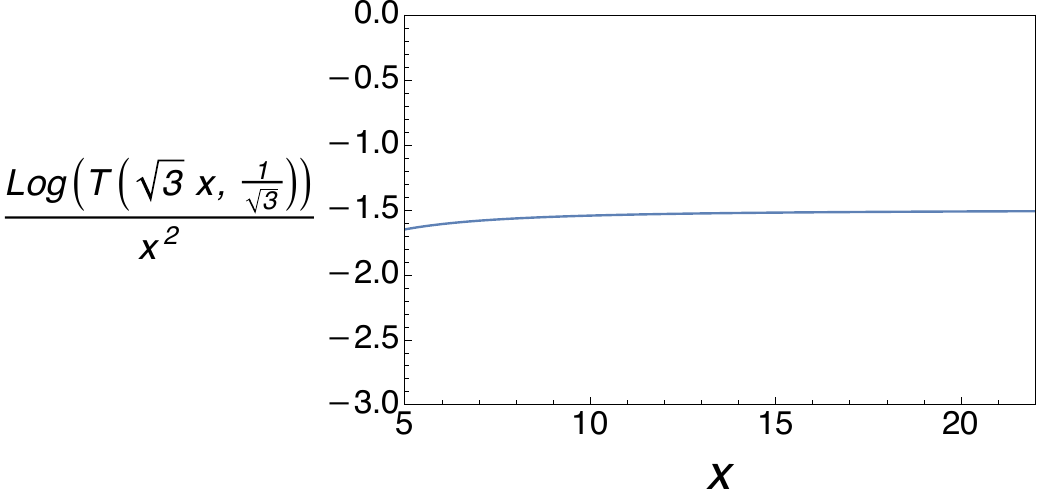}
    \caption{We plot $\frac{1}{x^2}\operatorname{log}\left(T\left(\sqrt{3}x, \frac{1}{\sqrt{3}}\right) \right)$ with respect to $x$. One can clearly see this fraction asymptotically goes to some negative constant close to $-1.6$. This means $T\left(\sqrt{3}x, \frac{1}{\sqrt{3}}\right) \simeq e^{-c_0 x^2}$ in the large $x$ limit with $c_0 \simeq 3/2$.}
    \label{fig:owen}
\end{figure}

The connected correlator has a well-known integral representation\footnote{Essentially a cyclic linear combination of $n!$ integrals, each of which is an $n$-dimensional Laplace transform of the product of $n$ so-called Airy kernels \cite{Moore:1990cn,Tracy:1992kc,Tracy:1992rf}
\begin{equation}
    \mathsf{K}(x_i,x_{i+1})\equiv\frac{\Ai(x_i)\Ai'(x_{i+1})-\Ai'(x_i)\Ai(x_{i+1})}{x_i-x_{i+1}},\quad i=1,\cdots,n\,.
\end{equation}
} given by \cite{Okounkov:2001usa} and is known to be a closed form only for $n=1, 2, 3$. We can easily check that  $\langle Z(\beta)^n \rangle_c \simeq\langle Z(n \beta)\rangle$ when $n=1,2,3$ for small temperatures, but proving this for $n >3$ is numerically difficult. We will content ourselves and assume $\langle Z(\beta)^n \rangle_c \simeq\langle Z(n \beta)\rangle$ is true for all $n$.

Meanwhile, even at the leading order of $\lambda$, the $T\oT$ deformation will give polynomial corrections in $\beta$ for low temperatures. Hence, the corrections in the approximation $\langle Z(\beta)^n \rangle_0 \simeq \langle Z(n\beta)\rangle_0$ can still be neglected even when we consider the $T\oT$ deformation.

There is one caveat, however, the all-genus density of states does not vanish for $E<0$. Instead, it is exponentially suppressed when $E<0$: 
%
%
%
\be \rho_{\text{Airy}}(E) =\Ai'\left(-E\right)^2 +  E\, \Ai\left(-E\right)^2~.
\ee
This means for both good and bad signs of $\lambda$, there will be states with complex-valued energy. We discuss the two cases separately and in each case, there will be a plausible non-perturbative contribution from the $f^+_\lambda (E)$ branch defined in \eqref{f(E)}.

The deformed quenched free energy is defined and computed as
\begin{equation}
\label{eq:deformfree}
    F_{q,\lambda}(\beta)=-T\langle \log Z\rangle_\lambda=T\int_0^{\infty}\frac{dx}{x}\left[e^{-\mathcal{Z}_\lambda(x)}-e^{-x\langle Z\rangle_\lambda}\right]-T\log\langle Z\rangle_\lambda \,.
\end{equation}

\subsection{Bad sign: $\lambda > 0$}
\label{sec:bad}
In this subsection, we study the quenched free energy of the deformed Airy model with $\lambda > 0$. In the deformed theory's spectrum, as explained in the introduction, the deformed energy $f^-_\lambda \left(E  > \frac{1}{8\lambda} \right)$ is complex-valued. Therefore, a cut-off in $E$ at $\frac{1}{8\lambda}$ is needed to remove the complex-valued energy states. Furthermore, there could be contributions arising from non-perturbative states with energy given by the other branch $f^+_\lambda(E)$. These kind of non-perturbative effects have been rigorously studied in \cite{Griguolo:2021wgy, Iliesiu:2020zld}. Their analyses lead us to conjecture what the non-perturbative contribution could be in the deformed Airy theory for the bad sign $\lambda > 0$. We will study the quenched free energy with and without the non-perturbative contributions. In both cases, we find the quenched free energy $F_{q,\lambda}(T)$ diverges for every $\lambda>0$ and $T$ in our low temperature approximation. It is unclear to us whether the low temperature approximation which causes $F_{q,\lambda}(T)$ to diverge or this is the feature of the deformation itself. Additionally, we do not know if repeating the same calculation in JT gravity will lead to the same problem. We will leave this for future investigations.

For the bad sign $\lambda > 0$, the issue with complex-valued energy is fixed by imposing a cut-off in the energy $E$ such that the deformed energy spectrum is real-valued. Furthermore, the resurgent analysis in \S 4.2 of \cite{Griguolo:2021wgy} indicates that there could be non-perturbative contributions to the partition function coming from the other branch $f^+_\lambda(E)$. More specifically, the deformed genus-zero partition function of JT gravity is given by 
\bea 
\label{eq:JTdisk}
\langle Z(\beta) \rangle^{g=0}_{\operatorname{JT},\lambda} &= \int_0^{1/8\lambda} dE \, \rho_{\operatorname{JT}}^{g=0}(E) \bigg(e^{-\beta f^-_\lambda (E)} - e^{-\beta f^+_\lambda (E)}\bigg)
\eea
where 
\be \rho_{\operatorname{JT}}^{g=0}(E) = \frac{\sinh{\left( 2\pi \sqrt{E}\right)}}{4\pi^2}
\ee
is just the usual JT gravity density of states for the disk. After a change of variables 
\be
\mathcal{E}=\frac{1}{4\lambda}\left(1\pm\sqrt{1-8\lambda E}\right)
\ee
for the two terms in \eqref{eq:JTdisk} respectively, they combine into a single integral (see for example \cite{Griguolo:2021wgy})
\bea
\langle Z(\beta) \rangle^{g=0}_{\operatorname{JT},\lambda}=\int_0^{1/2\lambda} d\mathcal{E} \, \rho_{\operatorname{JT},\lambda}^{g=0}(\mathcal{E}) e^{-\beta \mathcal{E}}\,,
\eea
where the deformed density of states is
\bea
\rho_{\operatorname{JT},\lambda}^{g=0}(E) &=  \frac{d\mathcal{E}^{-1}}{dE}\, \rho_{\operatorname{JT},0}^{g=0} \left(\mathcal{E}^{-1}\right) \\&=  (1-4\lambda E)\, \rho_{\operatorname{JT}}^{g=0}(E-2\lambda E^2)\,.
\eea
Motivated by this, and following our working definition of $T\overline{T}$ deformation in the beginning of this section, it is natural to expect that for the Airy model deformed by $\lambda > 0$, we should have the one-point function upon an identical change of variables as before 
\bea 
\label{eq:Airyguess}
\langle Z(\beta) \rangle_{\text{Airy},\lambda} &= \int_{-\infty}^{1/8\lambda} dE \, \rho_{\text{Airy}}(E) \, \bigg(e^{-\beta f^-_\lambda (E)} - e^{-\beta f^+_\lambda (E)}\bigg) \\
&= \int_{-\infty}^{\infty} d\mathcal{E} \, \rho_{\text{Airy},\lambda}(\mathcal{E}) e^{-\beta \mathcal{E}}\,,
\eea
where
\be \rho_{\text{Airy},\lambda}(E) = (1-4\lambda E) \rho_{\text{Airy}}(E-2\lambda E^2)\,,
\ee
and the second term in the first line signals non-perturbative effects. Note that here, instead of using the integration transformation for $\lambda < 0$, we must resort to our working definition introduced at the beginning of this section as the new prescription for the $T\overline{T}$ deformation which agrees with \eqref{eq:nonpert}. 

One can use this result together with the approximation $\langle Z(\beta)^n \rangle \simeq \langle Z(n\beta) \rangle$ to compute the quenched free energy. However, in this case, we find that the quenched free energy actually diverges. To see this, notice that the difference between the quenched free energy $F_q(\beta)$ and the annealed free energy $F_a(\beta)$ is given by
\be F_{q,\lambda} - F_{a,\lambda} = \int_0^\infty \frac{dx}{x} \left(e^{-\mathcal{Z}_\lambda(x)} -e^{-x \langle Z \rangle_\lambda}\right)\,,
\ee
where 
\begin{equation}
\label{eq:zcalbad}
\begin{aligned}
    \mathcal{Z}_\lambda(x) &=-\sum_{n=1}^{\infty}\frac{(-x)^n}{n!}\langle Z(n\beta)\rangle_{\text{Airy},\lambda} \\ 
    &=-\sum_{n=1}^{\infty}\frac{(-x)^n}{n!}\int_{-\infty}^{\frac{1}{8\lambda}}dE\, \rho_{\text{Airy}}(E)\left(e^{-n\beta f^-_\lambda (E)}  - e^{-n\beta f^+_\lambda (E)}\right) \\
    &= \int_{-\infty}^{\frac{1}{8\lambda}} dE \, \rho_{\text{Airy}}(E)\left[e^{-x e^{-\beta f^+_\lambda (E)}} - e^{-x e^{-\beta f^-_\lambda (E)}}\right]\,.
\end{aligned}
\end{equation}
Hence, in order for the $x$-integral to converge, the difference $D(x) \equiv e^{-\mathcal{Z}_\lambda(x)}-e^{-x \langle Z \rangle_\lambda}$ must at least go to zero as $x\rightarrow \infty$ but at least faster than $1/\ln x$. However, as we will show in the Appendix \ref{app:badsigndetail}, this is not the case for any $\lambda > 0$ and $T$. Here, we numerically plot $D(x)$ for $\lambda = 1/15$ and $T = 1/12$. As one can see, the function $D(x)$ is monotonically decreasing with $x$ at the beginning, but turns around and starts monotonically increasing at a very large value $x\sim 2\times 10^{19}$.

\begin{figure}[h]
    \centering
    \includegraphics[scale = 0.65]{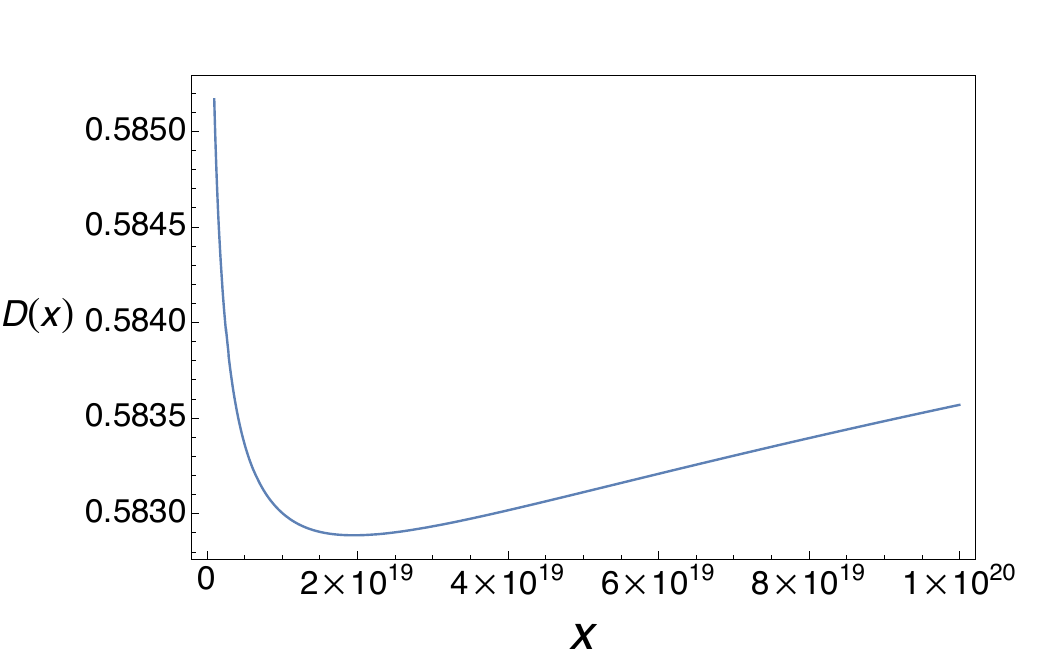}
    \caption{We plot $D(x)$ at $\lambda  = 1/15$ and $T = 1/12$ without the non-perturbative branch. We see the turning point is at $x\sim 2\times 10^{19}$.}
    \label{fig:diverge1}
\end{figure}

One might wonder if the non-perturbative branch with the energy $f^+_\lambda (E)$ causes the integral to diverge. We can certainly only include perturbative branch when computing the quenched free energy. However, as we show further justification in Appendix \ref{app:badsigndetail}, the integral still diverges. One can see this divergence numerically from $D(x)$ asymptotic to some finite number in the $x\rightarrow \infty$ limit.

\begin{figure}[h]
    \centering
    \includegraphics[scale = 0.65]{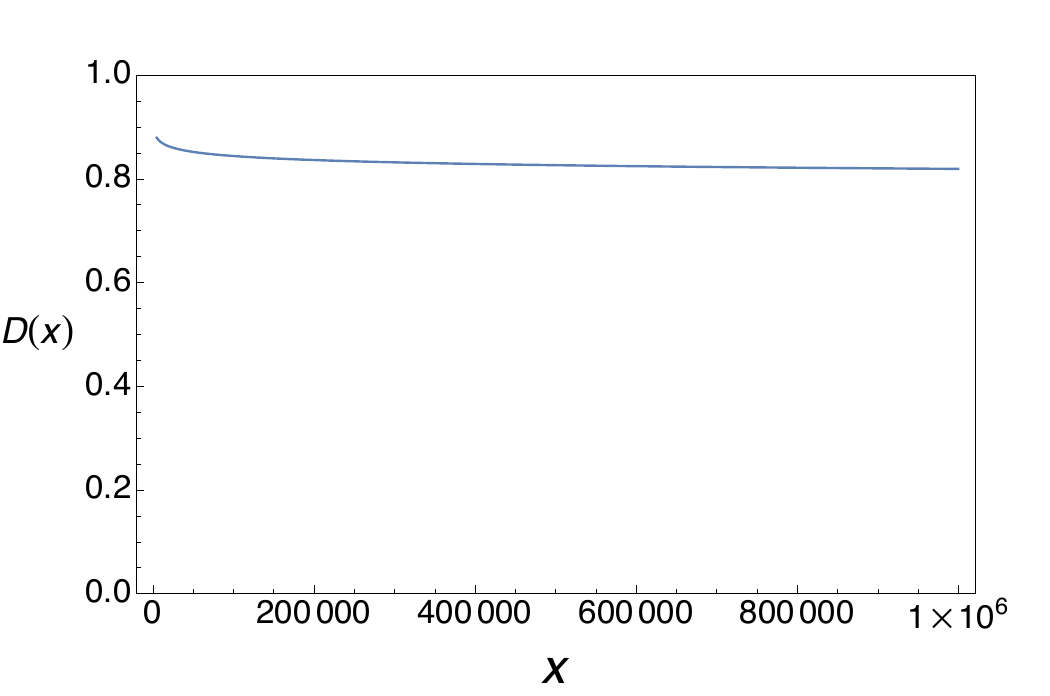}
    \caption{We plot $D(x)$ at $\lambda = 1/10$ and $T = 1/10$ without the non-perturbative branch. Though not as dramatic as Figure \ref{fig:diverge1}, as one can see $D(x)$ still does not go to zero as $x\rightarrow \infty$.}
    \label{fig:diverge2}
\end{figure}

It is unclear to us if such divergence is intrinsically physical, or due to any of our approximations. There are several possibilities. For instance, Okuyama's formula \eqref{OkuyamaF} may fail for the deformed theory in general. The derivation of \eqref{OkuyamaF} in \cite{Okuyama:2021pkf} requires one to exchange the integral with an infinite sum which is not absolutely convergent. This may lead to the failure of \eqref{OkuyamaF} in the deformed theory. Another possibility could be that the divergence is due to the non-perturbative instability of the Airy model. A reliable way to rule out some of these possibilities is to extend our work to JT gravity with the proper improvement of its non-perturbative behavior. We wish to investigate this in JT gravity per se in the future work to see if this divergence still persists.

\subsection{Good sign: $\lambda < 0$}\label{goodsign}
In this subsection, we compute the quenched free energy for the good sign $\lambda < 0$. Since non-perturbatively the density of states of Airy model extends to $E = -\infty$, the deformed theory will have unitarity issues caused by complex-valued energy which seems not to be discussed before in the previous literature of deformed JT gravity. All the densities of states in the literature have lower bounds, i.e., $\rho(E  \leq E_0 ) = 0$. Thus, by considering $\tilde{\rho}(E) = \rho(E-E_0)$, we can also have a well-defined spectrum for all $\lambda < 0$. This is why $\lambda < 0$ is referred as the \emph{good sign} in the literature\footnote{Originally called the ``wrong sign'', but suggested to be called the ``bad sign'' by \cite{Jiang:2019epa} instead.} \cite{Banerjee:2019ewu}. Here, the non-perturbative effect makes $\rho(E) \neq 0$ for all $E\in \mathbb{R}$. Therefore, the deformed energy spectrum will be complex-valued for $E < \frac{1}{8\lambda}$. There are two possible treatments. The first option is to impose a cut-off in the deformed energy spectrum up to when it becomes complex-valued. This cut-off resolves the unitarity issues, however, this leads to a violation of the flow equation \eqref{flow} of $\langle Z(\beta) \rangle_\lambda$ as a boundary term arises at the cut-off $E = \frac{1}{8\lambda}$. Alternatively, we may include these states with complex-valued energy, but we must then include their corresponding states from the non-perturbative sectors as well as to ensure that the partition function is real. We will refer this part as non-perturbative. By carefully choosing coefficients, the contributions in the boundary piece of this term cancels the boundary term \eqref{eq:bdry} in the first option. Thus, the flow equation of the one-point function $\langle Z(\beta) \rangle_\lambda$ will be satisfied in this case. We will study the quenched energy with and without the non-perturbative contribution. Excluding the non-perturbative contribution, we numerically confirm the quenched free energy is monotonically decreasing with temperature $T$ at a given $\lambda < 0$. We also find the quenched free energy monotonically decreases as we increase the absolute value of $\lambda$. Including the non-perturbative branch, unfortunately, we find that the quenched free energy computed from Okuyama's formula \eqref{OkuyamaF} diverges in general and we will illustrate this subtlety numerically in this subsection and include an analytical analysis in Appendix \ref{app:goodsigndetail}. 

We start with discussing in detail how to treat the complex-valued energy states in the deformed spectrum. One may expect that the correct answer is given by the exact recipe for the bad sign $\lambda > 0$, i.e., we cut off the spectrum below $E<\frac{1}{8\lambda}$ where the deformed energy becomes complex-valued and include the other branch for the remaining spectrum. However, there are two objections. The first objection is that if we consider the genus expansion, then the spectrum at genus-zero does not extend to $E<0$. As a result, the deformed spectrum remains unchanged and the good sign of the $T\oT$  deformation is well-defined making no additional branch required. If we included the other branch, through the genus expansion, the genus-zero partition function will receive corrections from the other branch as well which lead to inconsistencies. The second objection is that the Boltzmann weight $e^{-\beta f^+_\lambda (E)}$ diverges as $e^{\beta \sqrt{E/2|\lambda|}}$ when $E\rightarrow \infty$. Hence, the contribution from the other branch is divergent. These two reasons suggest we should not include the contribution from the other branch for the real-valued energy region. Thus, one might conclude the deformed partition function for the good sign $\lambda < 0$ is simply given by truncating the spectrum with complex-valued energy:
\be \langle Z(\beta) \rangle_{\lambda,\text{guess}} = \int_{\frac{1}{8\lambda}}^\infty dE \, \rho_{\text{Airy}}(E) e^{-\beta f_\lambda^-(E)} = \int_{\frac{1}{4\lambda}}^\infty dE\, \rho_{\text{Airy},\lambda}(E) e^{-\beta E}\,,
\ee
where 
\be \rho_{\text{Airy},\lambda}(E) = (1-4\lambda E)\rho_{\text{Airy}}(E(1-2\lambda E))\,.
\ee
However, there is one caveat. The deformed partition function should satisfy the differential equation \eqref{flow} derived in \cite{Iliesiu:2020zld}: 
\be 
\label{eq:flow}
\left[4\lambda \partial_\lambda \partial_\beta + 2\beta \partial_\beta^2 - \bigg(\frac{4\lambda}{\beta} - 1\bigg)\partial_\lambda\right] \langle Z(\beta) \rangle_\lambda = 0\,.
\ee
For convenience, we introduce the differential operator $\mathcal{F} \equiv 4\lambda \partial_\lambda \partial_\beta + 2\beta \partial_\beta^2 - \left(\frac{4\lambda}{\beta} - 1\right)\partial_\lambda$. Consider a change of variables $E = \tilde{E} + \frac{1}{8\lambda}$ so that the bound of the integral does not depend on $\lambda$:
\be \langle Z(\beta) \rangle_{\lambda, \text{guess}} = \int_0^\infty d\tilde{E} \, \rho_{\text{Airy}}\left(\tilde{E} + 1/8\lambda\right) e^{-\beta f_\lambda^-\left(\tilde{E} + 1/8\lambda\right)}\,.
\ee
As one can show, $\mathcal{F}$ acting on the integrand leads to a total derivative
\be 
\begin{aligned}
&\mathcal{F}\left[\rho_{\text{Airy}}\left(\tilde{E} + \frac{1}{8\lambda}\right) e^{-\beta f_\lambda^-\left(\tilde{E} + 1/8\lambda\right)}\right] \\
&\quad\quad\quad\quad\quad\quad\quad\quad\quad\quad= \frac{d}{d\tilde{E}}\left[e^{-\beta f_\lambda^- \left(\tilde{E}+1/8\lambda\right)} \rho_{\text{Airy}}\left(\tilde{E}+\frac{1}{8\lambda}\right) \frac{4\lambda - \beta\sqrt{-8\lambda \tilde{E}}}{8\beta\lambda^2}\right]\,.
\end{aligned}
\ee
Therefore,
\bea 
\label{eq:bdry}
\mathcal{F}\left[\langle Z(\beta) \rangle_{\lambda,\text{guess}}\right] &= \left[e^{-\beta f_\lambda^-(\tilde{E} + 1/8\lambda)} \rho_{\text{Airy}} \left(\tilde{E}+\frac{1}{8\lambda}\right)\frac{4\lambda - \beta\sqrt{-8\lambda \tilde{E}}}{8\beta\lambda^2}\right]\Bigg|_{\tilde{E}=0}^\infty \\
&= -\frac{e^{-\beta f_\lambda^-(1/8\lambda)}}{2\beta\lambda}\rho_{\text{Airy}}\left(\frac{1}{8\lambda}\right)\,.
\eea
Now we see the problem: our guess $\langle Z(\beta) \rangle_{\lambda,\text{guess}}$ violates the flow equation \eqref{eq:flow} due to the appearance of the boundary term \eqref{eq:bdry}. This is just another manifestation of the non-perturbative effect of the Airy model. If $\rho_{\text{Airy}}(E)$ had been supported on $[E_0,\infty)$, we can shift the ground state energy such that $\tilde{\rho}_{\text{Airy}}(E) = \rho_{\text{Airy}}(E - E_0)$ to remove the complex-valued energy and make $\tilde{\rho}_{\text{Airy}}(\frac{1}{8\lambda}) = 0$ such that the flow equation is satisfied. However, this is not possible due to the non-perturbative effects as $\rho_{\text{Airy}}(E)$ has support on the entire real axis.

Thus, to make sure that $\langle Z(\beta) \rangle_{\lambda, \text{guess}}$ satisfies the flow equation \eqref{eq:flow} while keeping it finite, we must include the complex-valued energy region where $E \in (-\infty, 1/8\lambda)$ to cancel the unwanted boundary term. To make sure the deformed partition function is real, we must also add its complex conjugate, i.e., the contribution from the other branch. Therefore, the partition function $\langle Z(\beta) \rangle_\lambda$ for $\lambda < 0$ should be given by
\be 
\label{eq:good1pt}
\langle Z(\beta) \rangle_{\text{Airy},\lambda} = \int_{1/8\lambda}^\infty dE\, \rho_{\text{Airy}}(E) e^{-\beta f_\lambda^-(E)} + \int_{-\infty}^{1/8\lambda} dE\, \rho_{\text{Airy}}(E) \frac{e^{-\beta f_\lambda^-(E)} + e^{-\beta f_\lambda^+(E)}}{2}\,,
\ee
where the sum of exponentials in the second integrand can be rewritten in terms of cosine as
\be
\label{eq:cosine1}
e^{-\frac{1}{4\lambda T}}\int_{-\infty}^{1/8\lambda} dE\, \rho_{\text{Airy}}(E)\cos \left( \frac{\sqrt{8\lambda E-1}}{4\lambda T}\right)
\ee
and is a highly oscillatory integral when $\lambda$ or $T$ is small, but can still fairly easily be numerically evaluated to be finite. As one can check, the boundary terms indeed cancel with each other and the flow equation \eqref{eq:flow} is satisfied.

One can then use Okuyama's formula \eqref{OkuyamaF} to numerically compute the quenched free energy. In this case, we can express $\mathcal{Z}_\lambda(x)$ using \eqref{eq:good1pt} as the following:
\bea 
\label{eq:calZ}
\mathcal{Z}_\lambda(x) &\simeq  - \sum_{n=1}^\infty \frac{(-x)^n}{n!} \langle Z(n\beta)\rangle_{\text{Airy},\lambda} \\
&= -\int_{1/8\lambda}^\infty dE\, \rho_{\text{Airy}}(E) \sum_{n=1}^\infty \frac{(-x)^n}{n!}e^{-n\beta f_\lambda^-(E)} \\
&\quad\quad\quad\quad\quad\quad\quad\quad\quad\quad\,- \frac{1}{2}\int_{-\infty}^{1/8\lambda} dE\, \rho_{\text{Airy}}(E) \sum_{n=1}^\infty\frac{(-x)^n}{n!} \left(e^{-n\beta f_\lambda^-(E)} + e^{-n\beta f_\lambda^+(E)}\right) \\
&= \int_{1/8\lambda}^\infty dE\, \rho_{\text{Airy}}(E) \left(1 - e^{-x e^{-\beta f_\lambda^-(E)}}\right)\\
&\quad\quad\quad\quad\quad\quad\quad\quad\quad\quad\quad\,\,+ \frac{1}{2}\int_{-\infty}^{1/8\lambda} dE\, \rho_{\text{Airy}}(E)\left[2 - e^{-x e^{-\beta f_\lambda^+(E)}} - e^{-x e^{-\beta f_\lambda^-(E)}}\right]\,.
\eea
Since the second integral can be rewritten as
\begin{equation}
\label{eq:cosine2}
    \int_{-\infty}^{1/8\lambda} dE\, \rho_{\text{Airy}}(E)\Bigg[1-\underbrace{e^{-xe^{-\frac{1}{4 \lambda T}} \cos\frac{\sqrt{8\lambda E-1}}{4 \lambda T}} \cos \left(xe^{-\frac{1}{4\lambda T}}\sin\frac{\sqrt{8 \lambda E -1}}{4 \lambda T}\right)}_{\textstyle
    {\equiv I(E,x,\lambda,T)}}\Bigg],
\end{equation}
we will refer to it as $\mathcal{Z}_{\lambda}^{\text{cos}}(x)$.\footnote{Although \eqref{eq:cosine1} does not significantly change the numerical details.}, and the part $I(E,x,\lambda,T)$ will be the key ingredient in Appendix \ref{app:goodsigndetail}.

We can turn off the non-perturbative effect by including only the first term $\mathcal{Z}_\lambda^{\text{pert},<}(x)$ in $\mathcal{Z}_\lambda(x)\equiv\mathcal{Z}_\lambda^{\text{pert},<}(x)\mathcal{Z}_\lambda^{\text{cos}}(x)$ as well in the one-point function $\langle Z(\beta)\rangle_{\text{Airy},\lambda}$. Next, we will study the quenched free energy first with and then without the non-perturbative contribution.

\subsubsection{Without the non-perturbative contribution}
We first study the quenched free energy without the contribution from the non-perturbative part. In this case, the numerical calculation is straightforward without subtleties. We are able to numerically confirm that the deformed quenched free energy $F_{q,\lambda}(T)$ monotonically decreases as $T$ increases. Furthermore, we find the quenched free energy $F_{q,\lambda}(T)$ monotonically decreases as the absolute value of $\lambda$ increases and present our numerical results below in Figures \ref{fig:ncl} and \ref{fig:ncT}. 

\begin{figure}[h]
     \centering
     \begin{subfigure}[b]{0.48\textwidth}
         \centering
         \includegraphics[width=\textwidth]{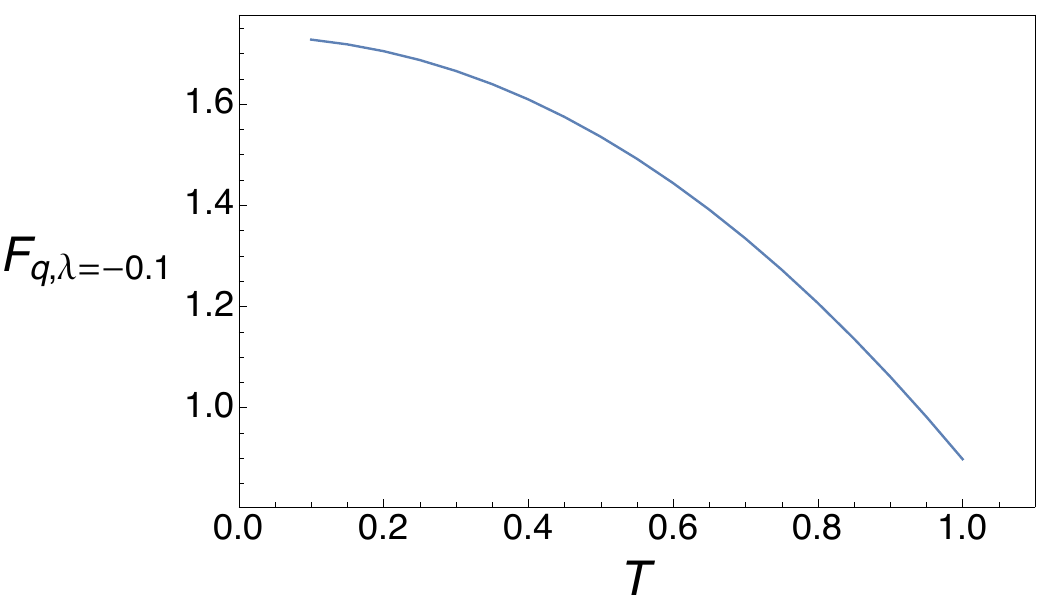}
         \caption{$\lambda = - 0.1$}
         \label{fig:ncl1}
     \end{subfigure}
     \hfill
     \begin{subfigure}[b]{0.48\textwidth}
         \centering
         \includegraphics[width=\textwidth]{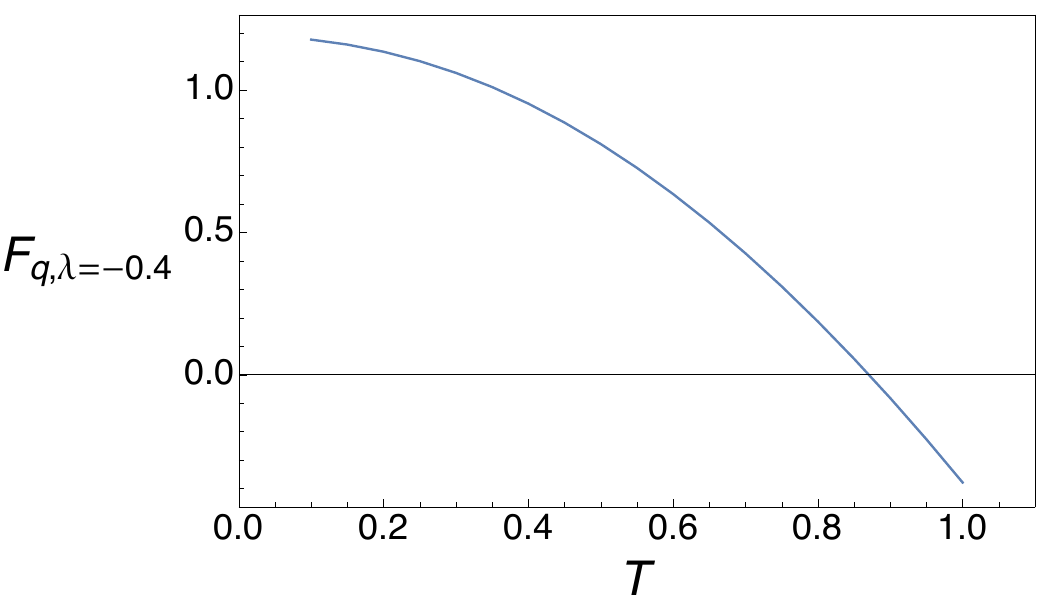}
         \caption{$\lambda = - 0.4$}
         \label{fig:ncl2}
     \end{subfigure}
     \hfill
     \begin{subfigure}[b]{0.48\textwidth}
         \centering
         \includegraphics[width=\textwidth]{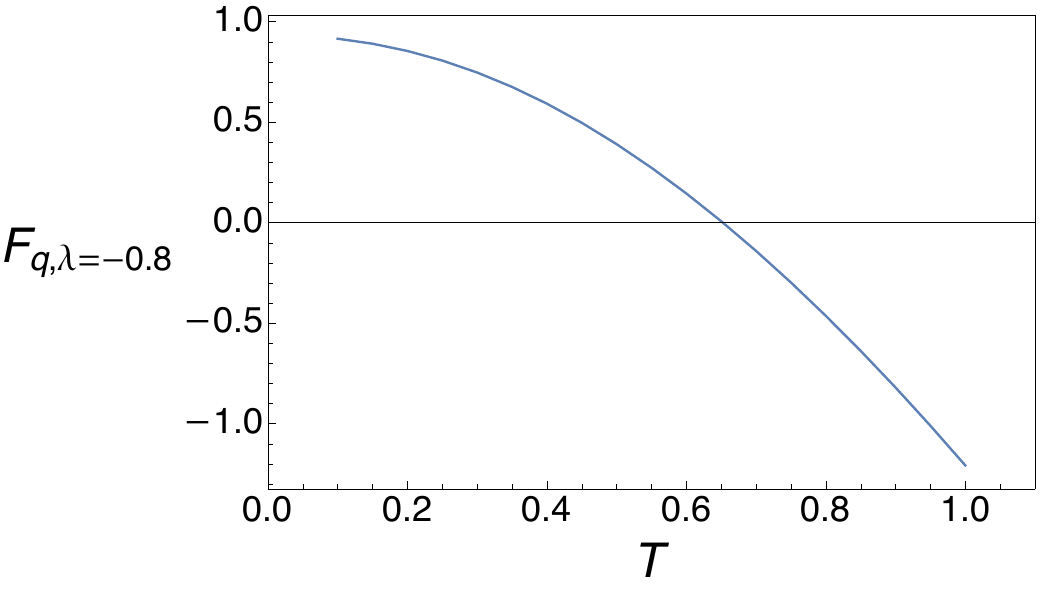}
         \caption{$\lambda = - 0.8$}
         \label{fig:ncl8}
     \end{subfigure}
     \hfill
     \begin{subfigure}[b]{0.48\textwidth}
         \centering
         \includegraphics[width=\textwidth]{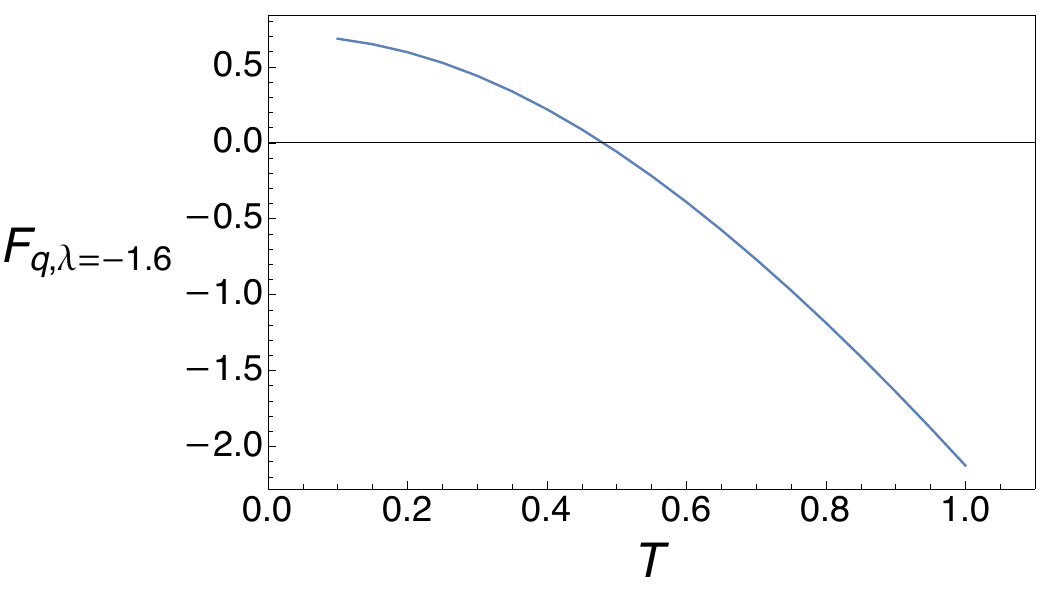}
         \caption{$\lambda = - 1.6$}
         \label{fig:ncl16}
     \end{subfigure}
     \hfill
     \begin{subfigure}[b]{0.48\textwidth}
         \centering
         \includegraphics[width=\textwidth]{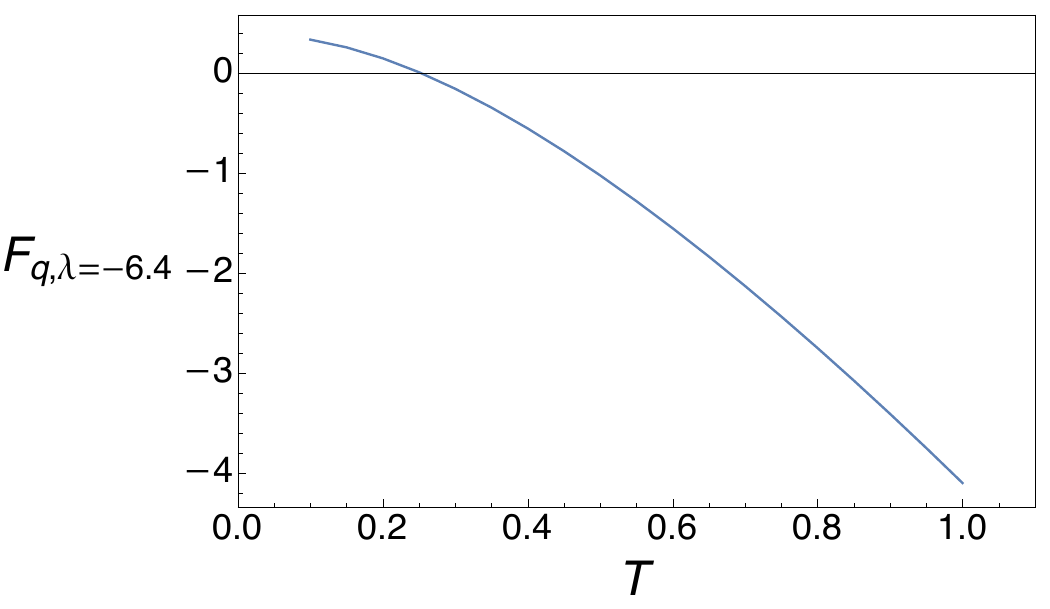}
         \caption{$\lambda = - 6.4$}
         \label{fig:ncl64}
     \end{subfigure}
     \hfill
     \begin{subfigure}[b]{0.48\textwidth}
         \centering
         \includegraphics[width=\textwidth]{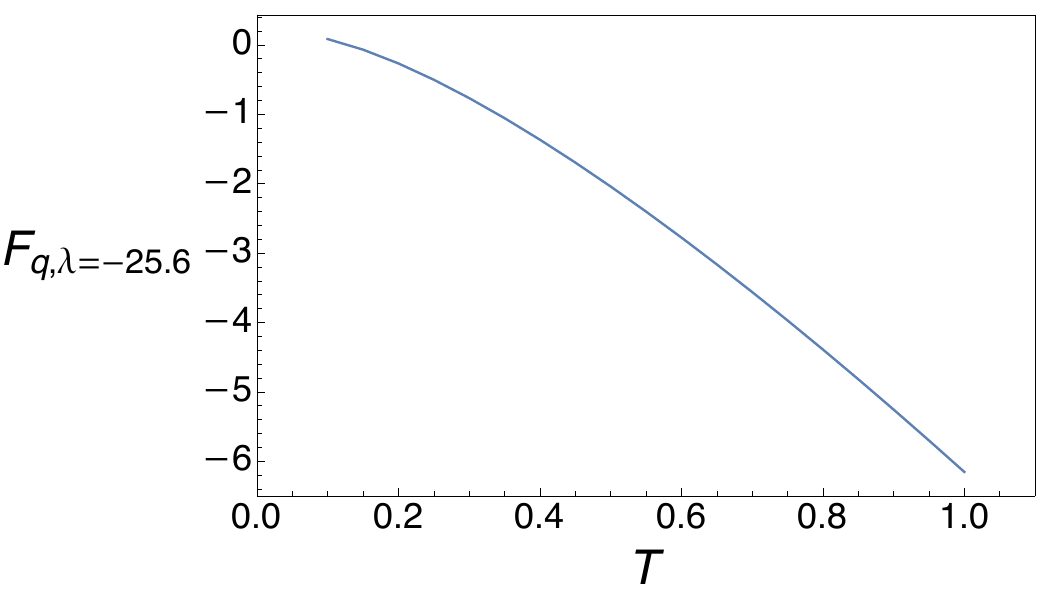}
         \caption{$\lambda = - 25.6$}
         \label{fig:ncl256}
     \end{subfigure}
     \caption{We plot the quenched free energy $F_{q,\lambda}(T)$ without the contribution from the non-perturbative branch as a function of $T$ for $\lambda = -0.1,-0.4,-0.8,-1.6,-6.4,-25.6$. As shown, $F_{q,\lambda}(T)$ is a monotonically decreasing function of $T$ in the deformed theory.} 
     \label{fig:ncl}
\end{figure}

\begin{figure}[h]
     \centering
     \begin{subfigure}[b]{0.48\textwidth}
         \centering
         \includegraphics[width=\textwidth]{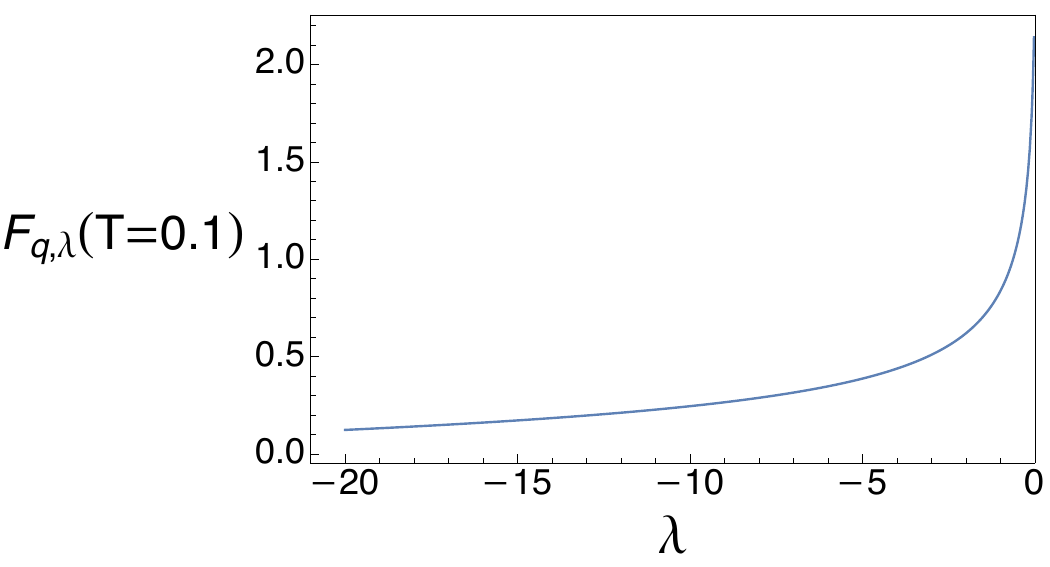}
         \caption{$T = 0.1$}
         \label{fig:ncT1}
     \end{subfigure}
     \hfill
     \begin{subfigure}[b]{0.48\textwidth}
         \centering
         \includegraphics[width=\textwidth]{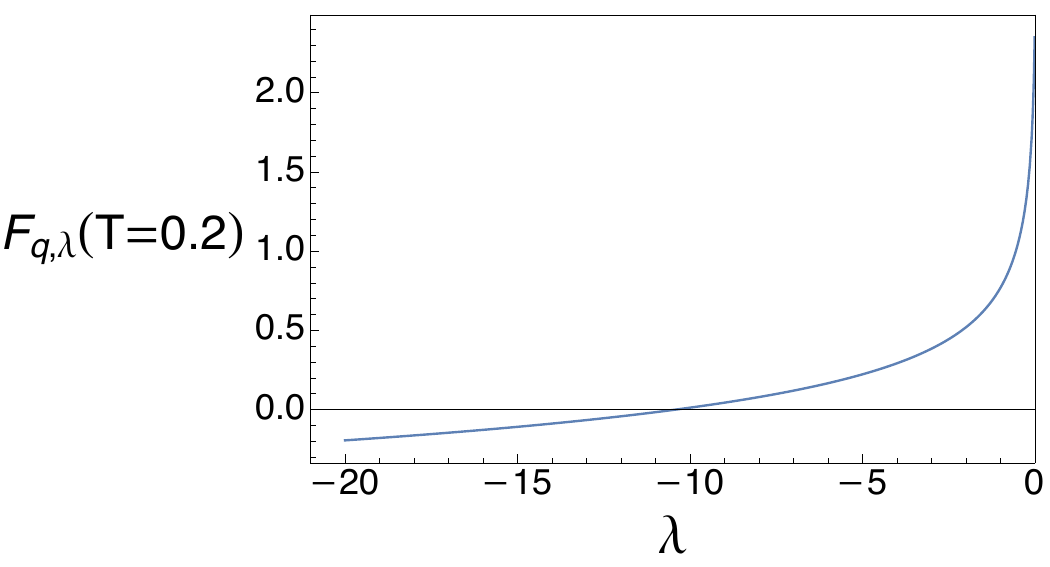}
         \caption{$T = 0.2$}
         \label{fig:ncT2}
     \end{subfigure}
     \hfill
     \begin{subfigure}[b]{0.48\textwidth}
         \centering
         \includegraphics[width=\textwidth]{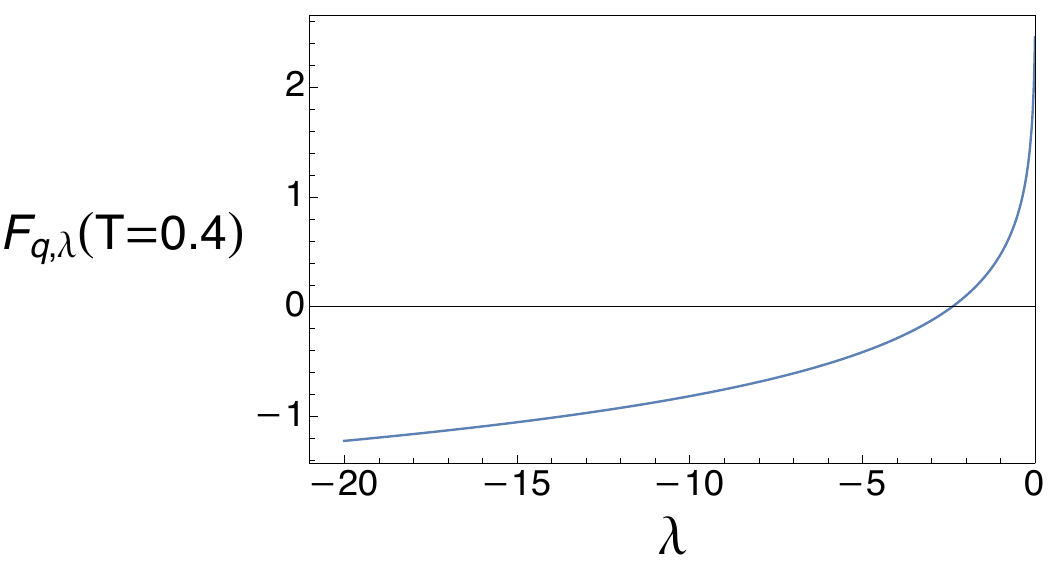}
         \caption{$T = 0.4$}
         \label{fig:ncT4}
     \end{subfigure}
     \hfill
     \begin{subfigure}[b]{0.48\textwidth}
         \centering
         \includegraphics[width=\textwidth]{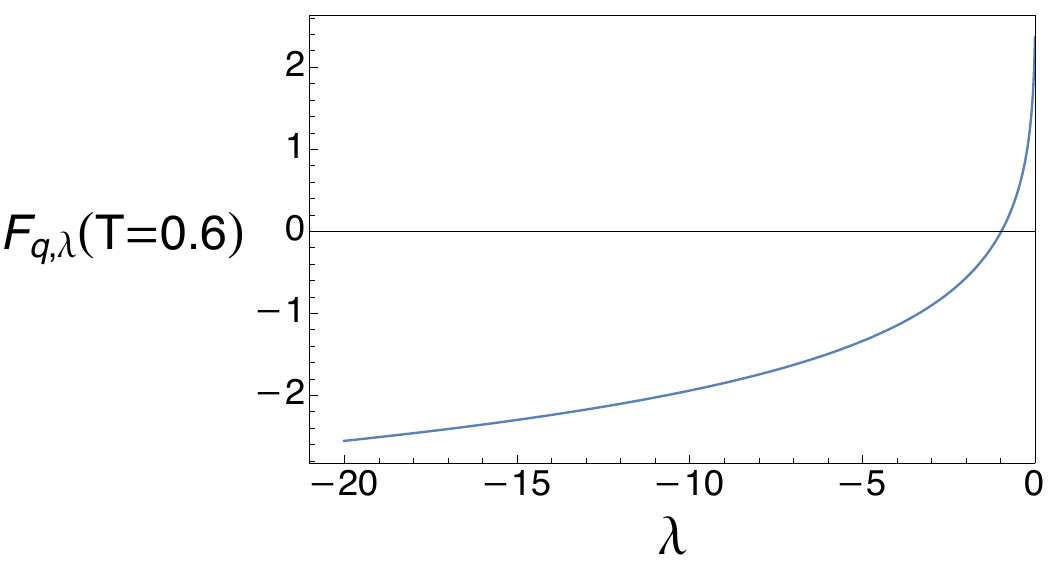}
         \caption{$T = 0.6$}
         \label{fig:ncT6}
     \end{subfigure}
     \caption{We plot the quenched free energy $F_{q,\lambda}(T)$ without the contribution from the non-perturbative branch as a function $\lambda$ for fixed $T = 0.1,0.2,0.4,0.6$. As shown, $F_{q,\lambda}(T)$ is monotonically decreasing as $|\lambda|$ increases.}
     \label{fig:ncT}
\end{figure}

\subsubsection{With the non-perturbative contribution}
\begin{figure}[H]
    \centering
    \includegraphics[scale = 0.7]{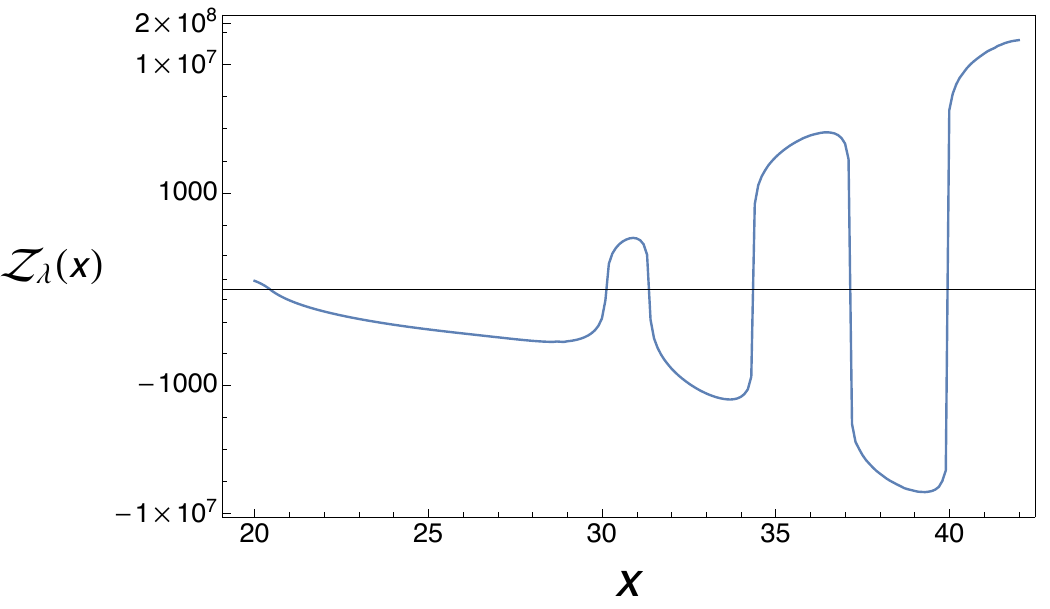}
    \caption{We plot $\mathcal{Z}_\lambda(x)$ against $x$ at $T = 0.3$ and $\lambda = -2$ in the symmetric logarithmic scale, i.e., $\left[\text{sgn}\,\mathcal{Z}_\lambda(x)\right]\log\left[|\mathcal{Z}_\lambda(x)|+1\right]$. One can see that the valleys' depths deepening towards the right makes $e^{-\mathcal{Z}_\lambda(x)}$ [and therefore $D(x)$ and $F_{q,\lambda}(T)$] unbounded in the large $x$ limit. Yet, it is not obvious how to rigorously derive the locations of these valleys in terms of $x$ and this could be an interesting exercise in (numerical) analysis.}
    \label{fig:cosine0}
\end{figure}

Here, similarly to what has been done in \S \ref{sec:bad} (Figures \ref{fig:diverge1} and \ref{fig:diverge2} in particular), we numerically demonstrate the integral diverges by showing the difference $D(x) \equiv e^{-\mathcal{Z}_\lambda(x)}-e^{-x \langle Z \rangle_\lambda}$ does not vanish in the large $x$ limit. Notice that $e^{-x \langle Z\rangle_\lambda}\rightarrow 0$ in the large $x$ limit and we can show that $D(x)$ diverges as $x\rightarrow\infty$ by proving $\mathcal{Z}_\lambda(x)$ is oscillating with its amplitude growing rapidly with $x$. All the analytical detail will be provided in Appendix \ref{app:goodsigndetail}. Here, instead, we will numerically plot Figure \ref{fig:cosine0} to illustrate this fact by showing the depth of the three valleys\footnote{They are not necessarily the first three valleys from the left because for small $x$, the valleys are not obvious and we did not attempt to analytically predict positions of valleys in terms of $x$.} deepening as $x$ increases which shows the divergence of $D(x)$. 

Incidentally, we can also resort to a commonly used trick for numerical improper integrals \cite{burden1997numerical}. We already know that the divergence comes from the $x\in[1,\infty)$ part of the $x$-integral \eqref{eq:deformfree}. On the other hand, because \eqref{eq:deformfree} is of the form $\int_a^{\infty}\frac{f(x)}{x}dx$, we can convert the integral to $\int_0^{1/a}\frac{f(1/t)}{t}dt$ by a change of variables $t=1/x$. One can apply this to \eqref{eq:deformfree} with $a=1$ to observe its divergence over a compact interval $t=1/x\in [0,1]$.

\section{Conclusion}
\label{Conclusion}
In this paper, we studied the $T\overline{T}$-deformed correlators for JT gravity and its dual matrix model. Additionally, we computed the quenched free energy of the Airy model under the $T\overline{T}$ deformation for both signs of $\lambda$ and their non-perturbative features. 

We briefly summarize our numerical results. At genus-zero and at the leading order of perturbation theory in $\lambda$, we confirmed the quenched free energy $F_{q,\lambda}(T)$ is a monotonic function in $T$ for a given $\lambda$ within the validity domain of the leading order approximation. We also find the good sign $\lambda<0$ deformation decreases $F_{q,\lambda}(T)$ while the bad sign $\lambda > 0$ increases $F_{q,\lambda}(T)$. 

For all genus and in the low temperature approximation, we computed the quenched free energy $F_{q,\lambda}(T)$ using Okuyama's formula \eqref{OkuyamaF} which diverges regardless whether we include the non-perturbative contribution of the $T\oT$ deformation or not when $\lambda > 0$. For the good sign $\lambda < 0$, we are able to numerically compute $F_{q,\lambda}(T)$ without including the potential non-perturbative contributions and confirm the monotonicity of $F_{q,\lambda}(T)$ at low temperature. Additionally, we find $F_{q,\lambda}(T)$ decreases by the deformation and matches the result in perturbation theory at genus-zero. However, when including the possible contributions from the non-perturbative branch, we find $F_{q,\lambda}(T)$ computed from Okuyama's formula \eqref{OkuyamaF} diverges again.

We conclude with a few open questions and future directions relevant to extending this work.

\begin{itemize}
    \item The first question would be to understand the source of the divergence in the deformed quenched free energy $F_{q,\lambda} (T)$. While it is possible that Okuyama's formula \eqref{OkuyamaF} may fail or require modification for the deformed theory. We suspect that the analytical continuation or exchanging order of the infinite sum and integral in the derivation may not hold for the $T\oT$-deformed theory. Additionally, it is also likely that the divergence is caused by the non-perturbative instability suffered from the Airy model. There have been works on how to improve the non-perturbative feature of JT gravity to remove this undesired feature \cite{Saad:2019lba,Johnson:2019eik,Johnson:2020exp,Johnson:2020heh,Johnson:2020lns,Johnson:2021rsh,Johnson:2021tnl,Johnson:2022wsr}. An interesting extension of our work is to study the deformed JT gravity at general temperatures and see if the same divergence we found in the Airy model appears as well as check if the non-perturbative instability would affect the quenched free energy qualitatively. For the good sign $\lambda < 0$ without the non-perturbative contributions from the $T\oT$ deformation, we expect the finite $F_{q,\lambda} (T)$ to decrease hold in JT gravity at general temperatures.
    \item As we have pointed out, there are non-perturbative effects in the Airy model which endow the density of states $\rho_{\text{Airy}}(E)$ with support on the entire real axis making the integral transformation \eqref{Direct} no longer hold. However, if we consider the genus expansion
\be \langle Z(\beta) \rangle_{\text{Airy}} = \sum_{g=0}^\infty \langle Z(\beta) \rangle_{\text{Airy}}^{g} \equiv \sum_{g=0}^\infty \frac{1}{g! \sqrt{4\pi \beta^3}}\bigg(\frac{\beta^3}{12}\bigg)^g\,,
\ee
then it is straightforward to compute the deformed one-point correlator. For instance, when $\lambda < 0$, we have the deformed one-point correlator relevant to the genus expansion
\be \langle Z(\beta) \rangle^g_{\text{Airy},\lambda}= \frac{e^{-\beta/4\lambda}}{g! \sqrt{-8\lambda} \pi \beta}  \bigg(\frac{\beta^3}{12}\bigg)^g K_{2-3g}\bigg(-\frac{\beta}{4\lambda}\bigg)\,.
\ee
Now, a natural question to ask is if one can sum over all $\langle Z(\beta) \rangle_{\text{Airy},\lambda}^g$ to produce the deformed one-point correlator $\langle Z(\beta) \rangle_{\text{Airy},\lambda}$? Unfortunately, as one can easily check, $\sum_{g=0}^\infty \langle Z(\beta) \rangle_{\text{Airy},\lambda}^g$ diverges. We naively expect a resurgent analysis to possibly help understand this. 

\item As we have explained, our version of the $T\oT$ deformation for the matrix model seems to require the measure in the matrix integral to be deformed as well. It would be important to further explore and figure out how to treat the branch cut in the half change of variables \eqref{half-diff1}-\eqref{half-diff2}. 

\end{itemize}
\newpage
\acknowledgments
We particularly thank Alexander Frenkel and Jiabao Yang for numerous discussions on JT gravity and matrix models. We thank Christian Ferko, David Gross, Ken Intriligator, Andreas Karch, Per Kraus and Ruben Monten for a careful read of the manuscript as well as providing useful suggestions. We thank Ho-Tat Lam and John McGreevy for helpful conversations. S.E. is supported from the Bhaumik Institute. H.-Y.S. is supported from the Simons Collaborations on Ultra-Quantum Matter, which is a grant from the Simons Foundation (651440, AK). Z.S. is supported from the US Department of Energy (DOE) under cooperative research agreement DE-SC0009919 and Simons Foundation award No. 568420 (K.I.). 

\appendix 

\section{Full rigorous details on the quenched free energy with bad sign $\lambda>0$}
\label{app:badsigndetail}
Here we will show that the quenched free energy defined in \eqref{eq:deformfree} always diverges when $\lambda>0$.

\subsection{Perturbative and nonperturbative branches combined}

As already shown numerically near Figures \ref{fig:diverge1} and \ref{fig:diverge2}, the difference $D(x)=e^{-\mathcal{Z}_\lambda(x)}-e^{-x \langle Z \rangle_\lambda}$ between the two numerators in the integrand \eqref{eq:deformfree} does not asymptote to zero as $x\rightarrow\infty$. This renders the $x$-integral to be divergent. In this appendix, we analytically examine the large-$x$ behavior of the integrand in \eqref{eq:deformfree} and show that the integral diverges when $\lambda>0$. We will see that this conclusion does not change regarding the fine details of the undeformed density of states $\rho_{\text{Airy}}(E)$.

The second term $e^{-x \langle Z \rangle_\lambda}$ in \eqref{eq:deformfree} surely vanishes as $x\rightarrow\infty$, and we will show that the exponent $-\mathcal{Z}_\lambda(x)$ in the first term vanishes as $x\rightarrow\infty$. We define the portion of the $E$-integrand in \eqref{eq:zcalbad} as
\begin{equation}
    h(E,x,\lambda,T)\equiv e^{-xe^{-\frac{1+\sqrt{1-8\lambda E}}{4\lambda T}}}-e^{-xe^{-\frac{1-\sqrt{1-8\lambda E}}{4\lambda T}}}\,,
\end{equation}
where $\beta$ has been replaced by the inverse temperature $1/T$, and we will work with $T$ in both Appendices \ref{app:badsigndetail} and \ref{app:goodsigndetail}.

First, we notice that, because $x,\lambda,T>0$ then
\begin{equation}
\lim_{E\rightarrow-\infty}h(E,x,\lambda,T)=1
\end{equation}
and
\begin{equation}
    \lim_{E\rightarrow1/8\lambda}h(E,x,\lambda,T)=0\,.
\end{equation}
Next, it is straightforward to see that the function $h$ is monotonically decreasing with $E$:
\begin{equation}
\label{eq:slope}
    \frac{\partial}{\partial E}h(E,x,\lambda,T)=-\frac{x e^{-\frac{\sqrt{1-8\lambda E}+1}{4\lambda T}} \left(e^{\frac{\sqrt{1-8 \lambda E}}{2\lambda T}-x e^{\frac{\sqrt{1-8\lambda E}-1}{4 \lambda T}}}+e^{-xe^{-\frac{\sqrt{1-8 \lambda E}+1}{4\lambda  T}}}\right)}{T \sqrt{1-8 \lambda E}}<0\,.
\end{equation}
Now let us examine how the value $h(0,x,\lambda,T)=e^{ -xe^{-\frac{1}{2 \lambda T}}}-e^{-x}$ changes with $x$:
\begin{equation}
    \frac{\partial}{\partial x}h(0,x,\lambda,T)=e^{-x}-e^{x \left(-e^{-\frac{1}{2 \lambda  T}}\right)-\frac{1}{2 \lambda T}}\,,
\end{equation}
and when $x>\frac{1}{2\lambda T\big(1-e^{-\frac{1}{2 \lambda  T}}\big)}>0$ (generically a very large value at low temperature), the value $h(0,x,\lambda,T)$ monotonically decreases as $x$ increases. For a fixed $\lambda$ and $T$, at large $x$, this value\footnote{One warning is that when $\lambda T$ is sufficiently small, around $10^{-3}$, numerical calculations (by Mathematica) are no longer reliable. To improve precision, we use
\texttt{MinRecursion}$\rightarrow9$ or \texttt{WorkingPrecision}$\rightarrow10$ for \texttt{NIntegrate}.} is $1-1=0$. This implies that as $x$ increases, the graph of $h(E,x,\lambda,T)$ against $E$ is shifting to the left monotonically.
Using these qualitative characteristics of $h(E,x,\lambda,T)$ and with some help from numerical plots for generic $x,\lambda,T$, we can see it is essentially a step function reflected (across the vertical axis) and translated to the left with corners smoothed out. 

Finally we examine the effect of $\lambda,T$ on $h(E,x,\lambda,T)$. For small $\lambda$ and $T$ and $x$ close to $1$, the slope of the function $h$ at $E=0$ changes with $T$ and $\lambda$ as
\begin{equation}
\begin{aligned}
    \frac{\partial}{\partial T}\left(\frac{\partial h}{\partial E}\bigg|_{E=0}\right)&=\frac{2\lambda T e^{\frac{1}{2\lambda T}}\left(e^{x e^{-\frac{1}{2 \lambda T}}+\frac{1}{2\lambda T}}+e^x\right)+\left(x-e^{\frac{1}{2\lambda T}}\right)e^x}{2 \lambda T^3e^x}x e^{-\frac{1}{\lambda T} -xe^{-\frac{1}{2 \lambda  T}}}\\
    &=\frac{2\lambda Te^xe^{\frac{1}{2\lambda T}}+e^{\frac{1}{2\lambda T}}\left(e^{xe^{-\frac{1}{2\lambda T}}}e^{\frac{1}{2\lambda T}-\log\frac{1}{2\lambda T}}-e^x\right)}{2 \lambda T^3e^x}xe^{-\frac{1}{\lambda T} -xe^{-\frac{1}{2 \lambda  T}}}\\
    &>\frac{e^{\frac{1}{2\lambda T}}\left(e^{\frac{1}{2\lambda T}-\log\frac{1}{2\lambda T}}-e^x\right)}{2 \lambda T^3e^x}xe^{-\frac{1}{\lambda T} -xe^{-\frac{1}{2 \lambda  T}}}\\
    &>\frac{e^{\frac{1}{2e\lambda T}}-e^x}{2 \lambda T^3e^x}xe^{-\frac{1}{2\lambda T} -xe^{-\frac{1}{2 \lambda  T}}}>0
    \end{aligned}
\end{equation}
and
\begin{equation}
\label{eq:lambdaslope}
    \frac{\partial}{\partial \lambda}\left(\frac{\partial}{\partial E}h\bigg|_{E=0}\right)=\frac{x-e^{\frac{1}{2 \lambda  T}}}{2 \lambda ^2 T^2}x e^{-\frac{1}{\lambda  T}-xe^{-\frac{1}{2 \lambda T}}}<0\,,
\end{equation}
respectively, so as temperature $T$ decreases, $h$ becomes steeper and steeper at $E=0$ and the other way around (although very inconspicuously) when $\lambda$ decreases. 

\begin{figure}[h]
    \centering
    \includegraphics[width = 15cm]{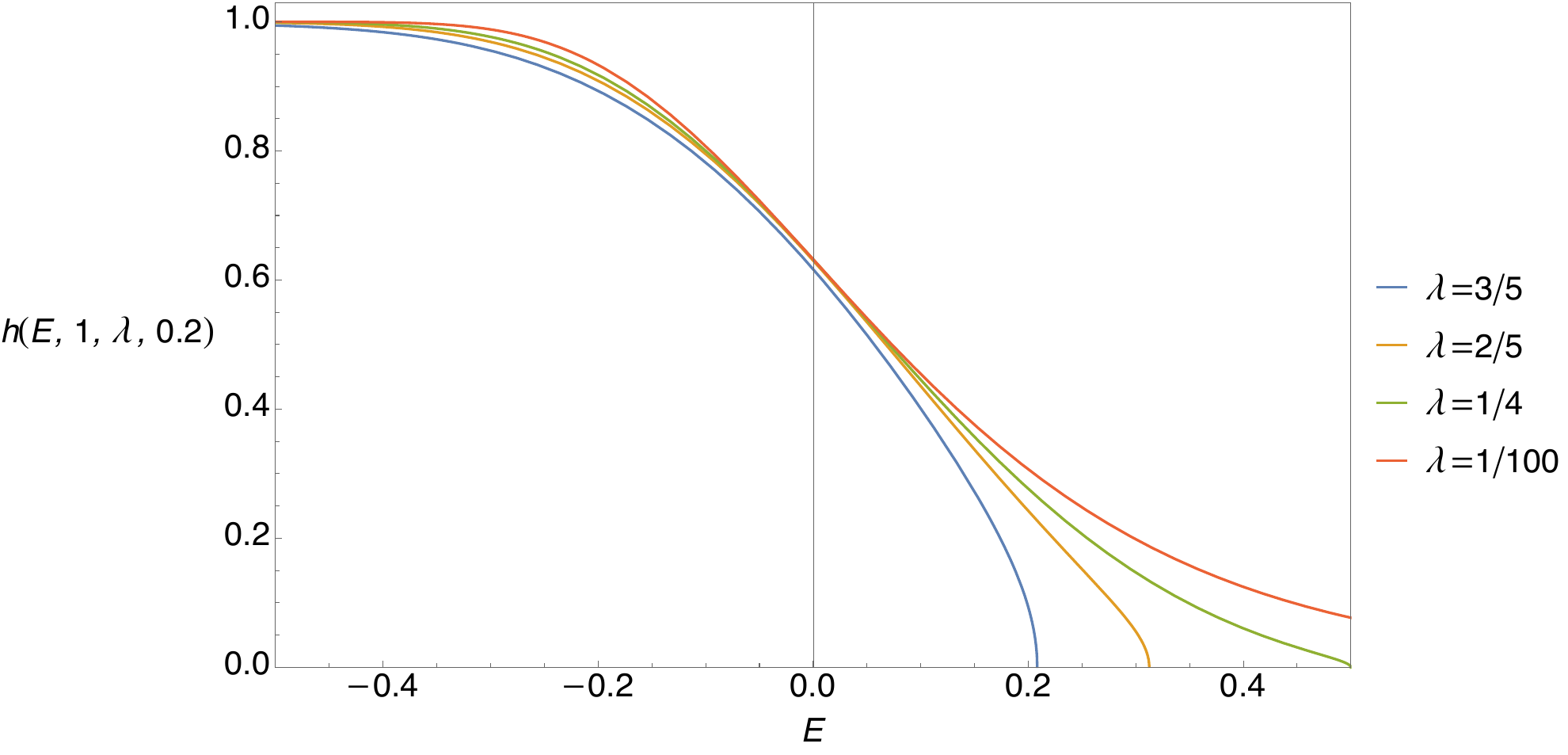}
    \caption{We plot the function $h(E,x,\lambda,T)$ near $x=1$ and at $T=0.2$. The slope of $h(E,x,\lambda,T)$ changes inconspicuously with respect to the deformation strength $\lambda$.}
    \label{fig:slopelambda}
\end{figure}

\begin{figure}[h]
    \centering
    \includegraphics[width = 15.1cm]{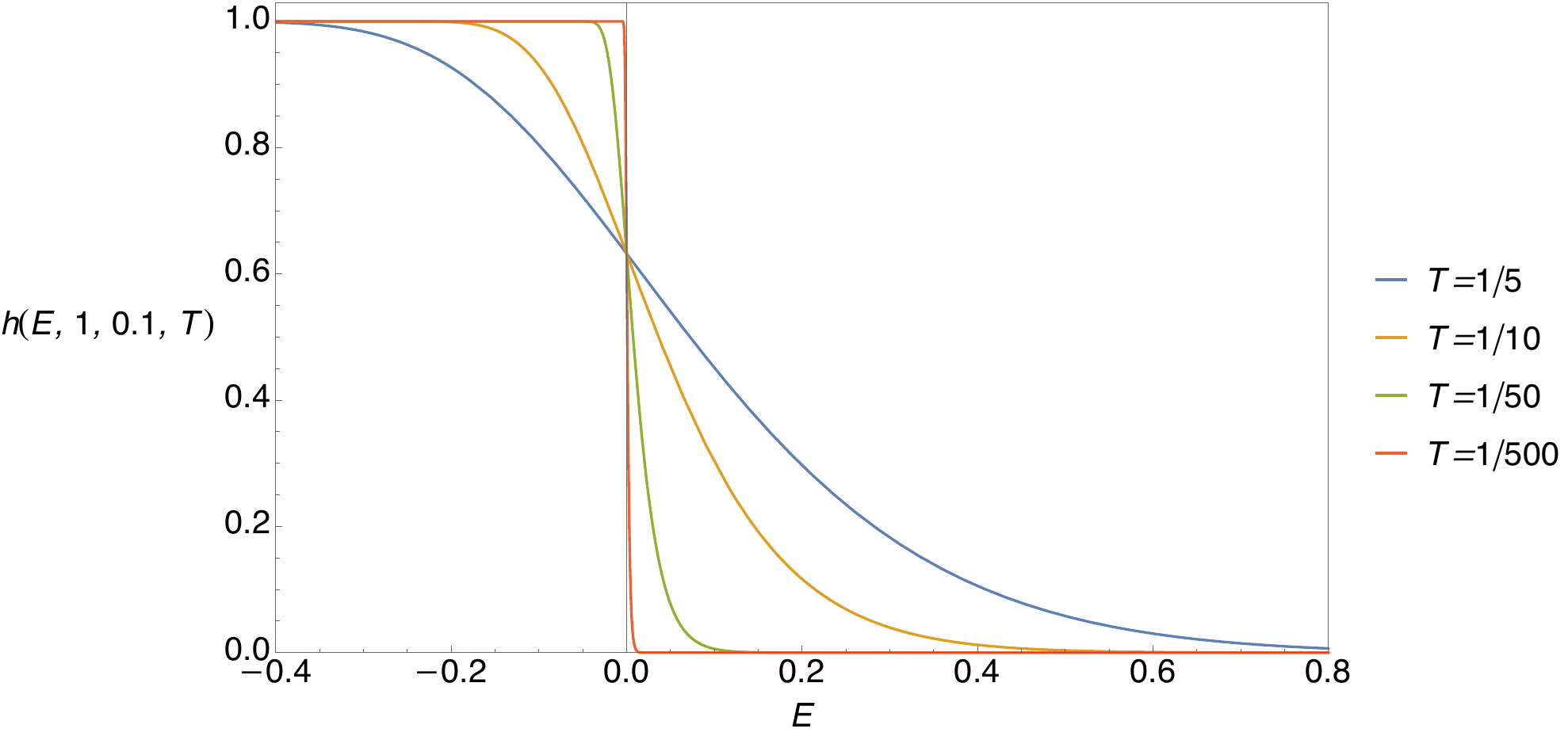}
    \caption{We plot $h(E,x,\lambda,T)$ near $x=1$ and at $\lambda=0.1$. Here $h(E,x,\lambda,T)$ is increasingly steeper as temperature $T$ decreases.}
    \label{fig:slopeT}
\end{figure}

Note that according to \eqref{eq:slope}, the slope $\partial h/\partial E$ has a limit $-\frac{xe^{-x}}{T}$ as $\lambda\rightarrow0$ and because of the exponentially suppressing factor $e^{-\frac{1}{\lambda  T}-xe^{-\frac{1}{2 \lambda  T}}}$ in \eqref{eq:lambdaslope}, when $x=1$, the graph of $h$ becoming steeper near $E=0$ as $\lambda$ decreases is not obvious at all as shown in Figure \ref{fig:slopelambda}. In contrast, the slope is not bounded as $T$ decreases as shown in Figure \ref{fig:slopeT}.

Overall, since we are restricting to small $T$ and $\lambda$, we can determine a rough picture of the integral in \eqref{eq:zcalbad} as follows:
\begin{itemize}
    \item Starting from $x=1$, and the value $h(0,1,\lambda,T)$ at $E=0$ is roughly $1-1/e$ (with small $T$ and $\lambda$);
    \item Then if we decrease $T$ or increase $\lambda$, we see clearly that the slope becomes steeper;
    \item Most importantly, if we keep increasing $x$ beyond a sufficiently large value (which parametrically depends on $T$ and $\lambda$), $h$ will eventually keep shifting to the left;
    \item Now \eqref{eq:zcalbad} is an $E$-integral over the product of $h(E,x,\lambda,T)$ and $\rho_{\text{Airy}}(E)$, which is essentially a ``convolution'' between $h$ and $\rho_{\text{Airy}}(E)$ [despite the changing shape of $h$, and the integration range here being $E\in(-\infty,1/8\lambda)$ instead of $E\in(-\infty,+\infty)$]. Since $\rho_{\text{Airy}}(E)$ has a fixed shape, this ``convolution'' only depends on the shape and position of $h(E,x,\lambda,T)$. We can see that as $x\rightarrow\infty$, $h(E,x,\lambda,T)$ moves to the far left, and consequently the overlap between $h(E,x,\beta,\lambda)$ and $\rho_{\text{Airy}}(E)$ approaches zero making the $E$-integral in \eqref{eq:zcalbad} vanish.
\end{itemize}

\begin{figure}[h]
    \centering
    \includegraphics[width = 15cm]{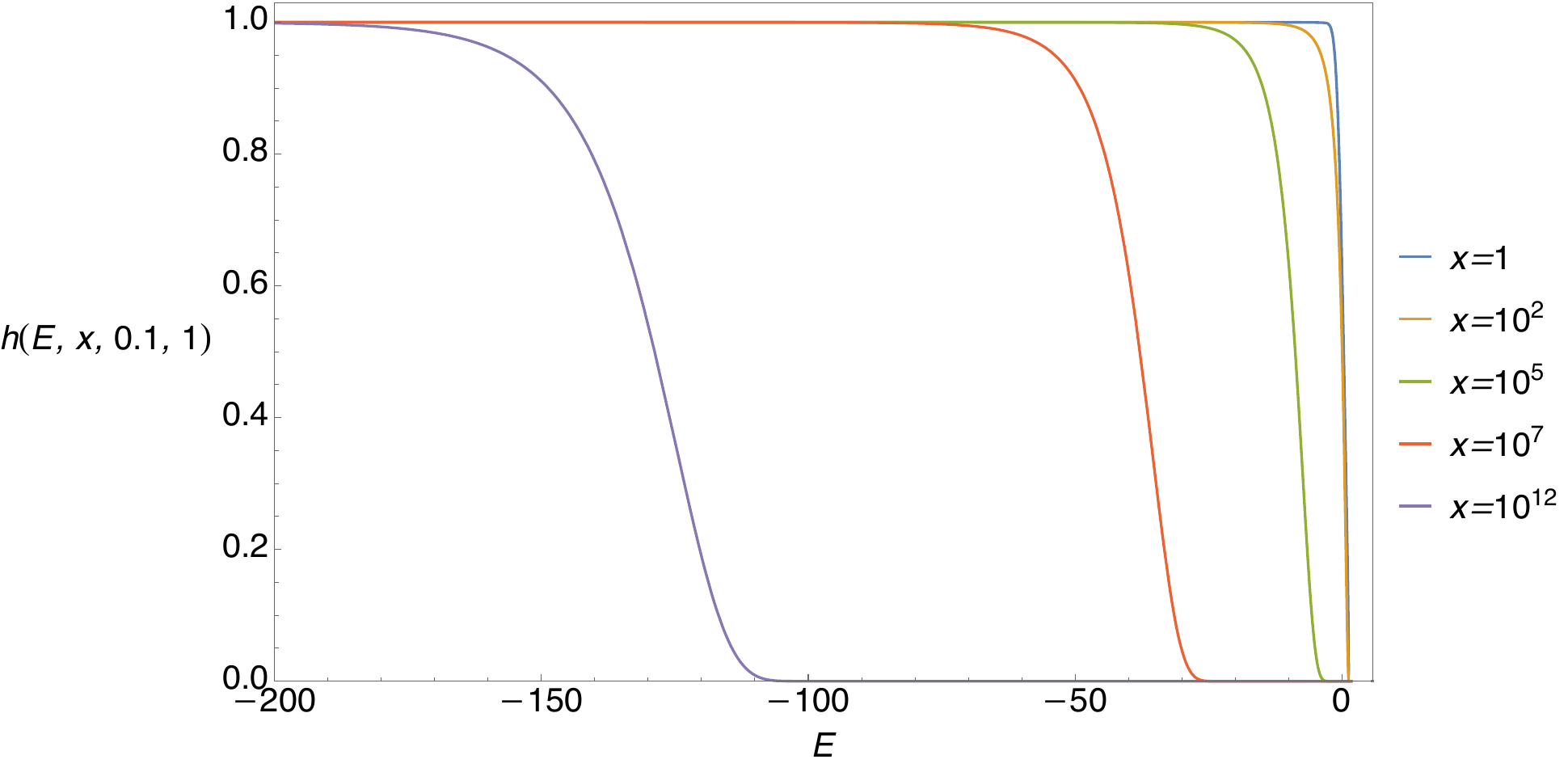}
    \caption{The graph of $h$ against $E$ monotonically shifts to the left when $x$ is sufficiently large. We set $\lambda=0.1$ and $T=1$.}
    \label{fig:step}
\end{figure}

\subsection{Perturbative branch alone}
Now let us take a step back and examine if we only include the perturbative branch, will the quenched free energy $F_{q,\lambda}(T)$ still diverge? Below we run an analysis in parallel with the previous subsection's, and we will see that the answer is a resounding yes.

\begin{figure}[h]
    \centering
    \includegraphics[width = 15.1cm]{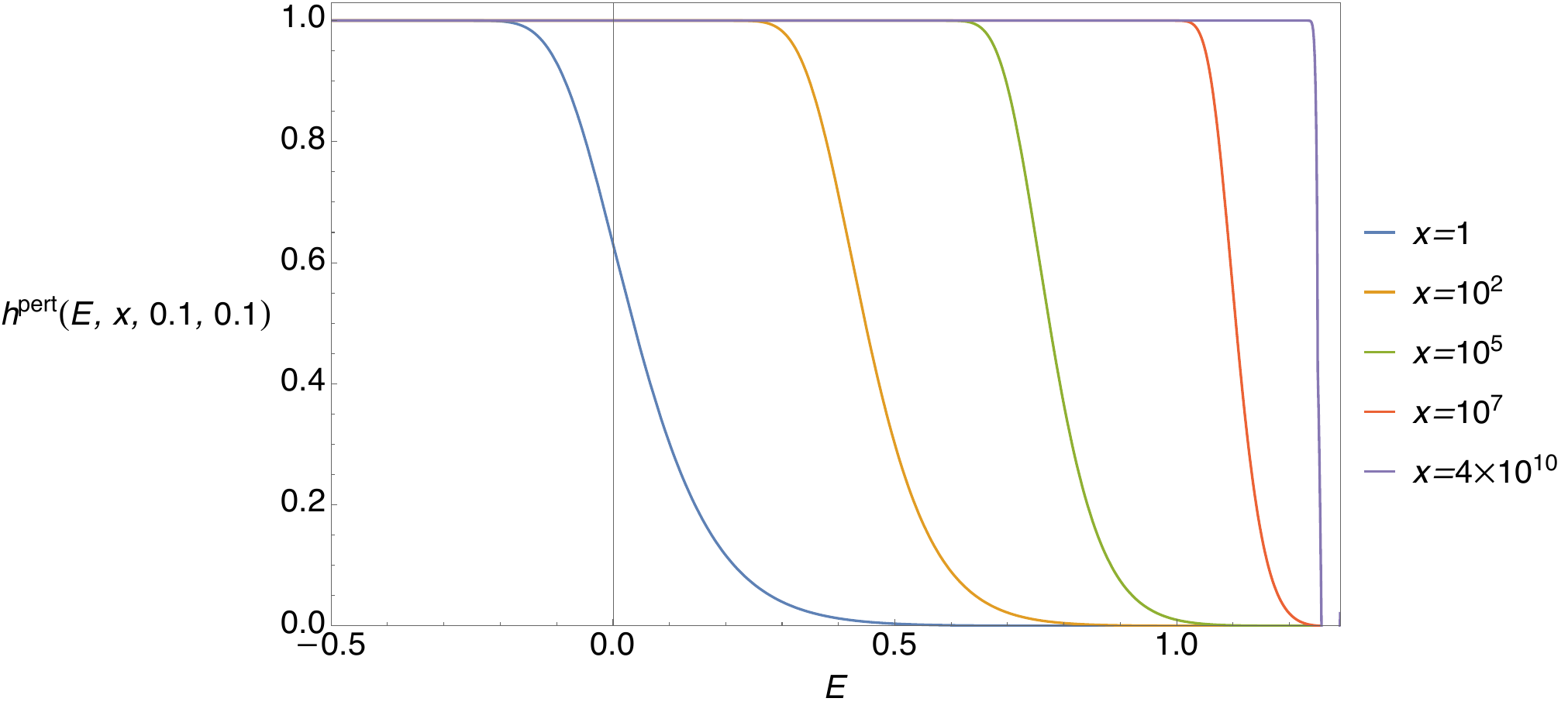}
    \caption{We show the graphs of $h^{\text{pert}}$ against $E$ monotonically shifts to the right when $x$ is sufficiently large. We set $\lambda=T=0.1$.}
    \label{fig:perturbative}
\end{figure}

This time we only include the first term from the integrand in \eqref{eq:zcalbad}:
\begin{equation}
\label{eq:pertZcal}
    \mathcal{Z}_{\lambda}^{\text{pert},>}(x)=\int^{\frac{1}{8\lambda}}_{-\infty}dE\rho_{\text{Airy}}(E)\left(1-e^{-xe^{-f_{\lambda}^-(E)/\lambda}}\right)
\end{equation}
and we define
\begin{equation}
    h^{\text{pert}}(E,x,\lambda,T)\equiv 1-e^{-xe^{-\frac{1-\sqrt{1-8\lambda E}}{4\lambda T}}}\,.
\end{equation}
We notice that 
\begin{equation}
\lim_{E\rightarrow-\infty}h^{\text{pert}}(E,x,\lambda,T)=1\,,
\end{equation}
and
\begin{equation}
    \lim_{E\rightarrow1/8\lambda}h^{\text{pert}}(E,x,\lambda,T)=1-e^{-xe^{-\frac{1}{4\lambda T}}}\,.
\end{equation}
Next, it is straightforward to see that the $E$-integrand is monotonically decreasing with $E$:
\begin{equation}
\label{eq:slopepert}
    \frac{\partial}{\partial E}h^{\text{pert}}(E,x,\lambda,T)=-\frac{x e^{\frac{\sqrt{1-8 \lambda E}-1}{4 \lambda T}-x e^{\frac{\sqrt{1-8  \lambda E}-1}{4 \lambda T}}}}{T\sqrt{1-8 \lambda E}}<0\,.
\end{equation}
Now let us examine how the value of $h^{\text{pert}}(0,x,\lambda,T)=1-e^{-x}$ changes with $x$:
\begin{equation}
    \frac{\partial}{\partial x}h^{\text{pert}}(0,x,\lambda,T)=e^{-x}>0\,,
\end{equation}
so as $x$ increases, the graph of $h$ against $E$ keeps moving to the right (different from the previous subsection), but it eventually reaches a limit at $E=1/8\lambda$ as shown in Figure \ref{fig:perturbative}. Since this limiting shape, a reflected and shifted Heaviside step function $1-\theta\left(E-\frac{1}{8\lambda}\right)$, does not depend on either $\lambda$ or $T$ there are no need for the plots similar to Figures \ref{fig:slopelambda} or \ref{fig:slopeT}.

Finally, as $x\rightarrow\infty$, the perturbative part $\mathcal{Z}_{\lambda}^{\text{pert},>}(x)$ of $\mathcal{Z}_\lambda(x)$ in \eqref{eq:zcalbad} asymptotes to
\begin{equation}
\begin{aligned}
    \int_{-\infty}^{\frac{1}{8\lambda}} dE \rho_{\text{Airy}}(E)&\left(1-\theta\left(E-\frac{1}{8\lambda}\right)\right)=\int_{-\infty}^{\frac{1}{8\lambda}} dE \rho_{\text{Airy}}(E)\\
    &\quad\quad\quad\quad\,\,\,=\frac{\text{Ai}\left(-\frac{1}{8 \lambda }\right)^2+8 \lambda  \text{Ai}'\left(-\frac{1}{8 \lambda }\right)^2-32 \lambda ^2 \text{Ai}\left(-\frac{1}{8 \lambda }\right) \text{Ai}'\left(-\frac{1}{8 \lambda }\right)}{96 \lambda ^2}>0\,,
    \end{aligned}
\end{equation}
because $\text{Ai}'(x < 0)<0$. Then the quenched free energy $F_{q,\lambda}(T)=\log\langle Z\rangle_\lambda-\int_0^{\infty}\frac{dx}{x}e^{-\mathcal{Z}_\lambda^{\text{pert},>}(x)}$ is still divergent.

\section{Proof of the divergence of the good-sign quenched free energy with non-perturbative effects}
\label{app:goodsigndetail}

\begin{figure}[h]
    \centering
    \includegraphics[scale = 0.4]{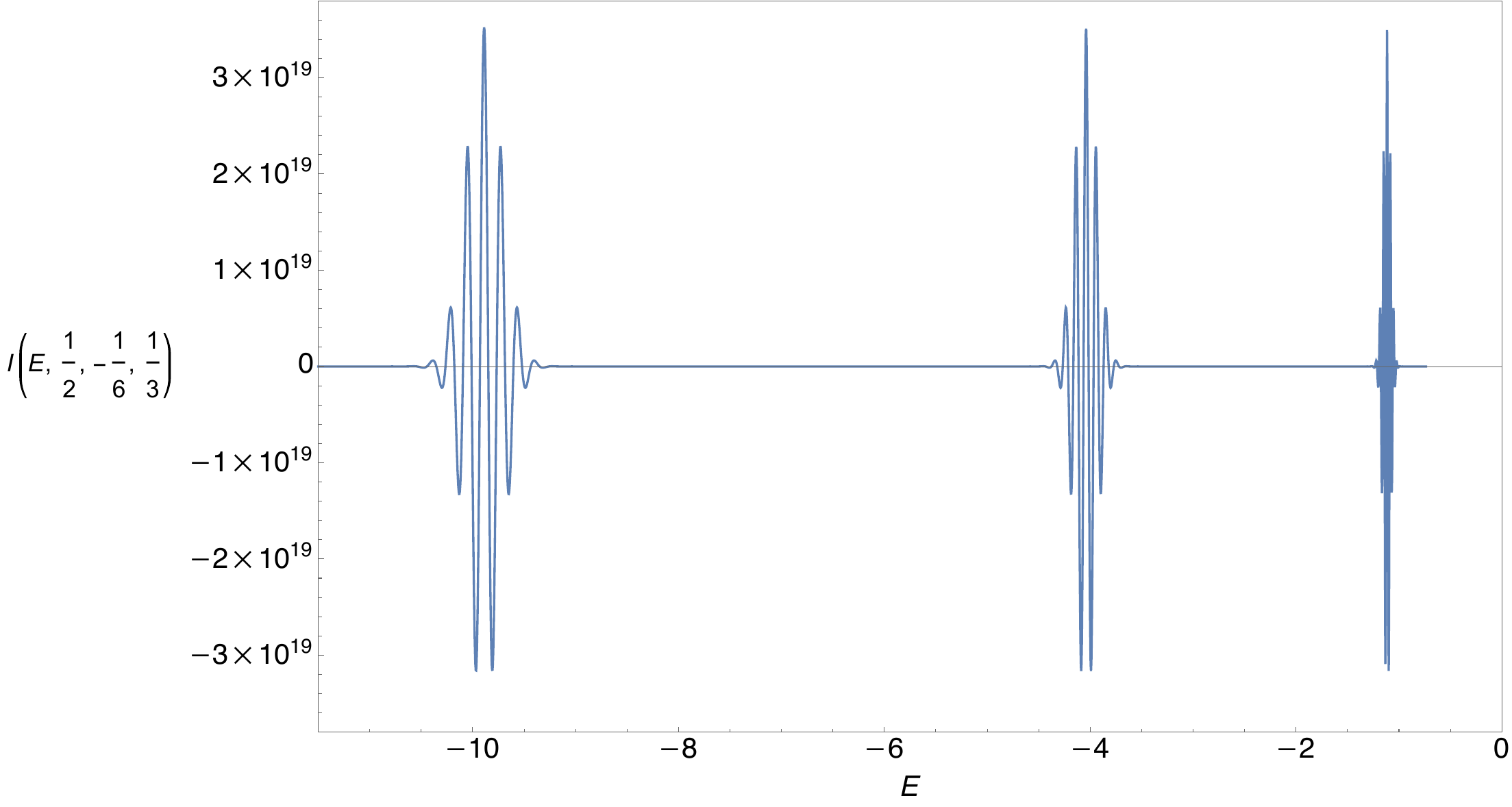}
    \caption{We plot $I \left(E < 0, x=\frac{1}{2}, \lambda=-\frac{1}{6}, T=\frac{1}{3} \right)$ and show the first three ``mountains'', each of which contain several sharp peaks. All ``mountains'' have the same height $e^{xe^{-\frac{1}{4\lambda T}}}$, and they are all symmetric with respect to their vertical axes. Here we choose small $x$ merely to make plotting easy, but yet the peaks are already sharp and localized.} 
    \label{fig:bumpy}
\end{figure}

Since inside the $x$-integral of \eqref{eq:deformfree}, $e^{-x\langle Z\rangle_\lambda}\rightarrow0$ as $x\rightarrow\infty$ due to the finiteness of $\langle Z\rangle_\lambda$ even with its nonperturbative part in \eqref{eq:good1pt} included, to discuss the convergence of this $x$-integral, again we only need to examine $e^{-\mathcal{Z}_\lambda(x)}$. The first term in its exponent $\mathcal{Z}_\lambda(x)$ as in \eqref{eq:calZ} is benign, so we focus our attention on the second term \eqref{eq:cosine2}, namely a highly oscillatory $E$-integral because of $I(E,x,\lambda,T)$. In Figure \ref{fig:bumpy}, the plot of $I(E,x,\lambda,T)$ against $E$ shows that it has infinitely many ``mountains'' of the same height, and within each ``mountain'', there are many peaks.\footnote{The state-of-the-art methods for infinite highly oscillatory integrals are Longman's method \cite{longman1960method} and double exponential quadrature method by Ooura and Mori \cite{ooura1991double}. Although we use neither of them, our treatment from now on is inspired by the former.}

We then notice that $I(E,x,\lambda,T)$ is a product of a ``waveform'' $\cos \left(xe^{-\frac{1}{4\lambda T}}\sin\frac{\sqrt{8 \lambda E -1}}{4 \lambda T}\right)$ and an ``envelope'' $e^{-xe^{-\frac{1}{4 \lambda T}} \cos\frac{\sqrt{8\lambda E-1}}{4 \lambda T}}$, and the number and height of peaks increase with $x$. For large $x$, the ``mountains'' (i.e. maxima of the ``envelope'') are well localized around the local minima of $\cos \left( \frac{\sqrt{8\lambda E-1}}{4 \lambda T} \right)$, namely at
\begin{equation}
\label{eq:localize}
    E_n=\frac{[4\lambda T(2n+1)\pi]^2+1}{8\lambda}, \quad n\in\mathbb{Z}_{\geq0}\,. 
\end{equation}
On the other hand, the asymptotics of $\rho_{\text{Airy}}(E)$ in the $E$-integral \eqref{eq:cosine2} can be determined from the asymptotics of the Airy function
\bea \text{Ai}(z) \sim \frac{e^{-\frac{2}{3} z^{\frac{3}{2}}}}{2\sqrt{\pi}z^{\frac{1}{4}}} \left[\sum_{m=0}^\infty \frac{(-1)^m \Gamma(m+ \frac{5}{6}) \Gamma(m + \frac{1}{6})\big(\frac{3}{4}\big)^m}{2\pi m! z^{3m/2}}\right],\quad\text{as}\,\,|z|\rightarrow\infty,\,\,\text{when}\,\,|\arg(z)|<\pi\,.
\eea
This means $\text{Ai}(-z)$ and $\text{Ai}'(-z)$ behave as $e^{\frac{2}{3} z^{\frac{3}{2}}}$ when $z\rightarrow -\infty$. This implies that $\rho_{\text{Airy}}(E) \sim e^{\frac{4}{3}E^3}$ as $E\rightarrow -\infty $. Therefore, for a given $x$, it quickly wins over the amplitude of $I(E,x,\lambda,T)$ in \eqref{eq:cosine2} as $E$ decreases. Hence, to work out the asymptotics of $\mathcal{Z}_\lambda(x)$ as $x\rightarrow\infty$, we only need to focus on the first or rightmost mountain centered at $E^*$ where $n=0$ in \eqref{eq:localize}. We define the horizontal range of the mountain to be between two adjacent maxima of $\cos \left( \frac{\sqrt{8\lambda E-1}}{4 \lambda T} \right)$ in the exponent of the ``envelope'', namely from $E_L=\frac{(8\lambda T\pi)^2+1}{8\lambda}$ to $E_R=\frac{1}{8\lambda}$. Notice that all these treatment of that ``mountain'' is independent of the ``waveform''.

We furthermore write $E$ as $\frac{(4\lambda T\pi)^2+1}{8\lambda}+\epsilon$ with a very small $\epsilon$, then in the ``envelope'' and the ``waveform'',
\begin{equation}
\begin{aligned}
    \sin \left( \frac{\sqrt{8\lambda E-1}}{4\lambda T} \right)&\sim \frac{\epsilon }{4 \pi  \lambda  T^2}-\frac{\epsilon ^2}{32\pi ^3 \lambda ^2 T^4},\\
    \cos \left( \frac{\sqrt{8\lambda E-1}}{4\lambda T} \right)&\sim \frac{\epsilon ^2}{32 \pi ^2 \lambda ^2 T^4}-1\,.
\end{aligned}
\end{equation}
Now we approximate the area under the peak of $e^{xe^{-\frac{1}{4 \lambda T}}\left(1-\frac{\epsilon ^2}{32 \pi ^2 \lambda ^2 T^4}\right)}$ by truncating its base at the values $e^{-8}\sim 3.35\times10^{-4}$ (which is a arbitrary) fraction of the peak value $e^{xe^{-\frac{1}{4\lambda T}}}$. This corresponds to the interval $[E^*+\epsilon,E^*-\epsilon]$ with
\begin{equation}
\label{eq:range}
    \epsilon=\frac{16 \pi  \lambda  T^2  e^{\frac{1}{8 \lambda  t}}}{\sqrt{x}}\,.
\end{equation}
Due to large $x$, this interval is much narrower than the range (between $E_L$ and $E_R$) of the first ``mountain'' just defined above. We will adopt this new truncation hereafter. Furthermore, because the first ``mountain'' becomes infinitely narrow as $x\rightarrow\infty$, the density of states $\rho_{\text{Airy}}(E)$ in \eqref{eq:cosine2} can be approximated by $\rho_{\text{Airy}}(E^*)$. As a result, up to a multiplicative factor, the $E$-integrand there becomes
\begin{equation}
\label{eq:approx}
    1-e^{xe^{-\frac{1}{4\lambda T}}}\underbrace{e^{-\frac{xe^{-\frac{1}{4\lambda T}}\epsilon ^2}{32 \pi ^2 \lambda ^2 T^4}}\cos\left[xe^{-\frac{1}{4\lambda T}}\left(\frac{\epsilon }{4 \pi  \lambda  T^2}-\frac{\epsilon ^2}{32\pi ^3 \lambda ^2 T^4}\right)\right]}_{\text{enters}\,\, \eqref{eq:approx2}}\,.
\end{equation}
Empirically, can we ignore the $\epsilon^2$ term in cosine and yet approximate well? From Figure \eqref{fig:waveform} below, we see that the first-order is a bit off so we include up to the second order in $\epsilon^2$. 
\begin{figure}[h]
    \centering
    \includegraphics[width = 15.13cm]{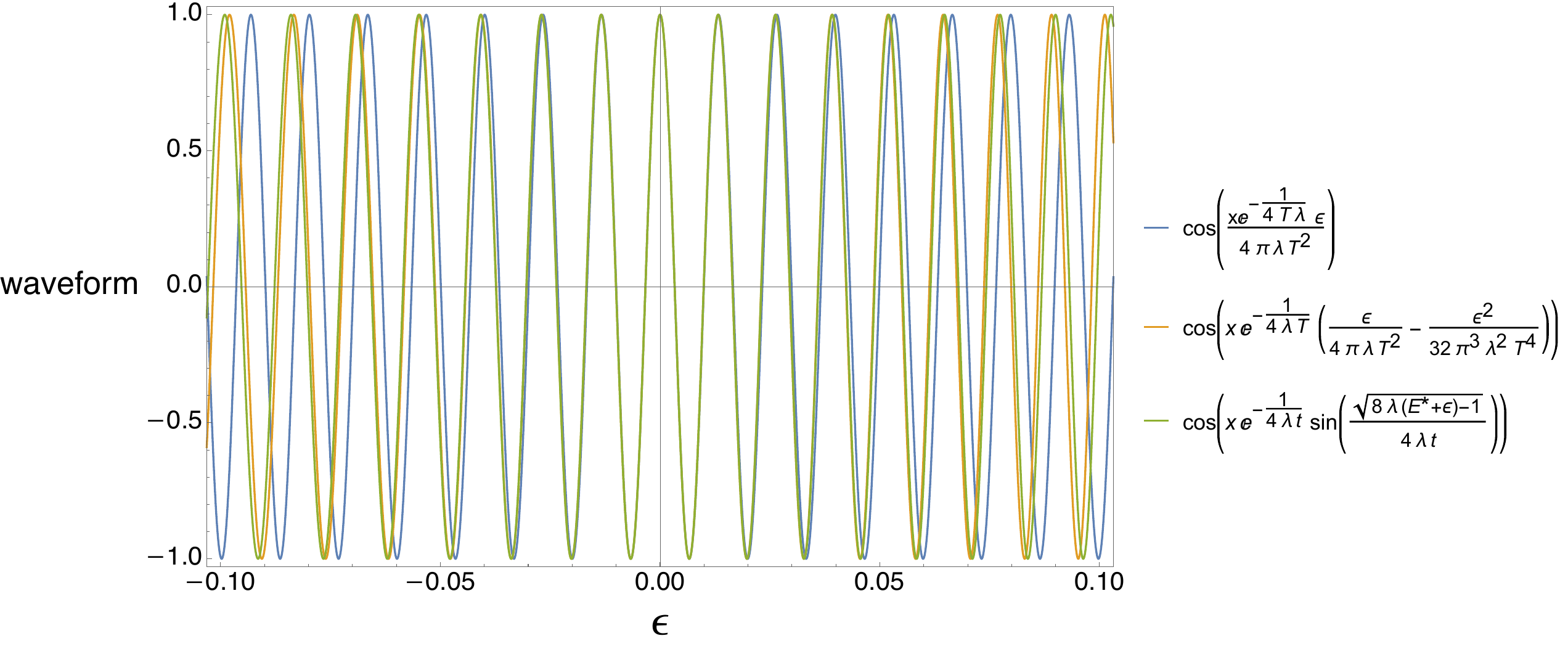}
    \caption{We show the comparison between the plots of the original waveform (green) and its first-order (blue) and second order (green) approximations in $\epsilon$, where $\lambda=-1/10,T=1/2,x=1$ (again, to make his figure visually discernible, a small $x$ is chosen). We see that the first-order one clearly deviates from the other two plots.}
    \label{fig:waveform}
\end{figure}

After setting $xe^{-\frac{1}{4\lambda T}}=a$ and $4 \pi\lambda  T^2=b$, we only need to integrate\footnote{Conveniently, if we have to include one further order in $\epsilon$, we cannot obtain a compact result in terms of common special functions for an integral such as:
\begin{equation}
\label{eq:peak}
    \int dx\,x^{k}e^{-dx^2}\cos\left(ax+bx^2+cx^3\right),\quad k\in\mathbb{Z}\,.
\end{equation}
} the underbraced part in \eqref{eq:approx} over the ``mountain'' range $[\epsilon,-\epsilon]$ is:
\begin{equation}
\label{eq:approx2}
  \int^{\frac{4b}{\sqrt{a}}}_{-\frac{4b}{\sqrt{a}}} d\epsilon\, e^{-\frac{a}{2b^2}\epsilon^2}\cos\left[a\left(\frac{\epsilon}{b}-\frac{\epsilon^2}{2\pi b^2}\right)\right]\,.
\end{equation}
The antiderivative of the $\epsilon$-integrand \eqref{eq:approx} is then
\begin{equation}
\label{eq:anti}
    \epsilon+\frac{\pi  e^{\frac{a}{1+\pi ^2}} b}{\sqrt{2a} \left(1+\pi ^2\right)}\, \text{Im}\left[\sqrt{\pi -i} (\pi +i) e^{\frac{\pi  a}{2 \pi +2 i}} \text{erfi}\left(\frac{\sqrt{a} (\pi  b-(1+i \pi )\epsilon)}{\sqrt{2 \pi } \sqrt{\pi -i} b}\right)\right]\,,
\end{equation}
where $\text{erfi}(\cdot)$ is the imaginary error function, defined as $\text{erfi}\,z=-i\text{erf}\,iz$. Then the second term, i.e., $\mathcal{Z}_{\lambda}^{\text{cos}}(x)$, in \eqref{eq:calZ} is obtained as:
\begin{equation}
    -\rho_{\text{Airy}}\left(\frac{(4\lambda T\pi)^2+1}{8\lambda}\right)\left(P+\frac{8b}{\sqrt{a}}\right)\,,
\end{equation}
where 
\begin{equation}
\begin{aligned}
\label{eq:realpart}
    P=&\frac{\pi e^{\frac{a}{1+\pi ^2}} b}{\sqrt{2a} \left(1+\pi ^2\right)}\\
    &\times\text{Re}\,\left\{\sqrt{\pi -i} (1-i \pi )e^{\frac{\pi  a}{2 \pi +2 i}} \left[\text{erfi}\left(\frac{\pi  \left(\sqrt{a}-4 i\right)-4}{\sqrt{2 \pi  (\pi -i)}}\right)-\text{erfi}\left(\frac{\pi  \left(\sqrt{a}+4 i\right)+4}{\sqrt{2 \pi  (\pi -i)}}\right)\right]\right\}\,.
    \end{aligned}
\end{equation}
We convert the factor inside the braces and outside of the square brackets to
\be
\label{eq:reim}
\sqrt[4]{1+\pi ^2}\,e^{A\pi}\{[\cos(A+\theta)-\pi\sin(A+\theta))-i(\sin(A+\theta)+\pi\cos(A+\theta)]\}\,,
\ee
where
\begin{equation}
\begin{aligned}
\label{eq:param}
A=\frac{\pi a}{2+2 \pi ^2},\quad
\theta=\frac{1}{2} \tan ^{-1}\frac{1}{\pi}\,,
\end{aligned}
\end{equation}
and terms between the square brackets to
\be
\label{eq:erfi}
\text{erfi}(x_1+iy_1)-\text{erfi}(x_2+iy_2)=i\text{erf}(ix_2-y_2)-i\text{erf}(ix_1-y_1)\,,
\ee
with
\begin{equation}
    \begin{aligned}
        x_{1,2}&=\frac{1}{\sqrt[4]{1+\pi ^2}}\left(\sqrt{\frac{a\pi}{2} } \cos \theta\pm2 \sqrt{2 \pi } \sin \theta\mp2 \sqrt{\frac{2}{\pi }} \cos\theta\right)\,,\\
        y_{1,2}&=\frac{1}{\sqrt[4]{1+\pi ^2}}\left(\sqrt{\frac{a\pi}{2}}\sin\theta\mp2 \sqrt{2 \pi } \cos \theta\mp2 \sqrt{\frac{2}{\pi }} \sin \theta\right)\,.
    \end{aligned}
\end{equation}
We again use the asymptotic formula \eqref{eq:error}, which is valid over the complex plane, to approximate \eqref{eq:erfi}.\footnote{We remark that other famous estimates, which are resoundingly accurate numerically, such as Identity 7.1.29 in  \cite{abramowitz1964handbook}):
\begin{equation}
\begin{aligned}
\label{eq:estimate}
    \text{erf}(x+iy)=&\,\text{erf}(x)+\frac{e^{-x^2}}{2\pi x}\left[(1-\cos2xy)+i\sin2xy\right]\\
    &+\frac{2}{\pi}e^{-x^2}\sum_{k=1}^\infty\frac{e^{-k^2/4}}{k^2+4x^2}[f_k(x,y)+ig_k(x,y)]+\epsilon(x,y)\,,
    \end{aligned}
\end{equation}
where
\begin{equation}
\begin{aligned}
    f_k(x,y)&=2x(1-\cos2xy\cosh ky)+k\sin2xy\sinh ky\,,\\
    g_k(x,y)&=2x\sin2xy\cosh ky+k\cos2xy\sinh ky\,,
    \end{aligned}
\end{equation}
and $|\epsilon(x,y)|\leq 10^{-16}|\text{erf}(x+iy)|$, is not useful here. The first line alone is terribly far from being sufficient because for quite a few small $k>0$, the absolute values of the real and imaginary parts in each summand of the second line grow as $\frac{e^{-x^2}}{2\pi x}e^{ky-k^2/4}$ when $x,y\rightarrow\infty$ makes it more dominant than the second term in the first line of \eqref{eq:estimate}.} Then we expand this difference multiplied by \eqref{eq:reim} in $a$ at infinity to the zeroth order (terms of order $O\big(a^{-3/2}\big)$ and higher are ignored), and it turns out both the real and imaginary parts of \eqref{eq:reim} contribute equally to this expansion. Therefore, the content within the braces in \eqref{eq:realpart} becomes
\begin{equation}
\label{eq:expansion}
\begin{aligned}
    &\quad4\exp\left[\frac{\pi  (2 \pi +i) \big(\pi ^2+1-\pi  \sqrt{1+\pi ^2}\big) a \csc ^2\theta}{4 (1+\pi ^2)^2}-\frac{2 \sin ^2\theta \big(2i\pi ^2+2 i+\sqrt{1+\pi ^2} \csc^2\theta\big)^2}{\pi(\pi +i)^2 \sqrt{1+\pi ^2}}\right]\\
    &\quad\quad\quad\quad\quad\quad\quad\quad\quad\times\cos\frac{2 \big(\sqrt{1+\pi ^2}-\pi\big) \sqrt{a} \csc^2\theta}{\sqrt{1+\pi ^2}}
    \left[\sin \left(A+\theta\right)+\cos \left(A+\theta\right)\right]O\big(a^{-1/2}\big)\,.
    \end{aligned}
\end{equation}
Finally, we plug in $\theta$ from \eqref{eq:param} to take the real part of the exponential and arrive at
\begin{equation}
    \frac{\pi ^2 a}{1+\pi ^2}-8+i\left(\frac{\pi  a}{2+2 \pi ^2}+\frac{8}{\pi }\right)\,.
\end{equation}
Consequently, as $a\rightarrow\infty$, \eqref{eq:expansion} grows as
\begin{equation}
\quad\,\,4e^{\frac{\pi ^2 a}{1+\pi ^2}}\cos\left(4\sqrt{a}\right)\cos A\left[\sin \left(A+\theta\right)+\cos \left(A+\theta\right)\right]O\big(a^{-1/2}\big)\,.
\end{equation}
Hence, after multiplying the factor in the first line of \eqref{eq:realpart}, we obtain the asymptotics of $P$ when $a\rightarrow\infty$, i.e., $x\rightarrow\infty$, roughly as
\begin{equation}
\label{eq:asymptotics}
    \frac{0.8175b\,e^a}{a}\cos(0.1445a)\cos4\sqrt{a}\left[\sin (0.1445a+0.154)+\cos(0.1445a+0.154)\right]\,.
\end{equation}
We conclude that a multiplicative part $\mathcal{Z}_\lambda^{\text{cos}}(x)$ of $\mathcal{Z}_\lambda(x)$ in the $x$-integral \eqref{eq:deformfree} is oscillating with an exponentially diverging amplitude as $x\rightarrow\infty$. Since we know that the integral \eqref{eq:deformfree} converges if we restrict to $\mathcal{Z}_\lambda^{\text{pert},<}(x)$ in $\mathcal{Z}_\lambda(x)$, including $\mathcal{Z}_\lambda^{\text{cos}}(x)$ will make the new integral divergent, rendering no hope of numerically evaluating the quenched free energy with the non-perturbative effect.

We finish by remarking that in principle, one can find a better estimate on $\rho_{\text{Airy}}(E)$ over the narrow range $[E^*+\epsilon,E^*-\epsilon]$ by modelling it as a polynomial in $\epsilon$ instead of a constant $\rho_{\text{Airy}}(E^*)$ and discover more accurate asymptotics than \eqref{eq:asymptotics}.
To this end, below, we collect some indefinite integrals\footnote{Compact results in terms of the Dawson function can be obtained for monomials up to $\epsilon^7$.} of the products of monomials in $\epsilon$ up to degree $3$, Gaussian function and cosine functions:
\begin{equation}
\begin{aligned}
    \int \epsilon  e^{-\frac{a \epsilon ^2}{2 b^2}} &\cos \left(\frac{a\epsilon }{b}-\frac{a\epsilon ^2}{2 \pi  b^2}\right) \, d\epsilon\\
    &=\frac{\pi b^2}{2 a} \text{Re}\left\{\frac{\pi\sqrt{2a}e^{-\frac{\pi a}{2 (\pi +i)}} \text{erfi}\left(\frac{\sqrt{a} (\pi  b+i (\pi +i) \epsilon )}{\sqrt{2 \pi } \sqrt{\pi +i} b}\right)}{(\pi +i)^{3/2}}-\frac{2 e^{-\frac{a \epsilon  ((\pi +i) \epsilon -2 i \pi  b)}{2 \pi  b^2}}}{\pi +i}\right\}\,,
    \end{aligned}
\end{equation}
\begin{equation}
\begin{aligned}
    \int \epsilon ^2 e^{-\frac{a \epsilon ^2}{2 b^2}} &\cos \left(\frac{a\epsilon }{b}-\frac{a\epsilon ^2}{2 \pi  b^2}\right) \, d\epsilon\\
    &=\frac{\pi b^2}{2a^{3/2}} \text{Re}\Bigg\{\frac{2e^{-\frac{a \epsilon((\pi +i) \epsilon -2 i \pi  b)}{2 \pi  b^2}}}{(\pi +i)^{5/2}}\bigg[\sqrt{2\pi}(1+i\pi  (a-1)) b F\left(\frac{\sqrt{a} (\pi  b+i (i+\pi ) \epsilon )}{b \sqrt{2 \pi } \sqrt{i+\pi }}\right)\\
    &\quad\quad\quad\quad\quad\quad-i \sqrt{a(\pi +i)}(\pi  b-i \pi  \epsilon +\epsilon )\bigg]\Bigg\}\,,
    \end{aligned}
\end{equation}
\begin{equation}
\begin{aligned}
    \int \epsilon ^3 e^{-\frac{a \epsilon ^2}{2 b^2}} &\cos \left(\frac{a\epsilon }{b}-\frac{a\epsilon ^2}{2 \pi  b^2}\right) \, d\epsilon\\
    &=\frac{\pi b^2}{a^2}\text{Re}\Bigg\{\frac{e^{-\frac{a \epsilon  ((\pi +i) \epsilon -2 i \pi  b)}{2 \pi  b^2}}}{(\pi +i)^{7/2}} \bigg[\pi ^{3/2} \sqrt{2a} (3 i-\pi  (a-3)) b^2 F\left(\frac{\sqrt{a} (\pi  b+i (i+\pi ) \epsilon )}{b\sqrt{i+\pi }}\right)\\
    &\quad\quad\quad\quad\quad+\sqrt{\pi+i} \left(\pi  (\pi  (a-2)-2 i) b^2+\pi  (1-i \pi ) a b \epsilon -(\pi +i)^2 a \epsilon ^2\right)\bigg]\Bigg\}\,,
    \end{aligned}
\end{equation}
where $F(\cdot)$ denotes the Dawson function $F(x)\equiv e^{-x^2}\int_0^{x}e^{t^2}dt$. They are absolutely not necessary to our proof of divergence of \eqref{eq:deformfree} with $\mathcal{Z}^{\text{cos}}_\lambda(x)$ included.
\newpage 
\bibliographystyle{JHEP}
\bibliography{JT}

\providecommand{\href}[2]{#2}\begingroup\raggedright\begin{thebibliography}{10}

\bibitem{Zamolodchikov:2004ce}
A.~B. Zamolodchikov, {\it {Expectation value of composite field T$\overline{T}$
  in two-dimensional quantum field theory}},
  \href{http://arxiv.org/abs/hep-th/0401146}{{\tt hep-th/0401146}}.

\bibitem{Smirnov:2016lqw}
F.~A. Smirnov and A.~B. Zamolodchikov, {\it {On space of integrable quantum
  field theories}},  {\em Nucl. Phys. B} {\bf 915} (2017) 363--383,
  [\href{http://arxiv.org/abs/1608.05499}{{\tt arXiv:1608.05499}}].

\bibitem{Cavaglia:2016oda}
A.~Cavagli\`a, S.~Negro, I.~M. Sz\'ecs\'enyi, and R.~Tateo, {\it {$T
  \overline{T}$-deformed 2D Quantum Field Theories}},  {\em JHEP} {\bf 10}
  (2016) 112, [\href{http://arxiv.org/abs/1608.05534}{{\tt arXiv:1608.05534}}].

\bibitem{Jiang:2019epa}
Y.~Jiang, {\it {A pedagogical review on solvable irrelevant deformations of 2D
  quantum field theory}},  {\em Commun. Theor. Phys.} {\bf 73} (2021), no.~5
  057201, [\href{http://arxiv.org/abs/1904.13376}{{\tt arXiv:1904.13376}}].

\bibitem{MG_CERN}
M.~Guica, {\it {$T\overline{T}$ deformations and holography}},  {\em
  \href{https://indico.cern.ch/event/857396/contributions/3706292/attachments/2036750/3410352/ttbar_cern_v1s.pdf}{https://indico.cern.ch/event/857396/
  contributions/3706292/attachments/2036750/3410352/ttbar\_cern\_v1s.pdf}}.

\bibitem{Jackiw:1984je}
R.~Jackiw, {\it {Lower Dimensional Gravity}},  {\em Nucl. Phys. B} {\bf 252}
  (1985) 343--356.

\bibitem{Teitelboim:1983ux}
C.~Teitelboim, {\it {Gravitation and Hamiltonian Structure in Two Space-Time
  Dimensions}},  {\em Phys. Lett. B} {\bf 126} (1983) 41--45.

\bibitem{Gorbenko:2018oov}
V.~Gorbenko, E.~Silverstein, and G.~Torroba, {\it {dS/dS and $ T\overline{T}
  $}},  {\em JHEP} {\bf 03} (2019) 085,
  [\href{http://arxiv.org/abs/1811.07965}{{\tt arXiv:1811.07965}}].

\bibitem{Lewkowycz:2019xse}
A.~Lewkowycz, J.~Liu, E.~Silverstein, and G.~Torroba, {\it {$ T\overline{T} $
  and EE, with implications for (A)dS subregion encodings}},  {\em JHEP} {\bf
  04} (2020) 152, [\href{http://arxiv.org/abs/1909.13808}{{\tt
  arXiv:1909.13808}}].

\bibitem{Shyam:2021ciy}
V.~Shyam, {\it {$T\bar{T}+\Lambda_2$ Deformed CFT on the Stretched dS$_3$
  Horizon}},  \href{http://arxiv.org/abs/2106.10227}{{\tt arXiv:2106.10227}}.

\bibitem{Coleman:2021nor}
E.~Coleman, E.~A. Mazenc, V.~Shyam, E.~Silverstein, R.~M. Soni, G.~Torroba, and
  S.~Yang, {\it {de Sitter Microstates from $T\bar T+\Lambda_2$ and the
  Hawking-Page Transition}},  \href{http://arxiv.org/abs/2110.14670}{{\tt
  arXiv:2110.14670}}.

\bibitem{Saad:2019lba}
P.~Saad, S.~H. Shenker, and D.~Stanford, {\it {JT gravity as a matrix
  integral}},  \href{http://arxiv.org/abs/1903.11115}{{\tt arXiv:1903.11115}}.

\bibitem{Okounkov:2001usa}
A.~Okounkov, {\it {Generating functions for intersection numbers on moduli
  spaces of curves}},  \href{http://arxiv.org/abs/math/0101201}{{\tt
  math/0101201}}.

\bibitem{Okuyama:2020ncd}
K.~Okuyama and K.~Sakai, {\it {Multi-boundary correlators in JT gravity}},
  {\em JHEP} {\bf 08} (2020) 126, [\href{http://arxiv.org/abs/2004.07555}{{\tt
  arXiv:2004.07555}}].

\bibitem{Johnson:2019eik}
C.~V. Johnson, {\it {Nonperturbative Jackiw-Teitelboim gravity}},  {\em Phys.
  Rev. D} {\bf 101} (2020), no.~10 106023,
  [\href{http://arxiv.org/abs/1912.03637}{{\tt arXiv:1912.03637}}].

\bibitem{Engelhardt:2020qpv}
N.~Engelhardt, S.~Fischetti, and A.~Maloney, {\it {Free Energy from Replica
  Wormholes}},  \href{http://arxiv.org/abs/2007.07444}{{\tt arXiv:2007.07444}}.

\bibitem{Okuyama:2021pkf}
K.~Okuyama, {\it {Quenched free energy from spacetime D-branes}},  {\em JHEP}
  {\bf 03} (2021) 073, [\href{http://arxiv.org/abs/2101.05990}{{\tt
  arXiv:2101.05990}}].

\bibitem{McGough:2016lol}
L.~McGough, M.~Mezei, and H.~Verlinde, {\it {Moving the CFT into the bulk with
  $ T\overline{T} $}},  {\em JHEP} {\bf 04} (2018) 010,
  [\href{http://arxiv.org/abs/1611.03470}{{\tt arXiv:1611.03470}}].

\bibitem{Kraus:2018xrn}
P.~Kraus, J.~Liu, and D.~Marolf, {\it {Cutoff AdS$_{3}$ versus the $
  T\overline{T} $ deformation}},  {\em JHEP} {\bf 07} (2018) 027,
  [\href{http://arxiv.org/abs/1801.02714}{{\tt arXiv:1801.02714}}].

\bibitem{Dubovsky:2018bmo}
S.~Dubovsky, V.~Gorbenko, and G.~Hernández-Chifflet, {\it {$ T\overline{T} $
  partition function from topological gravity}},  {\em JHEP} {\bf 09} (2018)
  158, [\href{http://arxiv.org/abs/1805.07386}{{\tt arXiv:1805.07386}}].

\bibitem{Datta:2018thy}
S.~Datta and Y.~Jiang, {\it {$T\overline{T}$ deformed partition functions}},
  {\em JHEP} {\bf 08} (2018) 106, [\href{http://arxiv.org/abs/1806.07426}{{\tt
  arXiv:1806.07426}}].

\bibitem{Aharony:2018bad}
O.~Aharony, S.~Datta, A.~Giveon, Y.~Jiang, and D.~Kutasov, {\it {Modular
  invariance and uniqueness of $T\overline{T}$ deformed CFT}},  {\em JHEP} {\bf
  01} (2019) 086, [\href{http://arxiv.org/abs/1808.02492}{{\tt
  arXiv:1808.02492}}].

\bibitem{Hartman:2018tkw}
T.~Hartman, J.~Kruthoff, E.~Shaghoulian, and A.~Tajdini, {\it {Holography at
  finite cutoff with a $T^2$ deformation}},  {\em JHEP} {\bf 03} (2019) 004,
  [\href{http://arxiv.org/abs/1807.11401}{{\tt arXiv:1807.11401}}].

\bibitem{Caputa:2019pam}
P.~Caputa, S.~Datta, and V.~Shyam, {\it {Sphere partition functions
  \textbackslash{}\& cut-off AdS}},  {\em JHEP} {\bf 05} (2019) 112,
  [\href{http://arxiv.org/abs/1902.10893}{{\tt arXiv:1902.10893}}].

\bibitem{Cardy:2019qao}
J.~Cardy, {\it {$T\bar T$ deformation of correlation functions}},  {\em JHEP}
  {\bf 12} (2019) 160, [\href{http://arxiv.org/abs/1907.03394}{{\tt
  arXiv:1907.03394}}].

\bibitem{He:2019vzf}
S.~He and H.~Shu, {\it {Correlation functions, entanglement and chaos in the $
  T\overline{T}/J\overline{T} $-deformed CFTs}},  {\em JHEP} {\bf 02} (2020)
  088, [\href{http://arxiv.org/abs/1907.12603}{{\tt arXiv:1907.12603}}].

\bibitem{He:2019ahx}
S.~He, J.-R. Sun, and Y.~Sun, {\it {The correlation function of (1,1) and (2,2)
  supersymmetric theories with $T\overline{T}$ deformation}},  {\em JHEP} {\bf
  04} (2020) 100, [\href{http://arxiv.org/abs/1912.11461}{{\tt
  arXiv:1912.11461}}].

\bibitem{Mazenc:2019cfg}
E.~A. Mazenc, V.~Shyam, and R.~M. Soni, {\it {A $T \overline{T}$ Deformation
  for Curved Spacetimes from 3d Gravity}},
  \href{http://arxiv.org/abs/1912.09179}{{\tt arXiv:1912.09179}}.

\bibitem{Brennan:2020dkw}
T.~D. Brennan, C.~Ferko, E.~Martinec, and S.~Sethi, {\it {Defining the $T
  \overline{T}$ Deformation on $\mathrm{AdS}_2$}},
  \href{http://arxiv.org/abs/2005.00431}{{\tt arXiv:2005.00431}}.

\bibitem{He:2020udl}
S.~He and Y.~Sun, {\it {Correlation functions of CFTs on a torus with a
  $T\overline{T}$ deformation}},  {\em Phys. Rev. D} {\bf 102} (2020), no.~2
  026023, [\href{http://arxiv.org/abs/2004.07486}{{\tt arXiv:2004.07486}}].

\bibitem{Ebert:2020tuy}
S.~Ebert, H.-Y. Sun, and Z.~Sun, {\it {$T\overline{T}$ deformation in SCFTs and
  integrable supersymmetric theories}},  {\em JHEP} {\bf 09} (2021) 082,
  [\href{http://arxiv.org/abs/2011.07618}{{\tt arXiv:2011.07618}}].

\bibitem{Caputa:2020lpa}
P.~Caputa, S.~Datta, Y.~Jiang, and P.~Kraus, {\it {Geometrizing $ T\overline{T}
  $}},  {\em JHEP} {\bf 03} (2021) 140,
  [\href{http://arxiv.org/abs/2011.04664}{{\tt arXiv:2011.04664}}].

\bibitem{He:2020qcs}
S.~He, {\it {Note on higher-point correlation functions of the $T\bar T$ or
  $J\bar T$ deformed CFTs}},  {\em Sci. China Phys. Mech. Astron.} {\bf 64}
  (2021), no.~9 291011, [\href{http://arxiv.org/abs/2012.06202}{{\tt
  arXiv:2012.06202}}].

\bibitem{Hirano:2020ppu}
S.~Hirano, T.~Nakajima, and M.~Shigemori, {\it {$ T\overline{T} $ Deformation
  of stress-tensor correlators from random geometry}},  {\em JHEP} {\bf 04}
  (2021) 270, [\href{http://arxiv.org/abs/2012.03972}{{\tt arXiv:2012.03972}}].

\bibitem{Ebert:2022cle}
S.~Ebert, E.~Hijano, P.~Kraus, R.~Monten, and R.~M. Myers, {\it {Field Theory
  of Interacting Boundary Gravitons}},
  \href{http://arxiv.org/abs/2201.01780}{{\tt arXiv:2201.01780}}.

\bibitem{Gross:2019ach}
D.~J. Gross, J.~Kruthoff, A.~Rolph, and E.~Shaghoulian, {\it {$T\overline{T}$
  in AdS$_2$ and Quantum Mechanics}},  {\em Phys. Rev. D} {\bf 101} (2020),
  no.~2 026011, [\href{http://arxiv.org/abs/1907.04873}{{\tt
  arXiv:1907.04873}}].

\bibitem{Gross:2019uxi}
D.~J. Gross, J.~Kruthoff, A.~Rolph, and E.~Shaghoulian, {\it {Hamiltonian
  deformations in quantum mechanics, $T\bar T$, and the SYK model}},  {\em
  Phys. Rev. D} {\bf 102} (2020), no.~4 046019,
  [\href{http://arxiv.org/abs/1912.06132}{{\tt arXiv:1912.06132}}].

\bibitem{Iliesiu:2020zld}
L.~V. Iliesiu, J.~Kruthoff, G.~J. Turiaci, and H.~Verlinde, {\it {JT gravity at
  finite cutoff}},  \href{http://arxiv.org/abs/2004.07242}{{\tt
  arXiv:2004.07242}}.

\bibitem{Stanford:2020qhm}
D.~Stanford and Z.~Yang, {\it {Finite-cutoff JT gravity and self-avoiding
  loops}},  \href{http://arxiv.org/abs/2004.08005}{{\tt arXiv:2004.08005}}.

\bibitem{Rosso:2020wir}
F.~Rosso, {\it {$T\overline{T}$ deformation of random matrices}},  {\em Phys.
  Rev. D} {\bf 103} (2021), no.~12 126017,
  [\href{http://arxiv.org/abs/2012.11714}{{\tt arXiv:2012.11714}}].

\bibitem{Griguolo:2021wgy}
L.~Griguolo, R.~Panerai, J.~Papalini, and D.~Seminara, {\it {Nonperturbative
  effects and resurgence in JT gravity at finite cutoff}},
  \href{http://arxiv.org/abs/2106.01375}{{\tt arXiv:2106.01375}}.

\bibitem{Ebert:2022xfh}
S.~Ebert, C.~Ferko, H.-Y. Sun, and Z.~Sun, {\it {$T \overline{T}$ Deformations
  of Supersymmetric Quantum Mechanics}},
  \href{http://arxiv.org/abs/2204.05897}{{\tt arXiv:2204.05897}}.

\bibitem{Ebert:2022ehb}
S.~Ebert, C.~Ferko, H.-Y. Sun, and Z.~Sun, {\it {$T\bar{T}$ in JT Gravity and
  BF Gauge Theory}},  \href{http://arxiv.org/abs/2205.07817}{{\tt
  arXiv:2205.07817}}.

\bibitem{Grumiller:2007ju}
D.~Grumiller and R.~McNees, {\it {Thermodynamics of black holes in two (and
  higher) dimensions}},  {\em JHEP} {\bf 04} (2007) 074,
  [\href{http://arxiv.org/abs/hep-th/0703230}{{\tt hep-th/0703230}}].

\bibitem{Grumiller:2020fbb}
D.~Grumiller and R.~McNees, {\it {Universal flow equations and chaos bound
  saturation in 2d dilaton gravity}},  {\em JHEP} {\bf 01} (2021) 112,
  [\href{http://arxiv.org/abs/2007.03673}{{\tt arXiv:2007.03673}}].

\bibitem{He:2022bbb}
S.~He, H.~Ouyang, and Y.~Sun, {\it {Note on $T\bar{T}$ deformed matrix models
  and JT supergravity duals}},  \href{http://arxiv.org/abs/2204.13636}{{\tt
  arXiv:2204.13636}}.

\bibitem{Forrester:1993vtx}
P.~J. Forrester, {\it {The spectrum edge of random matrix ensembles}},  {\em
  Nucl. Phys. B} {\bf 402} (1993) 709--728.

\bibitem{Ginsparg:1993is}
P.~H. Ginsparg and G.~W. Moore, {\it {Lectures on 2-D gravity and 2-D string
  theory}},  in {\em {Theoretical Advanced Study Institute (TASI 92): From
  Black Holes and Strings to Particles}}, pp.~277--469, 10, 1993.
\newblock \href{http://arxiv.org/abs/hep-th/9304011}{{\tt hep-th/9304011}}.

\bibitem{airy1838intensity}
G.~B. Airy, {\it On the intensity of light in the neighbourhood of a caustic},
  {\em Trans. Cambridge Philos. Soc.} {\bf 6} (1838) 379.

\bibitem{Mirzakhani:2006eta}
M.~Mirzakhani, {\it {Weil-Petersson volumes and intersection theory on the
  moduli space of curves}},  {\em J. Am. Math. Soc.} {\bf 20} (2007), no.~01
  1--24.

\bibitem{Dijkgraaf:2018vnm}
R.~Dijkgraaf and E.~Witten, {\it {Developments in Topological Gravity}},  {\em
  Int. J. Mod. Phys. A} {\bf 33} (2018), no.~30 1830029,
  [\href{http://arxiv.org/abs/1804.03275}{{\tt arXiv:1804.03275}}].

\bibitem{Eynard:2007fi}
B.~Eynard and N.~Orantin, {\it {Weil-Petersson volume of moduli spaces,
  Mirzakhani's recursion and matrix models}},
  \href{http://arxiv.org/abs/0705.3600}{{\tt arXiv:0705.3600}}.

\bibitem{Marolf:2020xie}
D.~Marolf and H.~Maxfield, {\it {Transcending the ensemble: baby universes,
  spacetime wormholes, and the order and disorder of black hole information}},
  {\em JHEP} {\bf 08} (2020) 044, [\href{http://arxiv.org/abs/2002.08950}{{\tt
  arXiv:2002.08950}}].

\bibitem{Alishahiha:2020jko}
M.~Alishahiha, A.~Faraji~Astaneh, G.~Jafari, A.~Naseh, and B.~Taghavi, {\it {On
  Free Energy for Deformed JT Gravity}},
  \href{http://arxiv.org/abs/2010.02016}{{\tt arXiv:2010.02016}}.

\bibitem{Mertens:2019tcm}
T.~G. Mertens and G.~J. Turiaci, {\it {Defects in Jackiw-Teitelboim Quantum
  Gravity}},  {\em JHEP} {\bf 08} (2019) 127,
  [\href{http://arxiv.org/abs/1904.05228}{{\tt arXiv:1904.05228}}].

\bibitem{Maxfield:2020ale}
H.~Maxfield and G.~J. Turiaci, {\it {The path integral of 3D gravity near
  extremality; or, JT gravity with defects as a matrix integral}},
  \href{http://arxiv.org/abs/2006.11317}{{\tt arXiv:2006.11317}}.

\bibitem{Witten:2020ert}
E.~Witten, {\it {Deformations of JT Gravity and Phase Transitions}},
  \href{http://arxiv.org/abs/2006.03494}{{\tt arXiv:2006.03494}}.

\bibitem{Witten:2020wvy}
E.~Witten, {\it {Matrix Models and Deformations of JT Gravity}},
  \href{http://arxiv.org/abs/2006.13414}{{\tt arXiv:2006.13414}}.

\bibitem{Okuyama:2019xvg}
K.~Okuyama, {\it {Replica symmetry breaking in random matrix model: a toy model
  of wormhole networks}},  {\em Phys. Lett. B} {\bf 803} (2020) 135280,
  [\href{http://arxiv.org/abs/1903.11776}{{\tt arXiv:1903.11776}}].

\bibitem{Johnson:2020exp}
C.~V. Johnson, {\it {Explorations of Non-Perturbative JT Gravity and
  Supergravity}},  \href{http://arxiv.org/abs/2006.10959}{{\tt
  arXiv:2006.10959}}.

\bibitem{Johnson:2020heh}
C.~V. Johnson, {\it {JT Supergravity, Minimal Strings, and Matrix Models}},
  \href{http://arxiv.org/abs/2005.01893}{{\tt arXiv:2005.01893}}.

\bibitem{Johnson:2020lns}
C.~V. Johnson and F.~Rosso, {\it {Solving Puzzles in Deformed JT Gravity: Phase
  Transitions and Non-Perturbative Effects}},  {\em JHEP} {\bf 04} (2021) 030,
  [\href{http://arxiv.org/abs/2011.06026}{{\tt arXiv:2011.06026}}].

\bibitem{Johnson:2021rsh}
C.~V. Johnson, {\it {On the Quenched Free Energy of JT Gravity and
  Supergravity}},  \href{http://arxiv.org/abs/2104.02733}{{\tt
  arXiv:2104.02733}}.

\bibitem{Johnson:2021tnl}
C.~V. Johnson, {\it {Consistency Conditions for Non-Perturbative Completions of
  JT Gravity}},  \href{http://arxiv.org/abs/2112.00766}{{\tt
  arXiv:2112.00766}}.

\bibitem{Stanford:2019vob}
D.~Stanford and E.~Witten, {\it {JT gravity and the ensembles of random matrix
  theory}},  {\em Adv. Theor. Math. Phys.} {\bf 24} (2020), no.~6 1475--1680,
  [\href{http://arxiv.org/abs/1907.03363}{{\tt arXiv:1907.03363}}].

\bibitem{Coleman:2019dvf}
E.~A. Coleman, J.~Aguilera-Damia, D.~Z. Freedman, and R.~M. Soni, {\it {$
  T\overline{T} $ -deformed actions and (1,1) supersymmetry}},  {\em JHEP} {\bf
  10} (2019) 080, [\href{http://arxiv.org/abs/1906.05439}{{\tt
  arXiv:1906.05439}}].

\bibitem{Okuyama:2019xbv}
K.~Okuyama and K.~Sakai, {\it {JT gravity, KdV equations and macroscopic loop
  operators}},  {\em JHEP} {\bf 01} (2020) 156,
  [\href{http://arxiv.org/abs/1911.01659}{{\tt arXiv:1911.01659}}].

\bibitem{gradshteyn2014table}
I.~S. Gradshteyn and I.~M. Ryzhik, {\em Table of integrals, series, and
  products}.
\newblock Academic press, 2014.

\bibitem{Faber:2000ma}
C.~Faber and R.~Pandharipande, {\it {Logarithmic series and hodge integrals in
  the tautological ring (with an appendix by D. Zagier)}},
  \href{http://arxiv.org/abs/math/0002112}{{\tt math/0002112}}.

\bibitem{Moore:1990cn}
G.~W. Moore, {\it {Matrix models of 2-D gravity and isomonodromic
  deformation}},  {\em Prog. Theor. Phys. Suppl.} {\bf 102} (1990) 255--286.

\bibitem{Tracy:1992kc}
C.~A. Tracy and H.~Widom, {\it {Level spacing distributions and the Airy
  kernel}},  {\em Phys. Lett. B} {\bf 305} (1993) 115--118,
  [\href{http://arxiv.org/abs/hep-th/9210074}{{\tt hep-th/9210074}}].

\bibitem{Tracy:1992rf}
C.~A. Tracy and H.~Widom, {\it {Level spacing distributions and the Airy
  kernel}},  {\em Commun. Math. Phys.} {\bf 159} (1994) 151--174,
  [\href{http://arxiv.org/abs/hep-th/9211141}{{\tt hep-th/9211141}}].

\bibitem{Banerjee:2019ewu}
A.~Banerjee, A.~Bhattacharyya, and S.~Chakraborty, {\it {Entanglement Entropy
  for $TT$ deformed CFT in general dimensions}},  {\em Nucl. Phys. B} {\bf 948}
  (2019) 114775, [\href{http://arxiv.org/abs/1904.00716}{{\tt
  arXiv:1904.00716}}].

\bibitem{burden1997numerical}
R.~L. Burden and J.~D. Faires, {\it Numerical analysis, 9th editions, brooks},
  {\em Cole, Thomson Learning Inc} {\bf 14} (1997) 190--192.

\bibitem{Johnson:2022wsr}
C.~V. Johnson, {\it {The Microstate Physics of JT Gravity and Supergravity}},
  \href{http://arxiv.org/abs/2201.11942}{{\tt arXiv:2201.11942}}.

\bibitem{longman1960method}
I.~Longman, {\it A method for the numerical evaluation of finite integrals of
  oscillatory functions},  {\em Mathematics of Computation} {\bf 14} (1960),
  no.~69 53--59.

\bibitem{ooura1991double}
T.~Ooura and M.~Mori, {\it The double exponential formula for oscillatory
  functions over the half infinite interval},  {\em Journal of Computational
  and Applied Mathematics} {\bf 38} (1991), no.~1-3 353--360.

\bibitem{abramowitz1964handbook}
M.~Abramowitz and I.~A. Stegun, {\em Handbook of mathematical functions with
  formulas, graphs, and mathematical tables}, vol.~55.
\newblock US Government printing office, 1964.

\end{thebibliography}\endgroup


\providecommand{\href}[2]{#2}\begingroup\raggedright\endgroup

\end{document}